\documentstyle[12pt,a4wide,artcustom,epsf,rotate]{article}

\newcommand{\dS}{\Delta S=1}
\newcommand{\dB}{\Delta B=1}
\newcommand{\dC}{\Delta C=1}
\newcommand{\dI}{\Delta I=1/2}
\newcommand{\dIth}{\Delta I=3/2}

\newcommand{\ord}{{\cal O}}

\newcommand{\as}{\alpha_{\rm s}}
\newcommand{\aem}{\alpha}

\newcommand{\hU}{\hat{U}}
\newcommand{\hR}{\hat{R}}
\newcommand{\hJ}{\hat{J}}
\newcommand{\hV}{\hat{V}}
\newcommand{\hK}{\hat{K}}
\newcommand{\hM}{\hat{M}}

\newcommand{\gf}{\gamma_5}

\newcommand{\hg}{\hat{\gamma}}
\newcommand{\hG}{\hat{\Gamma}}

\newcommand{\ndr}{{\rm NDR}}
\newcommand{\ndrb}{\overline{{\rm NDR}}}

\newcommand{\gs}{\hat{\gamma}_{\rm s}^{(0)}}
\newcommand{\gem}{\hat{\gamma}_{\rm e}^{(0)}}

\newcommand{\gss}{\hat{\gamma}_{\rm s}^{(1)}}
\newcommand{\gse}{\hat{\gamma}_{\rm se}^{(1)}}

\newcommand{\gst}{\hat{\gamma}_{\rm s}^{(0) T}}
\newcommand{\gemt}{\hat{\gamma}_{\rm e}^{(0) T}}
\newcommand{\gsst}{\hat{\gamma}_{\rm s}^{(1) T}}
\newcommand{\gset}{\hat{\gamma}_{\rm se}^{(1) T}}

\newcommand{\vQ}{\vec{Q}}
\newcommand{\vC}{\vec{C}}

\newcommand{\eps}{\varepsilon}
\newcommand{\epe}{\varepsilon'/\varepsilon}

\newcommand{\re}{\hat{r}_{\rm e}}
\newcommand{\rs}{\hat{r}_{\rm s}}
\newcommand{\dre}{\Delta\hat{r}_{\rm e}}
\newcommand{\drs}{\Delta\hat{r}_{\rm s}}

\newcommand{\hJe}{\hat{J}_{\rm e}}
\newcommand{\hJs}{\hat{J}_{\rm s}}
\newcommand{\hJse}{\hat{J}_{\rm se}}
\newcommand{\vardre}{\delta\hat{r}_{\rm e}}
\newcommand{\vardrs}{\delta\hat{r}_{\rm s}}

\newcommand{\mt}{m_{\rm t}}
\newcommand{\mb}{m_{\rm b}}
\newcommand{\mc}{m_{\rm c}}
\newcommand{\ms}{m_{\rm s}}
\newcommand{\mup}{m_{\rm u}}
\newcommand{\md}{m_{\rm d}}
\newcommand{\mw}{M_{\rm W}}

\newcommand{\gev}{\; {\rm GeV}}
\newcommand{\mev}{\; {\rm MeV}}

\newcommand{\Heff}{{\cal H}_{\rm eff}}
\newcommand{\Lf}{\Lambda_f}
\newcommand{\Lms}{\Lambda_{\overline{\rm MS}}}

\newcommand{\V}[1]{V_{\rm #1}}

\newcommand{\RE}{{\rm Re}}
\newcommand{\IM}{{\rm Im}}

\newcommand{\Kpipi}{K \rightarrow \pi\pi}

\newcommand{\gsim}{~{}_{\textstyle\sim}^{\textstyle >}~}

\newcommand{\svs}{\vbox{\vskip 5mm}}
\newcommand{\mvs}{\vbox{\vskip 8mm}}

\newcommand{\nn}{\nonumber}
\newcommand{\eqn}[1]{(\ref{#1})}

\newcommand{\newsection}[1]{\section{#1}\setcounter{equation}{0}}

\renewcommand{\baselinestretch}{1.2}

\begin{document}


\paperid{
{\sf MPI-Ph/93-11}\\
{\sf CERN-TH-6821/93}\\
{\sf TUM-T31-35/93}
}

\date{\sf March 1993}

\author{\\
{\normalsize Andrzej J. BURAS${}^{1,2}$, Matthias JAMIN${}^{3}$
and Markus E. LAUTENBACHER${}^{1}$}\\
\ \\
{\small\sl ${}^{1}$ Physik Department, Technische Universit\"at
M\"unchen,
                    D-8046 Garching, FRG.}\\
{\small\sl ${}^{2}$ Max-Planck-Institut f\"ur Physik
                    -- Werner-Heisenberg-Institut,}\\
{\small\sl P.O. Box 40 12 12, D-8000 M\"unchen, FRG.}\\
{\small\sl ${}^{3}$ Division TH, CERN, 1211 Geneva 23, Switzerland.}}

\title{
{\LARGE\sf
The Anatomy of $\epe$ Beyond Leading Logarithms with Improved Hadronic
Matrix Elements}\footnote{Supported by the German Bundesministerium f\"ur
Forschung und Technologie under contract 06 TM 732 and by the CEC
Science project SC1-CT91-0729.}
}

\maketitle

\newpage
\thispagestyle{empty}
\begin{abstract}
\noindent
We use the recently calculated two--loop anomalous dimensions of
current-current operators, QCD and electroweak penguin operators to
construct the effective Hamiltonian for $\dS$ transitions beyond the
leading logarithmic approximation. We solve the renormalization group
equations involving $\as$ and $\aem$ up to two--loop level and we give
the numerical values of Wilson coefficient functions $C_i(\mu)$ beyond
the leading logarithmic approximation in various renormalization
schemes.  Numerical results for the Wilson coefficients in $\dB$ and
$\dC$ Hamiltonians are also given.  We discuss several aspects of
renormalization scheme dependence and demonstrate the scheme
independence of physical quantities.  We stress that the scheme
dependence of the Wilson coefficients $C_i(\mu)$ can only be cancelled
by the one present in the hadronic matrix elements $\langle Q_i(\mu)
\rangle$.  This requires also the calculation of $\ord(\aem)$
corrections to $\langle Q_i(\mu) \rangle$. We propose a new
semi-phenomenological approach to hadronic matrix elements which
incorporates the data for $CP$-conserving $\Kpipi$ amplitudes and
allows to determine the matrix elements of all $(V-A)\otimes (V-A)$
operators in any renormalization scheme. Our renormalization group
analysis of all hadronic matrix elements $\langle Q_i(\mu) \rangle$
reveals certain interesting features.  We compare critically our
treatment of these matrix elements with those given in the literature.
When matrix elements of dominant QCD penguin ($Q_6$) and electroweak
penguin ($Q_8$) operators are kept fixed the effect of next-to-leading
order corrections is to lower considerably $\epe$ in the
't~Hooft--Veltman (HV) renormalization scheme with a smaller effect
in the dimensional regularization scheme with anticommuting
$\gamma_{5}$ ($\ndr$). Taking $\mt=130\gev$, $\Lms=300\mev$ and
calculating $\langle Q_6 \rangle$ and $\langle Q_8 \rangle$ in the
$1/N$ approach with $\ms(1\gev)=175\mev$, we find in the $\ndr$ scheme
$\epe = (6.7 \pm 2.6)\times 10^{-4}$ in agreement with the experimental
findings of E731.  We point out however that the increase of $\langle
Q_6 \rangle$ by only a factor of two gives $\epe = (20.0 \pm 6.5)\times
10^{-4}$ in agreement with the result of NA31. The dependence of $\epe$
on $\Lms$, $\mt$ and $\langle Q_{6,8} \rangle$ is presented .A detailed
anatomy of various contributions and comparison with the analyses of
Rome and Dortmund groups are given.

\end{abstract}

\newpage

\setcounter{page}{1}
\renewcommand{\thepage}{\roman{page}}
\tableofcontents
\newpage

\renewcommand{\thepage}{\arabic{page}}
\setcounter{page}{1}


\newsection{Introduction}
The $CP$-violating ratio $\epe$ is governed by penguin contributions
\cite{gilman:79b,guberina:80}.  Until 1989 there had been a general
belief \cite{gilmanhagelin:83,burasgerard:88} that $\epe$ is mainly
described by the QCD--penguin contributions with the electroweak
penguins ($\gamma$- and $Z^0$-penguins) playing only a secondary role.
A new development in this field came in 1989 through the increase of
$\mt$ and the work of Flynn and Randall \cite{flynn:89} who pointed out
that for large $\mt$ the electroweak penguin contributions could become
important and, having opposite sign to QCD-penguin contributions, could
considerably suppress $\epe$. A detailed anatomy of $\epe$ in the
presence of a heavy top quark has been done by Buchalla, Harlander and
one of the authors \cite{buchallaetal:90}.  It has been found that
with increasing $\mt$ the electroweak penguins could indeed compete
with QCD penguins to cancel their contribution completely for $\mt
\approx \ord(220\gev)$ so that $\epe \approx 0$ like in superweak
theories. This rather surprising result has been confirmed in
subsequent years by Paschos and Wu \cite{paschos:91} and by Lusignoli,
Maiani, Martinelli and Reina \cite{lusignoli:92}.

On the experimental side, after heroic efforts on both sides of the
Atlantic the situation is not yet conclusive
\cite{barrwinstein:91,gibbons:93},

\begin{equation}
\RE(\epe) = \left\{
\begin{array}{ll}
(23 \pm 7)    \cdot 10^{-4} & {\rm NA31} \\
(7.4 \pm 6.0) \cdot 10^{-4} & {\rm E731}
\end{array} \right. \, .
\label{eq:1.1}
\end{equation}

While the result of NA31 clearly indicates a non-zero $\epe$, the value
of E731 is compatible with superweak theories in which $\epe \equiv 0$
\cite{wolfenstein:64}. From the point of view of the analyses quoted
above the NA31 result points toward a top quark mass of $\mt \approx
\ord(100\gev)$ whereas the E731 result points toward $\mt \gsim
\ord(150\gev)$. Hopefully, in the next five years the experimental
situation on $\epe$ will be clarified through the new measurements by
the two collaborations at the $10^{-4}$ level and by experiments at the
$\Phi$-factory in Frascati.

Since the outcome of the fight between QCD and electroweak penguins is
rather sensitive to various approximations used in
refs.~\cite{buchallaetal:90,paschos:91,lusignoli:92}, it is important
to improve the theoretical calculations of short distance (Wilson
coefficient functions $C_i(\mu)$) and also the estimates of long
distance (hadronic matrix elements $\langle Q_i(\mu) \rangle$)
contributions. In the present paper we will make progress on both parts
of the problem by calculating the Wilson coefficients beyond the
leading logarithmic approximation and by extracting several hadronic
matrix elements from the existing very accurate data on $CP$-conserving
$\Kpipi$ amplitudes.  Needless to say that many results presented here
are not only important for $\epe$ but can also be applied to all $\dS$,
$\dC$ and $\dB$ decay processes.

The present paper culminates our extensive studies of $\Delta F=1$
effective non-leptonic Hamiltonians beyond the leading
logarithmic approximation
\cite{burasweisz:90}--\nocite{burasetal:92a,burasetal:92b}\cite{burasetal:92c}.
In ref.~\cite{burasweisz:90} two--loop anomalous dimensions
$\ord(\as^2)$ of current-current operators have been calculated,
confirming the previous results of Altarelli et al.~\cite{altarelli:81}. This
calculation has been extended to include QCD penguin operators and
electroweak penguin operators in
ref.~\cite{burasetal:92a,burasetal:92b} where a two--loop $10\times 10$
anomalous dimension matrix $\ord(\as^2)$ has been presented.
Subsequently, the corresponding two--loop matrix of order $\ord(\aem\,\as)$
has been calculated \cite{burasetal:92c}. The latter is necessary for a
consistent treatment of electroweak penguin operators beyond the
leading logarithmic approximation.

One of the aims of the present paper is to construct the effective Hamiltonian
for $\dS$ transitions which incorporates the results of
refs.~\cite{burasweisz:90}--\nocite{burasetal:92a,burasetal:92b}
\cite{burasetal:92c}. This Hamiltonian will take the form
\begin{equation}
\Heff = \frac{G_F}{\sqrt{2}} \sum_{i=1}^{10} Q_i(\mu) C_i(\mu)
      \equiv \frac{G_F}{\sqrt{2}} \vQ(\mu)^T \vC(\mu)\,,
\label{eq:1.2}
\end{equation}
where the Wilson coefficients $C_i(\mu)$ include leading and
next-to-leading QCD corrections and leading order corrections in the
electromagnetic coupling constant $\aem$. Here $Q_{1,2}$ stand for
current-current operators, $Q_{3-6}$ for QCD penguin operators and
$Q_{7-10}$ for electroweak penguin operators. Explicit expressions for
these operators will be given in section~2. A truncated effective
Hamiltonian for $Q_{1-6}$ with $\aem=0$ has been presented already by
us in ref.~\cite{burasetal:92a}, where an application to the $\dI$ rule
and $\epe$ has been made. The present paper can be considered on one
hand as a generalization of ref.~\cite{burasetal:92a} to include
electroweak penguin operators. On the other hand it can be considered
as a generalization of the analyses
\cite{flynn:89,buchallaetal:90,paschos:91,lusignoli:92} to include
next-to-leading logarithmic effects. Whereas in
refs.~\cite{flynn:89,buchallaetal:90,paschos:91,lusignoli:92} the
logarithms $(\as t)^n$ and $\aem t (\as t)^n$ with $t=\ln(\mw^2/\mu^2)$
have been summed, the present analysis includes also the logarithms
$\as (\as t)^n$ and $\aem (\as t)^n$. Moreover, we improve the
treatment of hadronic matrix elements.

The central questions which one would like to address in a paper like
this one are as follows:
\begin{itemize}
\item
Do the complete next-to-leading corrections enhance or suppress the
Wilson coefficients of current-current operators ?
\item
Do the complete next-to-leading corrections enhance or suppress the
Wilson coefficient functions of QCD penguin operators ?
\item
Do these corrections enhance or suppress the Wilson coefficients of
electroweak penguin operators relative to the ones of QCD penguin
operators ?
\item
What is the impact of all these effects on $\epe$ and the $\dI$ rule ?
\item
How does $\epe$ depend on $\Lms$ ?
\end{itemize}

Unfortunately, the answers to these questions are complicated by the
fact that the coefficients $C_i(\mu)$ depend at the next-to-leading
level on the renormalization scheme for the operators $Q_i$, and only
after the hadronic matrix elements $\langle Q_i(\mu) \rangle$ have been
calculated at the same level, can this scheme dependence be removed.
This requires also the calculation of $\ord(\aem)$ corrections to
$\langle Q_i(\mu) \rangle$. Thus the size of next-to-leading order
corrections to $C_i(\mu)$ calculated in the $\ndr$ scheme
(anticommuting $\gf$ in $D \not= 4$ space-time dimensions) differs from
the corresponding corrections found in the HV scheme (non-anticommuting
$\gf$ in $D \not= 4$ space-time dimensions).  In order to cancel these
dependences one would have to calculate $\langle Q_i(\mu) \rangle$ in a
method sensitive to such a scheme dependence. To our knowledge only QCD
sum rules could be useful in this respect at present but future lattice
calculations also could overcome this difficulty, at least in
principle. In spite of this our short distance calculation provides an
important and necessary step towards a more accurate estimate of weak
decay amplitudes than the one given merely by the leading logarithmic
approximation. Moreover, in order to consistently study the
$\mt$-dependence of weak amplitudes it is mandatory to go beyond the
leading logarithmic approximation as will be evident from what
follows.

The scheme dependence just discussed is an important point which we
would like to address here. To this end we will evaluate the Wilson
coefficient functions in three different schemes and we will investigate
how the scheme dependence affects the answers to the set of questions
stated above if no proper scheme dependence is incorporated in the
matrix elements $\langle Q_i(\mu) \rangle$. In particular, we find as a
rather unexpected result that the next-to-leading QCD corrections to
Wilson coefficients of current-current operators $Q_1$ and $Q_2$
calculated in the renormalization schemes considered in fact {\em suppress}
slightly the $\dI$ transitions and {\em enhance} the $\Delta I = \frac{3}{2}$
transitions. We explain why the authors of refs.~\cite{burasweisz:90}
and \cite{altarelli:81} who performed correct calculations reached
conclusions opposite to ours.

Other not unrelated issues are the $\mu$-dependence of $\langle
Q_i(\mu) \rangle$ and the actual values of these matrix elements.  For
this reason we have made a critical comparison of the existing
estimates of $\langle Q_i(\mu) \rangle$ based on the vacuum insertion,
the $1/N$-expansion and lattice methods. In this connection, we point
out that the evolution of $\langle Q_i(\mu) \rangle$ with $\mu$ can be
calculated within renormalization group improved perturbation theory.
Performing to our knowledge the first detailed renormalization group
analysis of $\langle Q_i(\mu) \rangle$ we find that due to the
asymmetry of the anomalous dimension matrices the $\mu$-dependences of
$\langle Q_i(\mu) \rangle$ are quite different from the ones of
$C_i(\mu)$. This allowed us to answer the following question:

\begin{itemize}
\item
Should the parameters $B_i$ entering various $\langle Q_i(\mu) \rangle$
be $\mu$-dependent in order to be consistent with the true QCD
evolution ?
\end{itemize}

Rather unexpectedly, we have found that the $B_i$ parameters of {\em
all} $(V-A)\otimes (V+A)$ operators $Q_5$ -- $Q_8$ show only very weak
$\mu$-dependence for $\mu \gsim 1\gev$ supporting to some extent the
structure of these matrix elements which is common to the three
approaches in question. On the other hand the $B_i$ parameters of
$(V-A)\otimes (V-A)$ operators show sizeable $\mu$-dependence which is
fully missing in the vacuum insertion approach, partly present in the
$1/N$ approach and not yet calculable in the lattice approach.

In view of the latter problem and the fact that the present methods are
not yet capable of addressing properly the renormalization scheme
dependence of $\langle Q_i(\mu) \rangle$, nor fully explain the known
data on $CP$-conserving $\Kpipi$ decays, we have developed a
phenomenological framework in which the matrix elements $\langle
Q_i(\mu) \rangle$ necessary for the calculation of $\epe$ are extracted
as far as possible from the data on $CP$-conserving $\Kpipi$ decays. In
this framework the central role is played by the calculable scheme and
$\mu$-dependent coefficients $C_i(\mu)$ and the scheme and $\mu$-{\em
independence} of measured $\Kpipi$ amplitudes gives us automatically
the scheme and $\mu$-dependence of $\langle Q_i(\mu) \rangle$ which are
in addition consistent with the $CP$-conserving $\Kpipi$ data. Thus the
$\dI$ rule is incorporated in our approach from the beginning. We
verify that the scheme and $\mu$-dependences of $\langle Q_i(\mu)
\rangle$ found this way are to a very good approximation consistent
with the ones given by the renormalization group.

Our approach has the four basic parameters
\begin{equation}
\Lms \, , \qquad
B_{2}^{(1/2)}(\mc) \, , \qquad
B_{6}^{(1/2)}(\mc) \, , \qquad
B_{8}^{(3/2)}(\mc) \, . \qquad \label{eq:1.3}
\end{equation}
The choice $\mu = \mc$ turns out to be very convenient but is not
necessary.  The last three parameterize hadronic matrix elements for
$\Kpipi$, and will be defined in section~5.2. $B_2^{(1/2)}(\mc)$
parameterizes all matrix elements $\langle Q_i \rangle_0$ of
$(V-A)\otimes (V-A)$ operators. $B_6^{(1/2)}(\mc)$ and
$B_8^{(3/2)}(\mc)$ parameterize the $(V-A)\otimes (V+A)$ operators.  In
our phenomenological approach all matrix elements are given in terms of
these four parameters in such a way that for each set of these
parameters all data on $CP$-conserving $\Kpipi$ amplitudes are
reproduced. Moreover, all the matrix elements $\langle Q_i(\mu)
\rangle_2$ of $(V-A)\otimes(V-A)$ operators $(i=1,2,9,10)$ are given in
our approach as numerical functions of only $\Lms$, $\mu$ and the
renormalization scheme considered. With all these matrix elements at
hand we can predict the $CP$-violating quantity
\begin{equation}
\frac{\eps'}{\eps} \; = \; 10^{-4}\,\left[\frac{\IM\lambda_{\rm t}}
{1.7\cdot 10^{-4}}\right]\, \left[\,P^{(1/2)}-P^{(3/2)}\,\right] \, ,
\label{eq:1.4}
\end{equation}
where $\lambda_{\rm t}=\V{ts}^* \V{td}$ and
\begin{eqnarray}
P^{(1/2)} & = & a_0^{(1/2)} + a_2^{(1/2)}\,B_2^{(1/2)} +
                  a_6^{(1/2)}\,B_6^{(1/2)} \, ,
\label{eq:1.5} \\
\mvs
P^{(3/2)} & = & a_0^{(3/2)} + a_8^{(3/2)}\,B_8^{(3/2)} \, .
\label{eq:1.6}
\end{eqnarray}
We calculate the coefficients $a_i$ as functions of $\Lms$ and $\mt$
for the leading order and the two renormalization schemes ($\ndr$, HV)
considered.

We hope very much that this novel anatomy of various aspects of $\epe$
including next-to-leading order corrections and an improved treatment
of hadronic matrix elements will shed some light on the present
uncertainties and on the theoretical prediction of $\epe$ in the
Standard Model (SM). By doing this we will attempt to clarify the
discrepancy between recent theoretical estimates of $\epe$ which bears
some similarity in it to the discrepancy concerning experimental
results on $\epe$ given in eq.~\eqn{eq:1.1}. Whereas the Dortmund group
\cite{froehlich:91,heinrichetal:92} found last year $\epe$ to be close
to the NA31 result, the very recent analysis of the Rome group
\cite{ciuchini:92} favours the experimental result of E731.

Our paper is organized as follows: In section~2, we will present
general formulae for $C_i(\mu)$ beyond the leading logarithmic
approximation.  These formulae generalize the ones given in
refs.~\cite{burasetal:92a} and \cite{buchallaetal:90}. In section~3,
the renormalization scheme dependence will be discussed. In section~4,
we calculate the coefficients $C_i(\mw)$ and use the known one--loop
\cite{guberina:80},
\cite{gaillard:74}--\nocite{altarelli:74,vainshtein:77,gilman:79,
guberina:80,bijnenswise:84,burasgerard:87,sharpe:87}\cite{lusignoli:89}
and two--loop anomalous dimension matrices
\cite{burasweisz:90}--\nocite{burasetal:92a,burasetal:92b}\cite{burasetal:92c}
to find $C_i(\mu)$ in various renormalization schemes.  In section~5,
we collect and compare various existing theoretical estimates of
hadronic matrix elements. In section~6, we present the details of our
phenomenological approach to the matrix elements, we calculate the
matrix elements as functions of the parameters in eq.~\eqn{eq:1.3} and
compare  the results with the ones obtained in section~5. Here, we also
present the renormalization group evolution of $\langle Q_i(\mu)
\rangle$ advertised earlier in this paper. In section~7, we recall the
CKM-parameters necessary for our numerical analysis of $\epe$ and we
give the $\Delta S=2$ Hamiltonian which we used to determine the
CKM-phase $\delta$ from the experimental value of the parameter $\eps$
describing indirect $CP$-violation in the $K$-system. In section~8, we
give the basic expressions for $\epe$ in terms of $C_i(\mu)$ and
$\langle Q_i(\mu) \rangle$, and we address all the issues mentioned
above. In particular we give the results for $\epe$ as a function of
$\Lms$, $\mt$ and the matrix elements $\langle Q_6 \rangle_0$ and
$\langle Q_8 \rangle_2$. For completeness, we give in section~9, the
numerical values for $C_i(\mu)$ relevant for $\dB$ transitions. We end
our paper with a list of main messages coming from our new anatomy of
$\epe$.

\newpage

\newsection{Basic Formulae for Wilson Coefficient Functions}
\subsection{Operators}
In order to illustrate the scheme dependence of Wilson coefficient
functions we will use two bases of operators. Our main basis is given
as follows
\begin{eqnarray}
Q_{1} & = & \left( \bar s_{\alpha} u_{\beta}  \right)_{\rm V-A}
            \left( \bar u_{\beta}  d_{\alpha} \right)_{\rm V-A}
\, , \nn \\
Q_{2} & = & \left( \bar s u \right)_{\rm V-A}
            \left( \bar u d \right)_{\rm V-A}
\, , \nn \\
Q_{3} & = & \left( \bar s d \right)_{\rm V-A}
   \sum_{q} \left( \bar q q \right)_{\rm V-A}
\, , \nn \\
Q_{4} & = & \left( \bar s_{\alpha} d_{\beta}  \right)_{\rm V-A}
   \sum_{q} \left( \bar q_{\beta}  q_{\alpha} \right)_{\rm V-A}
\, , \nn \\
Q_{5} & = & \left( \bar s d \right)_{\rm V-A}
   \sum_{q} \left( \bar q q \right)_{\rm V+A}
\, , \nn \\
Q_{6} & = & \left( \bar s_{\alpha} d_{\beta}  \right)_{\rm V-A}
   \sum_{q} \left( \bar q_{\beta}  q_{\alpha} \right)_{\rm V+A}
\, , \label{eq:2.1n} \\
Q_{7} & = & \frac{3}{2} \left( \bar s d \right)_{\rm V-A}
         \sum_{q} e_{q} \left( \bar q q \right)_{\rm V+A}
\, , \nn \\
Q_{8} & = & \frac{3}{2} \left( \bar s_{\alpha} d_{\beta} \right)_{\rm V-A}
         \sum_{q} e_{q} \left( \bar q_{\beta}  q_{\alpha}\right)_{\rm V+A}
\, , \nn \\
Q_{9} & = & \frac{3}{2} \left( \bar s d \right)_{\rm V-A}
         \sum_{q} e_{q} \left( \bar q q \right)_{\rm V-A}
\, , \nn \\
Q_{10}& = & \frac{3}{2} \left( \bar s_{\alpha} d_{\beta} \right)_{\rm V-A}
         \sum_{q} e_{q} \left( \bar q_{\beta}  q_{\alpha}\right)_{\rm V-A}
\, , \nn
\end{eqnarray}
where $\alpha$, $\beta$ denote colour indices ($\alpha,\beta
=1,\ldots,N$) and $e_{q}$  are quark charges. We omit the colour
indices for the colour singlet operators. $(V\pm A)$ refer to
$\gamma_{\mu} (1 \pm \gf)$. This basis closes  under QCD and QED
renormalization.

For $\mu < \mc$ the sums over quark flavours in \eqn{eq:2.1n} run over
$u$, $d$ and $s$. However, when  $\mb > \mu > \mc$ is considered also
$q=c$ has to be included. Moreover, in this case two additional
current--current operators have to be taken into account,
\begin{equation}
Q_1^c = \left(\bar s_\alpha c_\beta  \right)_{\rm V-A}
        \left(\bar c_\beta  d_\alpha \right)_{\rm V-A}
\, , \qquad
Q_2^c = \left(\bar s c \right)_{\rm V-A}
        \left(\bar c d \right)_{\rm V-A} \, .
\label{eq:2.2n}
\end{equation}

It should be stressed that this basis differs from the one used by
Gilman and Wise \cite{gilman:79} and by the Rome group
\cite{ciuchini:92} in the form for $Q_1$, $Q_2$, $Q_1^c$ and $Q_2^c$
where the corresponding Fierz conjugates have been adopted
\begin{equation}
\widetilde{Q}_1 = \left(\bar s d \right)_{\rm V-A}
                  \left(\bar u u \right)_{\rm V-A}
\, , \qquad
\widetilde{Q}_2 = \left(\bar s_\alpha d_\beta  \right)_{\rm V-A}
                  \left(\bar u_\beta  u_\alpha \right)_{\rm V-A}
\, ,
\label{eq:2.3n}
\end{equation}
\begin{equation}
\widetilde{Q}_1^c = \left(\bar s d \right)_{\rm V-A}
                    \left(\bar c c \right)_{\rm V-A}
\, , \qquad
\widetilde{Q}_2^c = \left(\bar s_\alpha d_\beta  \right)_{\rm V-A}
                    \left(\bar c_\beta  c_\alpha \right)_{\rm V-A}
\, .
\label{eq:2.4n}
\end{equation}
We will refer to the basis of refs.~\cite{gilman:79,ciuchini:92} as the
second basis. Both are equally good although we somewhat prefer the
basis \eqn{eq:2.1n}, because there $Q_2$ is taken in the colour singlet
form as it appears in the tree level Hamiltonian.

Now, as we have stressed in
refs.~\cite{burasetal:92a,burasetal:92b,burasetal:92c} the two--loop
anomalous dimension matrices in the two bases differ when calculated in
the $\ndr$ scheme. Consequently, the corresponding coefficients $C_i(\mu)$
differ. We will refer to these two possibilities as schemes $\ndr$ and
$\ndrb$, respectively. On the other hand  in the HV scheme it is
immaterial whether the first or the second basis is used but the
corresponding HV Wilson coefficient functions differ from those obtained
in the $\ndr$ and $\ndrb$ schemes.

In what follows we will present general formulae for Wilson coefficients
valid in any scheme leaving the discussion of scheme dependence to
section~3.

We should also remark that we do not include in our analysis the
operators
\begin{equation}
\widetilde{Q}_{11} = \left(\bar s d \right)_{\rm V-A}
                     \left(\bar b b \right)_{\rm V-A}
\, , \qquad
          {Q}_{12} = \left(\bar s b \right)_{\rm V-A}
                     \left(\bar b d \right)_{\rm V-A} \, ,
\label{eq:2.5n}
\end{equation}
which have been taken into account in ref.~\cite{buchallaetal:90,paschos:91}.
Even at leading order their effect on $\epe$ is smaller than 1\% and in
order not to complicate the analysis further,we have decided to drop them.
This was also the strategy of the Rome group \cite{ciuchini:92}.

\subsection{Renormalization Group Equations}
The renormalization group equation for $\vC(\mu)$ is given by
\begin{equation}
\left[ \mu \frac{\partial}{\partial\mu} +
       \beta(g) \frac{\partial}{\partial g} \right]
\vC(\frac{\mw^2}{\mu^2},g^2,\aem) =
\hg^T(g^2,\aem) \; \vC(\frac{\mw^2}{\mu^2},g^2,\aem) \, ,
\label{eq:2.1}
\end{equation}
where $\beta(g)$ is the QCD beta function
\begin{equation}
\beta(g) =
   -\,\beta_0 \frac{g^3}{16\pi^2} - \beta_1 \frac{g^5}{(16\pi^2)^2}
   - \beta_{1e} \frac{e^2 \; g^3}{(16\pi^2)^2} \, ,
\label{eq:2.2}
\end{equation}
with
\begin{equation}
\beta_0 = 11 - \frac{2}{3} \, f \, , \qquad
\beta_1 = 102 - \frac{38}{3} \, f \, , \qquad
\beta_{1e} = -\,\frac{8}{9}\Big(u+\frac{d}{4}\Big) \, ,
\label{eq:2.3}
\end{equation}
and $f=u+d$ denoting the number of active flavours, $u$ and $d$ being
the number of $u$-type and $d$-type flavours respectively. To the order
considered in this paper, in principle we also have to take into
account the third term in eq.~\ref{eq:2.3}. However, we have checked
that its contribution to the coefficient functions is negligible, and
we will drop it for the rest of this work.

In what follows, we also neglect the running of the electromagnetic
coupling constant $\aem$. This is a very good approximation because only
scales $1\gev \leq \mu \leq \mw$ are involved in our analysis. In the
numerical analysis we will take $\aem = 1/128$. For the effective QCD
coupling constant we will use
\begin{equation}
\as^{(f)}(Q) = \frac{4\pi}{\beta_0 \ln(Q^2/\Lf^2)}
\left[ 1 - \frac{\beta_1}{\beta_0^2} \frac{\ln\ln(Q^2/\Lf^2)}{\ln(Q^2/\Lf^2)}
\right] \, ,
\label{eq:2.4}
\end{equation}
with $\Lf$ in the $\overline{\rm MS}$ scheme. Demanding the continuity
of $\as^{(f)}(Q)$ at quark thresholds in the form
\begin{equation}
\as^{(3)}(\mc) = \as^{(4)}(\mc) \, ,
\qquad
\as^{(4)}(\mb) = \as^{(5)}(\mb) \, ,
\label{eq:2.4b}
\end{equation}
gives relations between the various values for $\Lf$. In what follows we
will denote $\Lms \equiv \Lambda_4 \equiv \Lambda_{QCD}$.

Next, $\hg(g^2,\aem)$ is the full $10\times 10$ anomalous dimension matrix
which we expand in the following way
\begin{equation}
\hg(g^2,\aem) = \hg_{\rm s}(g^2) + \frac{\aem}{4\pi} \hG(g^2) \, ,
\label{eq:2.5}
\end{equation}
where
\begin{equation}
\hg_{\rm s}(g^2) = \frac{\as}{4\pi} \gs + \frac{\as^2}{(4\pi)^2} \gss  \, ,
\label{eq:2.6}
\end{equation}
and
\begin{equation}
\hG(g^2) = \gem + \frac{\as}{4\pi} \gse  \, .
\label{eq:2.7}
\end{equation}

Explicit expressions for $(\gs, \gss)$ and $(\gem, \gse)$ can be found in
refs.~\cite{burasetal:92b} and \cite{burasetal:92c}, respectively. They
will not be repeated here.

The solution of the renormalization group equation \eqn{eq:2.1} is given
by
\begin{equation}
\vC(\frac{\mw^2}{\mu^2},g^2,\aem) =
\left[ T_g \exp \int_{g(\mw)}^{g(\mu)} \!\! dg' \;
       \frac{\hg^T(g'^2,\aem)}{\beta(g')}
\right]
\vC(1,g^2(\mw),\aem) \, ,
\label{eq:2.8}
\end{equation}
where $T_g$ denotes the ordering in the QCD coupling constant such that the
couplings increase from right to left.

\subsection{Generalized Evolution Matrix}
Let us next introduce a generalized evolution matrix from $m_2$ down to
$m_1 < m_2$
\begin{equation}
\hU(m_1,m_2,\aem) \equiv T_g \exp
\int_{g(m_2)}^{g(m_1)} \!\! dg' \; \frac{\hg^T(g'^2,\aem)}{\beta(g')} \, ,
\label{eq:2.9}
\end{equation}
which is the generalization of eq.~(2.3) of ref.~\cite{burasetal:92a}.
Eq.~\eqn{eq:2.8} can be now in a short hand notation written as
\begin{equation}
\vC(\mu) =
\hU(\mu,\mw,\aem) \vC(\mw) \, .
\label{eq:2.10}
\end{equation}
The matrix $\hU(m_1,m_2,\aem)$ can be decomposed as follows
\begin{equation}
\hU(m_1,m_2,\aem) =
\hU(m_1,m_2) + \frac{\aem}{4\pi} \hR(m_1,m_2) \, ,
\label{eq:2.11}
\end{equation}
where $\hU(m_1,m_2)$ describes pure QCD evolution and $\hR(m_1,m_2)$
the additional evolution in the presence of the electromagnetic
interaction.  $\hU(m_1,m_2)$ sums the logarithms $(\as t)^n$ and $\as
(\as t)^n$ with $t=\ln(m_2^2/m_1^2)$ whereas $\hR(m_1,m_2)$
sums the logarithms $t (\as t)^n$ and $(\as t)^n$.  $\hU(m_1,m_2)$ has
been given in ref.~\cite{burasetal:92a}. The leading order formula for
$\hR(m_1,m_2)$ can be found in ref.~\cite{buchallaetal:90} except that
here we used a different overall normalization (an additional factor
$-4\pi$ in $\hR$). For completeness, we will now recall the formula for
$\hU(m_1,m_2)$ and subsequently, we will derive a corresponding
expression for $\hR(m_1,m_2)$.

\subsection{The Pure QCD Evolution Matrix $\hU(m_1,m_2)$}
As shown in \cite{burasetal:92a} the pure QCD evolution can be written as
\begin{equation}
\hU(m_1,m_2) =
\left( \hat{1} + \frac{\as(m_1)}{4\pi} \hJ \right)
\hU^{(0)}(m_1,m_2)
\left( \hat{1} - \frac{\as(m_2)}{4\pi} \hJ \right) \, ,
\label{eq:2.12}
\end{equation}
where $\hU^{(0)}(m_1,m_2)$ denotes the evolution matrix in the leading
logarithmic approximation and $\hJ$ summarizes the next-to-leading
correction to this evolution. If
\begin{equation}
(\gs)_D \equiv \hV^{-1} \gst \hV \, ,
\qquad
\hat{G} \equiv \hV^{-1} \gsst \hV \, ,
\label{eq:2.13}
\end{equation}
where $(\gs)_D$ denotes a diagonal matrix whose diagonal elements are the
components of the vector $\vec{\gamma}_{\rm s}^{(0)}$, then
\begin{equation}
\hU^{(0)}(m_1,m_2) =
\hV \left[\left( \frac{\as(m_2)}{\as(m_1)} \right)^{\vec{a}}\right]_D \hV^{-1}
\qquad {\rm with} \qquad
\vec{a} = \frac{\vec{\gamma}_{\rm s}^{(0)}}{2 \beta_0} \, .
\label{eq:2.14}
\end{equation}

For the matrix $\hJ$ one gets
\begin{equation}
\hJ = \hV \; \hat{S} \; \hV^{-1} \, ,
\label{eq:2.15}
\end{equation}
where the elements of $\hat{S}$ are given by
\begin{equation}
S_{ij} = \delta_{ij} \; \gamma_{s,i}^{(0)} \; \frac{\beta_1}{2 \beta_0^2}
- \frac{G_{ij}}{2\beta_0 + \gamma_{s,i}^{(0)} - \gamma_{s,j}^{(0)}} \, ,
\label{eq:2.16}
\end{equation}
with $\gamma_{s,i}^{(0)}$ denoting the components of
$\vec{\gamma}_{\rm s}^{(0)}$ and $G_{ij}$ the elements of $\hat{G}$. Although
eq.~\eqn{eq:2.16} can develop singularities for certain combinations of
the $\gamma_{s,i}^{(0)}$, the physically relevant evolution matrix
\eqn{eq:2.12} always remains finite after proper combination of relevant
terms.

\subsection{The Evolution Matrix $\hR(m_1,m_2)$}
Inserting \eqn{eq:2.5} into \eqn{eq:2.9}, we find a general expression
for $\hR(m_1,m_2)$,
\begin{equation}
\hR(m_1,m_2) = \int_{g(m_2)}^{g(m_1)} \!\! dg' \;
\frac{\hU(m_1,m') \Gamma^T(g') \hU(m',m_2)}{\beta(g')} \, ,
\label{eq:2.17}
\end{equation}
where $\hU(m_1,m_2)$ is given by \eqn{eq:2.12}, $\Gamma(g)$ has
been defined in \eqn{eq:2.7} and $g' \equiv g(m')$.

It is instructive to discuss the leading order in \eqn{eq:2.17} which is
obtained by keeping only the first terms in \eqn{eq:2.2} and
\eqn{eq:2.7} and setting $\hJ \equiv 0$ in $\hU(m_1,m_2)$ of
eq.~\eqn{eq:2.12}. We then find
\begin{equation}
\hR^{(0)}(m_1,m_2) =
-\frac{2\pi}{\beta_0} \; \hV \; \hK^{(0)}(m_1,m_2) \hV^{-1} \, ,
\label{eq:2.18}
\end{equation}
where the matrix $\hK^{(0)}(m_1,m_2)$ is given by
\begin{equation}
(\hK^{(0)}(m_1,m_2))_{ij} = \hM^{(0)}_{ij} \int_{\as(m_2)}^{\as(m_1)}
\!\! d\as' \; \left( \frac{\as'}{\as(m_1)} \right)^{a_i}
\frac{1}{\as'^2}
\left(  \frac{\as(m_2)}{\as'} \right)^{a_j} \, ,
\label{eq:2.19}
\end{equation}
with $a_i$ being the components of the vector $\vec{a}$ defined in
eq.~\eqn{eq:2.14}. Moreover, the matrix $\hM^{(0)}$ is given by
\begin{equation}
\hM^{(0)} = \hV^{-1} \; \gemt \; \hV \, .
\label{eq:2.20}
\end{equation}

Straightforward integration gives
\begin{equation}
(\hK^{(0)}(m_1,m_2))_{ij} = \frac{\hM^{(0)}_{ij}}{a_i - a_j - 1}
\left[
\left( \frac{\as(m_2)}{\as(m_1)} \right)^{a_j} \frac{1}{\as(m_1)} -
\left( \frac{\as(m_2)}{\as(m_1)} \right)^{a_i} \frac{1}{\as(m_2)}
\right] \, ,
\label{eq:2.21}
\end{equation}
which is precisely eq.~(3.23) of ref.~\cite{buchallaetal:90}.

Similarly to eq.~\eqn{eq:2.16}, we note an apparent singularity in
the element $(7,8)$ for which $a_7=a_8 +1$ when $f=3$. However, the
expression in parenthesis also vanishes in this case and no singularity
is present. For numerical calculations it is safer to use in the case
$a_i = a_j + 1$ the following formula obtained directly from
eq.\eqn{eq:2.19}
\begin{equation}
(\hK^{(0)}(m_1,m_2))_{ij} = \hM^{(0)}_{ij} \frac{1}{\as(m_1)}
\left( \frac{\as(m_2)}{\as(m_1)} \right)^{a_j}
\ln\frac{\as(m_1)}{\as(m_2)} \, .
\label{eq:2.22}
\end{equation}

It is straightforward to generalize \eqn{eq:2.18} beyond the leading
logarithmic approximation. We find
\begin{equation}
\hR(m_1,m_2) = -\frac{2\pi}{\beta_0} \; \hV \; \hK(m_1,m_2) \hV^{-1}
\equiv \hR^{(0)}(m_1,m_2) + \hR^{(1)}(m_1,m_2) \, ,
\label{eq:2.23}
\end{equation}
where
\begin{equation}
\hK(m_1,m_2) = \hK^{(0)}(m_1,m_2) +
               \frac{1}{4\pi} \sum_{i=1}^{3} \hK_i^{(1)}(m_1,m_2) \, .
\label{eq:2.24}
\end{equation}

The next-to-leading corrections originating in $\hJ \not= 0$, $\gse
\not= 0$ and $\beta_1 \not= 0$ are represented by the matrices
$\hK_i^{(1)}(m_1,m_2)$ and more globally by $\hR^{(1)}(m_1,m_2)$.

Let us introduce
\begin{equation}
\hG^{(1)} = \gset - \frac{\beta_1}{\beta_0} \gemt \, ,
\label{eq:2.25}
\end{equation}
and
\begin{equation}
\hM^{(1)} =
\hV^{-1} \left( \hG^{(1)} + \left[ \gemt, \hJ \right] \right) \hV \, .
\label{eq:2.26}
\end{equation}

The matrices $\hK_i^{(1)}(m_1,m_2)$ are then given as follows
\begin{equation}
\left( \hK_1^{(1)}(m_1,m_2) \right)_{ij} =
\left\{
\begin{array}{ll}
\frac{M^{(1)}_{ij}}{a_i - a_j}
\left[ \left( \frac{\as(m_2)}{\as(m_1)} \right)^{a_j} -
       \left( \frac{\as(m_2)}{\as(m_1)} \right)^{a_i} \right] & i \not= j \\
\mvs
M^{(1)}_{ii} \left( \frac{\as(m_2)}{\as(m_1)} \right)^{a_i}
             \ln\frac{\as(m_1)}{\as(m_2)}             & i=j
\end{array} \, ,
\right.
\label{eq:2.27}
\end{equation}

\begin{eqnarray}
\hK_2^{(1)}(m_1,m_2) & = &
-\,\as(m_2) \; \hK^{(0)}(m_1,m_2) \; \hat{S} \, ,
\label{eq:2.28} \\
\hK_3^{(1)}(m_1,m_2) & = &
\phantom{-}\,\as(m_1) \; \hat{S} \; \hK^{(0)}(m_1,m_2) \, .
\label{eq:2.29}
\end{eqnarray}

Comparison of \eqn{eq:2.21} with \eqn{eq:2.27} and the additional
factors of $\as(m_i)$ in \eqn{eq:2.28} and \eqn{eq:2.29} make it clear
that $\hR^{(1)}(m_1,m_2)$ is by one logarithm lower or by one order in
$\as$ higher than $\hR^{(0)}(m_1,m_2)$. For later purposes it will be
useful to introduce ``$\as$-counting" in which $\hR^{(0)}(m_1,m_2)$ is
$\ord(1/\as)$ and $\hR^{(1)}(m_1,m_2)$ is $\ord(1)$.

\subsection{A Different Form of the Evolution Matrix}
The evolution matrix $\hU(m_1,m_2,\aem)$ can also be written in
a form which nicely generalizes the pure QCD evolution matrix of
eq.~(\ref{eq:2.12}),
\begin{equation}
\hU(m_1,m_2,\aem) \; = \; \hat W(m_1)\,\hU^{(0)}(m_1,m_2)\,
\hat W'(m_2) \,
\label{eq:2.29a}
\end{equation}
where
\begin{eqnarray}
\hat W(m_1)  & = & \Big(1+\frac{\aem}{4\pi}\hJ_{se}\Big) \Big(1+
\frac{\as(m_1)}{4\pi}\hJ_s\Big)\Big(1+\frac{\aem}{\as(m_1)}\hJ_e\Big) \, ,
\label{eq:2.29b1} \\
\hat W'(m_2) & = & \Big(1-\frac{\aem}{\as(m_2)}\hJ_e\Big)\Big(1-
\frac{\as(m_2)}{4\pi}\hJ_s\Big)\Big(1-\frac{\aem}{4\pi}\hJ_{se}\Big) \, .
\label{eq:2.29b2}
\end{eqnarray}
This is in fact the form used in ref.~\cite{ciuchini:92}, where only
implicit equations for the matrices $\hJ_s$, $\hJ_e$, and $\hJ_{se}$
have been given. Here, we give explicit expressions for the $\hJ_i$
which can be easily found from the previous subsections. $\hJ_s$ is
simply given by $\hJ$ of (\ref{eq:2.15}) and $\hJ_e$, $\hJ_{se}$ are
found to be
\begin{equation}
\hJ_e = \hV \; \hat{S}_e \; \hV^{-1} \,, \qquad
\hJ_{se} = \hV \; \hat{S}_{se} \; \hV^{-1} \,,
\label{eq:2.29c}
\end{equation}
where
\begin{equation}
(\hat S_e)_{ij} = \frac{1}{2\beta_0}\,\frac{\hM_{ij}^{(0)}}{(1+a_j-a_i)} \, ,
\qquad
(\hat S_{se})_{ij} = \frac{1}{2\beta_0}\,\frac{\hM_{ij}^{(1)}}{(a_j-a_i)} \, ,
\label{eq:2.29d}
\end{equation}
with $\hM_{ij}^{(0)}$ and $\hM_{ij}^{(1)}$ defined in eqs.~(\ref{eq:2.20})
and (\ref{eq:2.26}), respectively. We note that $\hat S_e$ can develop
singularities for certain combinations of $a_j$ and $a_i$, and
$\hat S_{se}$ is singular for $i=j$. All these singularities cancel
in the final expression for $\hU(m_1,m_2,\aem)$ when all contributing
terms are combined. This is seen for instance in (\ref{eq:2.27}). Thus
although formula (\ref{eq:2.29a}) is more elegant than the evolution
matrix presented in sections~2.4 and~2.5, the former formulation in
terms of the evolution matrices $\hU(m_1,m_2)$ and $\hR(m_1,m_2)$,
which is non-singular at all stages, is more suitable for numerical
calculations.

\subsection{The Initial Conditions $\vC(\mw)$}
In order to complete the analysis, we have to discuss the structure of
$\vC(\mw)$. This discussion generalizes the one given in section  2 of
ref.~\cite{burasetal:92a}. In order to find $\vC(\mw)$ the one--loop
current-current and penguin diagrams of fig.~\ref{fig:1a} with the full
$W$ and $Z$ propagators and internal top quark exchanges have to be
calculated first. Subsequently, the result of this calculation should
be expressed in terms of matrix elements $\langle\vQ(\mw)\rangle$. The
latter are found by inserting the operators $\vQ$ in the one--loop
current-current and penguin diagrams of fig.~\ref{fig:1b} and
calculating the finite contributions in some renormalization scheme. We
note the absence of $Z^{0}$-contributions to  $\langle\vQ(\mw)\rangle$.
Indeed, in the mass independent renormalization schemes used here such
contributions are zero.

If
\begin{equation}
T = \langle \vQ^{(0)^T} \rangle \left[
\vec{T}^{(0)} + \frac{\as(\mw)}{4\pi} \vec{T}_{s}^{(1)}
              + \frac{\aem}{4\pi} \vec{T}_{e}^{(1)} \right]
\equiv
\langle \vQ(\mw)^T \rangle \vC(\mw) \, ,
\label{eq:2.30}
\end{equation}
denotes the result of evaluating the diagrams of fig.~\ref{fig:1a} with
$\langle\vQ^{(0)}\rangle$ being the tree level matrix elements and
\begin{equation}
\langle\vQ(\mw)\rangle =
\left[ \hat{1} + \frac{\as(\mw)}{4\pi} \rs
               + \frac{\aem}{4\pi} \re \right]
\langle\vQ^{(0)}\rangle \, ,
\label{eq:2.31}
\end{equation}
obtained from fig.~\ref{fig:1b}, then
\begin{equation}
\vC(\mw) = \vec{T}^{(0)} +
\frac{\as(\mw)}{4\pi} \left[ \vec{T}_{\rm s}^{(1)} - \rs^T \vec{T}^{(0)}
 \right]
+ \frac{\aem}{4\pi} \left[ \vec{T}_{\rm e}^{(1)} - \re^T \vec{T}^{(0)}
 \right] \, .
\label{eq:2.32}
\end{equation}

It should be stressed that $\vec{T}_{\rm e}^{(1)}$,  $\vec{T}_{\rm
s}^{(1)}$, $\rs$ and $\re$ depend on the assumptions made about the
properties of the external lines in figs.~1 and 2, i.e.,~on the infrared
structure of the theory.  This dependence cancels however in
\eqn{eq:2.32} so that $\ord(\as)$ and $\ord(\aem)$ corrections in
\eqn{eq:2.32} do not depend on external states as it should be. They
depend however on the renormalization scheme through the matrices $\rs$
and $\re$.

The formulae \eqn{eq:2.30}-\eqn{eq:2.32} are general and apply also to
situations in which several operators are present in the limit
$\aem=\as=0$ represented by the first term in \eqn{eq:2.32}. In the
case at hand only the operator $Q_2$ is present in this limit and the
vector $\vec{T}^{(0)}$ is the transposed of $(0,1,0,\ldots,0)$. This
implies that only $\langle Q_2(\mw) \rangle$ has to be calculated in
order to find $\vC(\mw)$. This is evident from \eqn{eq:2.32} and the
discussion in appendix~C of ref.~\cite{burasetal:92a}. Yet, this more
general formulation is very useful for the analysis of scheme
dependences as we will see below.

\subsection{Final Formulae for Wilson Coefficient Functions}
Let us define
\begin{equation}
\vC(\mu) = \vC_{\rm s}(\mu) + \frac{\aem}{4\pi} \; \vC_{\rm e}(\mu) \, .
\label{eq:2.33}
\end{equation}

Inserting \eqn{eq:2.11} and \eqn{eq:2.32} into \eqn{eq:2.10}, we find
\begin{equation}
\vC_{\rm s}(\mu) =
\left( \hat{1} + \frac{\as(\mu)}{4\pi} \hJ \right)
\hU^{(0)}(\mu,\mw)
\left( \vec{T}^{(0)} + \frac{\as(\mw)}{4\pi} \vec{T}_{\rm s}^{(1)} -
       \frac{\as(\mw)}{4\pi} (\rs^T + \hJ) \vec{T}^{(0)} \right) \, ,
\label{eq:2.34}
\end{equation}
and
\begin{eqnarray}
\vC_{\rm e}(\mu) &=&
\hU^{(0)}(\mu,\mw) \left[ \vec{T}_{\rm e}^{(1)} - \re^T \vec{T}^{(0)} \right]
\nn \\ & &
+ \,\hR^{(0)}(\mu,\mw) \left[ \vec{T}^{(0)} +
  \frac{\as(\mw)}{4\pi} (\vec{T}_{\rm s}^{(1)} - \rs^T \vec{T}^{(0)}) \right]
\nn \\ & &
+ \,\hR^{(1)}(\mu,\mw) \vec{T}^{(0)} \, .
\label{eq:2.35}
\end{eqnarray}

\newsection{Renormalization Scheme Dependence}
\subsection{General Consistency Relations}
The two--loop anomalous dimensions depend on the renormalization scheme
for operators and in particular on the treatment of $\gf$ in $D\not= 4$
dimensions. In refs.~\cite{burasetal:92a} and \cite{burasetal:92c}, we have
derived two relations between $\hg^{(1)}_a$ and $\hg^{(1)}_b$ calculated
in two different renormalization schemes $a$ and $b$.
\begin{equation}
(\gss)_b = (\gss)_a + \left[\Delta\rs,\gs\right] + 2 \, \beta_0 \; \Delta\rs\,,
\label{eq:3.1}
\end{equation}
\begin{equation}
(\gse)_b = (\gse)_a + \left[\Delta\rs,\gem\right]+\left[\Delta\re,\gs\right]\,,
\label{eq:3.2}
\end{equation}
with
\begin{equation}
\Delta\hat{r}_i = (\hat{r}_i)_b - (\hat{r}_i)_a,
\qquad
i=s,e \, ,
\label{eq:3.3}
\end{equation}
and $\hat{r}_i$ defined in \eqn{eq:2.31}. It follows that the
combinations
\begin{equation}
\hat{W}_{\rm s} = \gss - \left[ \rs,\gs \right] - 2 \, \beta_0 \; \rs \, ,
\label{eq:3.4}
\end{equation}
and
\begin{equation}
\hat{W}_{\rm se} = \gse - \left[ \rs,\gem \right] - \left[ \re,\gs \right]
 \, ,
\label{eq:3.5}
\end{equation}
are independent of the renormalization scheme considered. We should also
point out that $\Delta\rs$ and $\Delta\re$ do not depend on the infrared
structure of the theory, whereas this is not the case for $\rs$ and
$\re$ as has been stressed in section 2.

Relations \eqn{eq:3.1}-\eqn{eq:3.4} turned out to be very useful in
testing the two--loop anomalous dimension matrices found in
refs.~\cite{burasetal:92b} and \cite{burasetal:92c}. They also play a
central role in demonstrating the scheme independence of physical
quantities. For the pure QCD case this has been discussed in
ref.~\cite{burasetal:92a}. Below we will extend these considerations to
$\ord(\aem)$ corrections.

\subsection{Relations Between Coefficient Functions}
The relations between coefficient functions obtained in two different
schemes $a$ and $b$ are entirely given in terms of the matrices $\Delta
\hat{r}_i$ of eq.~\eqn{eq:3.3} which follow from the evaluation of the
one--loop diagrams of fig.~\ref{fig:1b}.

One has
\begin{equation}
\vC_b(\mu) = \left[
\hat{1} - \frac{\as(\mu)}{4\pi} \; \drs^T - \frac{\aem}{4\pi} \dre^T
\right] \; \vC_a(\mu) \, ,
\label{eq:3.12}
\end{equation}
with the corresponding relation for the hadronic matrix elements
\begin{equation}
\langle \vQ(\mu) \rangle_b^T =
\langle \vQ(\mu) \rangle_a^T
\left[ \hat{1} + \frac{\as(\mu)}{4\pi} \; \drs^T + \frac{\aem}{4\pi} \dre^T
\right] \, .
\label{eq:3.13}
\end{equation}

The scheme independence of the effective Hamiltonian \eqn{eq:1.2}
follows directly from these relations.

The matrices $\drs$ and $\dre$ relating HV and $\ndr$ schemes can be
found in sections~3.3 and 3.4 of ref.~\cite{burasetal:92b} and in
section 6.2 of ref.~\cite{burasetal:92c}, respectively. The
corresponding matrices relating $\ndrb$ and $\ndr$ schemes are given in
eq.~(5.14) of ref.~\cite{burasetal:92c}. See however eq.~\eqn{eq:4.21b} of
the present paper.

\subsection{Scheme Independence of Physical Quantities}
Let us return to the general formulae \eqn{eq:2.34} and \eqn{eq:2.35}
for the coefficients $\vC(\mu)$. In ref.~\cite{burasetal:92a}, we have
shown that the combination $\rs^T + \hJ$ in eq.~\eqn{eq:2.34} is
renormalization scheme independent, implying that the next-to-leading
QCD corrections to $\vC_{\rm s}(\mu)$ are only scheme dependent through the
matrix $\hJ$ at the lower end of the evolution which involves
$\as(\mu)$. This scheme dependence is cancelled by the one of
$\ord(\as)$ corrections to $\langle \vQ(\mu) \rangle$ so that the
resulting decay amplitudes do not depend on the renormalization scheme
as it should be. We would like now to demonstrate that an analogous
proof can be made in the case of $\vC_{\rm e}(\mu)$ given by \eqn{eq:2.35}.
This can certainly be done, but it turns out that a much more elegant
proof can be made by using the form of \eqn{eq:2.29a}.

First we want to demonstrate the scheme independence of terms at the
upper end of the evolution. Inserting \eqn{eq:2.29a} and \eqn{eq:2.32}
into \eqn{eq:2.10} and recalling that $\vec{T}^{(0)}$, $\vec{T}^{(1)}$
and $\hJe$ are scheme independent we find that it is sufficient
to consider the combination
\begin{equation}
\hat{X} =
\left( \frac{\as(\mw)}{4\pi} \; \hat{1} + \frac{\aem}{4\pi} \; \hJe \right) \;
\left[ \rs^T + \hJs \right] + \frac{\aem}{4\pi}
\left[ \re^T + \hJse \right] \, .
\label{eq:3.6a}
\end{equation}
The first part of this expression is scheme independent due to scheme
independence of $\hJe$ and $\rs^T + \hJs$ as shown in
\cite{burasetal:92a}. It remains therefore to show that also $\re^T +
\hJse$ is scheme independent.

Now the scheme independence of $\hat{W}_{\rm se}$ and $\rs^T + \hJs$
allows to extract the scheme dependent part of $\hM^{(1)}$:
\begin{equation}
\hM^{(1)}_{ij}(\hbox{\rm \small scheme\ dependent}) =
\left( \gamma^{(0)}_{{\rm s}, i} - \gamma^{(0)}_{{\rm s}, j} \right)
(\hat{P}_{\rm e})_{ij}
\label{eq:3.6b}
\end{equation}
where $\hat{P}_{\rm e} \equiv \hV^{-1} \; \re^T \; \hV$. Therefore
using \eqn{eq:2.29c} and \eqn{eq:2.29d} we conclude that the scheme
dependent part of $\hJse$ is simply $-\re^T$ which ends the proof.

In a very similar way one can show that the scheme dependence around
the $\mu$ scale present in $\hat{W}(\mu)$ of \eqn{eq:2.29b1} is
cancelled by the scheme dependence of the $\ord(\as(\mu))$ and
$\ord(\aem)$ corrections to the matrix elements $\langle \vQ^T(\mu)
\rangle$. This ends the proof of the scheme independence of physical
amplitudes.

\newsection{$\dS$ Hamiltonian Beyond Leading Logarithms}
\subsection{General Remarks}
In the absence of strong and electromagnetic interactions the effective
tree-level Hamiltonian for $\dS$ non-leptonic decays can be written as follows
\begin{equation}
\Heff(\dS) = \frac{G_F}{\sqrt{2}} \V{ud} \V{us}^*
\left\{ (1-\tau) (Q_2^u - Q_2^c) + \tau (Q_2^u -Q_2^t) \right\} \, ,
\label{eq:4.1a}
\end{equation}
where
\begin{equation}
\tau \equiv -\frac{\V{td}\V{ts}^*}{\V{ud}\V{us}^*}
\qquad {\rm and } \qquad
Q_2^q = (\bar s q)_{V-A} (\bar q d)_{V-A}, \quad q=u,c,t \, .
\label{eq:4.1b}
\end{equation}
If the $\ndrb$ scheme is adopted the Fierz conjugates $\widetilde{Q}_2^q$ of
\eqn{eq:2.3n} and \eqn{eq:2.4n} are present in \eqn{eq:4.1a}. For the HV
scheme it is immaterial whether $Q_2^q$ or $\widetilde{Q}_2^q$ are used.

The inclusion of QCD corrections and the evolution of $\Heff$ from $\mw$
down to $\mu << \mw$ has been discussed in the leading logarithmic
approximation at length in \cite{buchallaetal:90}. The next-to-leading
order formulae involving $Q_1$ -- $Q_6$ have already been given in
ref.~\cite{burasetal:92a}. Here, we generalize
the results of \cite{burasetal:92a}  to include also the electroweak
penguin operators $Q_7$ -- $Q_{10}$ together with $\ord(\aem)$
corrections using the formalism developed in previous sections. A
similar though less explicit generalization has been made by the
authors of ref.~\cite{ciuchini:92}. In the latter paper however mainly
the Wilson coefficients relevant for $\epe$ have been considered. Here,
we present also the results for the coefficients relevant for the real
parts of $\Kpipi$ amplitudes which enter the discussion of the $\dI$
rule. Moreover, it will be useful to discuss the cases $\mu < \mc$ and
$\mc < \mu < \mb$ separately, as well as the interesting case $\mu=\mc$
on the border line of effective 4- and 3-flavour theories. This case is
simultaneously relevant for $\dC$ transitions after proper replacement
of quark flavours and is a useful renormalization point for our approach
to hadronic matrix elements. The case $\mu \approx \ord(\mb)$, relevant
for $\dB$ decays, is considered in section 9.

\subsection{Master Formulae for Wilson Coefficient Functions $(\mu < \mc)$}
The transformation of $\Heff$ of \eqn{eq:4.1a} from $\mw$ down to $\mu <
\mc$ involves 5-, 4- and 3-flavour effective theories and can be
compactly described as follows
\begin{equation}
Q_2^u - Q_2^c \rightarrow \sum_{i=1}^{10} z_i(\mu) \; Q_i \, ,
\qquad
Q_2^u - Q_2^t \rightarrow \sum_{i=1}^{10} v_i(\mu) \; Q_i \, ,
\label{eq:4.1c}
\end{equation}
where $Q_i$ are the operators given in \eqn{eq:2.1n}, with $q=u,d,s$.
Consequently, we find the familiar expression
\begin{equation}
\Heff(\dS) = \frac{G_F}{\sqrt{2}} \V{ud} \V{us}^* \sum_{i=1}^{10}
\left( z_i(\mu) + \tau \; y_i(\mu) \right) Q_i(\mu) \, ,
\label{eq:4.1}
\end{equation}
where
\begin{equation}
y_i(\mu) = v_i(\mu) - z_i(\mu) \, .
\label{eq:4.3}
\end{equation}

The coefficients $z_i$ and $v_i$ are the components of the ten
dimensional column vectors $\vec{z}$ and $\vec{v}$
\begin{equation}
\vec{z}(\mu) = \hU_3(\mu,\mc,\aem) \vec{z}(\mc) \, ,
\label{eq:4.4}
\end{equation}
and
\begin{equation}
\vec{v}(\mu) =
\hU_3(\mu,\mc,\aem) \hM(\mc) \hU_4(\mc,\mb,\aem) \hM(\mb) \hU_5(\mb,\mw,\aem)
\vec{C}(\mw) \, .
\label{eq:4.5}
\end{equation}
The full evolution matrices $\hU_f(m_1,m_2,\aem)$ are given by
\eqn{eq:2.11} with $f$ denoting the number of effective flavours.
$\hM(m_i)$ is the matching matrix at quark threshold $m_i$. Its origin
and structure is discussed separately in section~4.3.

Using \eqn{eq:4.4} and \eqn{eq:4.5}, care must be taken to
remove consistently higher order terms in $\as$ and $\aem$ resulting
from the multiplication of the evolution matrices and initial values.
To this end one should remember our discussion of the ``$\as$-counting"
in the case of the evolution matrix $\hR(m_1,m_2)$.
{}From this counting it is evident that e.g.~terms involving products
$\hJ \hR^{(1)}$  should be discarded to the order considered here.

The initial values $\vC(\mw)$ necessary for the evaluation of $v_i(\mu)$
are found according to the procedure outlined in section 2. For the case
$\aem=0$ they have already been given in \cite{burasetal:92a}.
For the $\ndr$ scheme we find
\begin{eqnarray}
C_1(\mw) &=& \frac{\as(\mw)}{4\pi} B_{\rm s,1}^{\rm \ndr} +
\frac{\aem}{4\pi} B_{\rm e,1}^{\rm \ndr} \, ,
\label{eq:4.6} \\
C_2(\mw) &=& 1 + \frac{\as(\mw)}{4\pi} B_{\rm s,2}^{\rm \ndr} +
\frac{\aem}{4\pi} B_{\rm e,2}^{\rm \ndr} \, ,
\label{eq:4.7} \\
C_3(\mw) &=& -\frac{\as(\mw)}{24\pi} \widetilde{E}(x_t)
             +\frac{\aem}{6\pi} \frac{1}{\sin^2\theta_W}
             \left[ 2 B(x_t) + C(x_t) \right] \, ,
\label{eq:4.8} \\
C_4(\mw) &=& \frac{\as(\mw)}{8\pi} \widetilde{E}(x_t) \, ,
\label{eq:4.9} \\
C_5(\mw) &=& -\frac{\as(\mw)}{24\pi} \widetilde{E}(x_t) \, ,
\label{eq:4.10} \\
C_6(\mw) &=& \frac{\as(\mw)}{8\pi} \widetilde{E}(x_t) \, ,
\label{eq:4.11} \\
C_7(\mw) &=& \frac{\aem}{6\pi} \left[ 4 C(x_t) + \widetilde{D}(x_t) \right]
\, ,
\label{eq:4.12} \\
C_8(\mw) &=& 0 \, ,
\label{eq:4.13} \\
C_9(\mw) &=& \frac{\aem}{6\pi} \left[ 4 C(x_t) + \widetilde{D}(x_t) +
             \frac{1}{\sin^2\theta_W} (10 B(x_t) - 4 C(x_t)) \right] \, ,
\label{eq:4.14} \\
C_{10}(\mw) &=& 0 \, ,
\label{eq:4.15}
\end{eqnarray}
where
\begin{equation}
\begin{array}{lclclcl}
B_{\rm s,1}^{\rm\ndr} &=& 11/2\,, &\qquad& B_{\rm s,2}^{\rm\ndr}&=&-11/6\,, \\
B_{\rm e,1}^{\rm\ndr} &=&  0\,,   &\qquad& B_{\rm e,2}^{\rm\ndr}&=&-35/18 \,,
\end{array}
\label{eq:4.16}
\end{equation}
\begin{equation}
\widetilde{E}(x_t) = E(x_t) - \frac{2}{3} \, ,
\qquad
\widetilde{D}(x_t) = D(x_t) - \frac{4}{9} \, ,
\label{eq:4.16a}
\end{equation}
\begin{equation}
x_t = \frac{\mt^2}{\mw^2} \, ,
\label{eq:4.17}
\end{equation}
and
\begin{eqnarray}
B(x) &=& \frac{1}{4} \left[ \frac{x}{1-x} + \frac{x \ln x}{(x-1)^2} \right]\,,
\label{eq:4.18} \\
C(x) &=&
  \frac{x}{8} \left[ \frac{x-6}{x-1} + \frac{3 x + 2}{(x-1)^2} \ln x \right]\,,
\label{eq:4.19} \\
D(x) &=& -\frac{4}{9} \ln x + \frac{-19 x^3 + 25 x^2}{36 (x-1)^3} +
         \frac{x^2 (5 x^2 - 2 x - 6)}{18 (x-1)^4} \ln x \, ,
\label{eq:4.20} \\
E(x) &=& -\frac{2}{3} \ln x + \frac{x (18 -11 x - x^2)}{12 (1-x)^3} +
         \frac{x^2 (15 - 16 x  + 4 x^2)}{6 (1-x)^4} \ln x \, .
\label{eq:4.21}
\end{eqnarray}

Comparing with eqs.~(3.26)-(3.35) of ref.~\cite{buchallaetal:90}, we
observe that the initial conditions for the penguin
operators given here differ from the ones used in
ref.~\cite{buchallaetal:90}.  There the leading order evolution matrices
have been used but in the initial conditions the $\mt$-dependent
next-to-leading terms have been already taken into account. These
$\mt$-dependent terms represented by the functions $B(x_t)$, $C(x_t)$,
$D(x_t)$ and $E(x_t)$ are renormalization scheme independent. Yet, the
initial conditions may also contain scheme dependent next-to-leading
terms which of course were not present in the analysis of
ref.~\cite{buchallaetal:90}. In the $\ndr$ scheme these scheme
dependent terms enter the initial conditions for $Q_1$ and $Q_2$ and
also modify the functions $E(x_t)$ and $D(x_t)$ by constant terms as
given in \eqn{eq:4.16a}.

The initial values for the $\ndrb$ and HV schemes can be found by using
the relations \eqn{eq:3.12}. It is then interesting to observe that in
these schemes the additional terms in \eqn{eq:4.16a} are absent and the
initial conditions for $Q_1$ and $Q_2$ are changed. The corresponding
two--loop anomalous dimensions also change so that the final result for
$\Heff$ remains scheme independent. This clarifies the observation made
by Flynn and Randall \cite{flynn:89} that the initial conditions for
penguin operators do depend on the form of the operators. As already
emphasized in \cite{burasetal:92a}, the inclusion of two--loop anomalous
dimensions cancels this dependence. Compared to
ref.~\cite{burasetal:92a}, we have however slightly changed our
treatment of the HV scheme. Since in the HV scheme the anomalous
dimension of the weak current $\gamma_{\rm J}$ is non-zero at
$\ord(\as^2)$ \cite{burasweisz:90}, an additional contribution arises
which can be either included in the anomalous dimension matrix as
$\hg_{\rm s} -2 \gamma_{\rm J} \; \hat{1}$ \cite{burasweisz:90} or in the
initial condition. In \cite{burasetal:92a} we choose to work with the
latter treatment, which however introduces an unnecessarily large
scheme dependence at the lower end of the evolution. Therefore, here we
have included the anomalous dimension of the weak current in the
anomalous dimension matrix. This also changes the initial conditions in
the HV scheme, as was already discussed in ref.~\cite{burasetal:92a}.
Moreover, $\drs$ given in \cite{burasetal:92a,burasetal:92b} has to be
changed to
\begin{equation}
\drs \rightarrow \drs - 4 \; C_{\rm F} \; \hat{1} \, .
\label{eq:4.21b}
\end{equation}

In order to calculate $z_i(\mu)$ one has to consider the difference
$Q_2^u - Q_2^c$. Then due to the GIM mechanism the coefficients
$z_i(\mu)$ of penguin operators $Q_i$, $i\not=1,2$ are zero in 5- and
4-flavour theories. The evolution in this range of $\mu$ ($\mu > \mc$)
involves then only current-current operators $Q_{1,2} \equiv Q_{1,2}^u$
and $Q_{1,2}^c$ with the initial conditions
\begin{eqnarray}
z_1(\mw) &=& \frac{\as(\mw)}{4\pi} B_{s,1}^{\ndr} +
             \frac{\aem(\mw)}{4\pi} B_{e,1}^{\ndr} \, ,
\label{eq:4.16b} \\
z_2(\mw) &=& 1 + \frac{\as(\mw)}{4\pi} B_{s,2}^{\ndr} +
                 \frac{\aem(\mw)}{4\pi} B_{e,2}^{\ndr} \, ,
\label{eq:4.16c}
\end{eqnarray}
as in \eqn{eq:4.6} and \eqn{eq:4.7}. $Q_{1,2}^u$ and $Q_{1,2}^c$ do not
mix under renormalization with each other and the initial conditions
for the coefficients of $Q_{1,2}^c$ to be denoted by $z_{1,2}^c(\mu)$
are identical to the ones given above. We then find
\begin{equation}
\left( \begin{array}{ll} z_1(\mc) \\ z_2(\mc) \end{array} \right) =
\hU_4(\mc,\mb,\aem) \; \hM(\mb) \; \hU_5(\mb,\mw,\aem) \;
\left( \begin{array}{ll} z_1(\mw) \\ z_2(\mw) \end{array} \right) \, ,
\label{eq:4.16d}
\end{equation}
where this time the evolution matrices $\hU_{4,5}$ contain only the
$(2,2)$ anomalous dimension matrices describing the mixing between
current-current operators. When the charm quark is integrated out the
operators $Q_{1,2}^c$ disappear from the effective Hamiltonian and the
coefficients $z_i(\mu)$, $i\not=1,2$ for penguin operators become
non-zero. In order to calculate $z_i(\mc)$ for penguin operators a
proper matching between effective 4- and 3-quark theories has to be
made. This matching has been already discussed in detail in
ref.~\cite{burasetal:92a}. However, there we have set $q^2 = -\mc^2$ in
evaluating the penguin diagrams of fig.~\ref{fig:1b}. Meanwhile, we
have realized that in the effective 3-quark theory, $q$ being related
to external light quarks must be much smaller than $\mc$ and the
consistent procedure within the operator product expansion is to set it
identically to zero in the full expression. Numerically, the difference
between coefficients calculated with $q^2 = -\mc^2$ and $q^2 = 0$ is
small, however, the resulting modified expressions represent this time
correct matching.  Proceeding this way and applying an analogous
procedure to QED penguin diagrams we are able to calculate
$\vec{z}(\mc)$ of \eqn{eq:4.4}. We have in the $\ndr$ scheme
\begin{equation}
\vec{z}^{\ndr}(\mc) =
\left( \begin{array}{c}
z_1^{\ndr}(\mc) \\ z_2^{\ndr}(\mc) \\
-\as/(24\pi) F_{\rm s}^{\ndr}(\mc) \\ \as/(8\pi) F_{\rm s}^{\ndr}(\mc) \\
-\as/(24\pi) F_{\rm s}^{\ndr}(\mc) \\ \as/(8\pi) F_{\rm s}^{\ndr}(\mc) \\
\aem/(6\pi) F_{\rm e}^{\ndr}(\mc) \\  0  \\
\aem/(6\pi) F_{\rm e}^{\ndr}(\mc) \\  0
\end{array} \right) \, ,
\label{eq:4.25}
\end{equation}
where
\begin{equation}
F_{\rm s}^{\ndr}(\mu) =
-\frac{2}{3} \left( \ln(\frac{\mc^2}{\mu^2})+1 \right) \; z_2(\mu) \,,
\label{eq:4.26}
\end{equation}
\begin{equation}
F_{\rm e}^{\ndr}(\mu) =
-\frac{4}{9} \left( \ln(\frac{\mc^2}{\mu^2}) + 1 \right) \;
\left( 3 z_1(\mu) + z_2(\mu) \right) \, ,
\label{eq:4.27}
\end{equation}
with $\mu \approx \ord(\mc)$. In the $\ndrb$ and HV schemes the ``1'' in
\eqn{eq:4.26} and \eqn{eq:4.27} is absent. Consequently, in the $\ndrb$
and HV schemes one has $z_i(\mc) = 0$ for $i\not=1,2$ but in the case of
the $\ndr$ scheme these coefficients are non-vanishing.

Due to the fact that $z_1$ and $z_2$ have opposite sign and $|z_1(\mc)|
\approx \frac{1}{3} |z_2(\mc)|$ the coefficients $z_7(\mc)$ and
$z_9(\mc)$ are very small. This is reminiscent of the situation in
exclusive two-body decays such as $D^0 \rightarrow \pi^0 K^0$, for which
such a cancellation between $Q_1$ and $Q_2$ contributions takes place.
In the context of the $\dS$ Hamiltonian such a cancellation has already
been pointed out in ref.~\cite{ciuchini:90}.

Again, when using the formulae above one has to take care that
$\ord(\aem^2)$ terms are omitted as well as terms $\ord(\as^2)$ and
$\ord(\aem\,\as)$

At this stage it is mandatory for us to compare the results in
\eqn{eq:4.26} and \eqn{eq:4.27} with the ones of \cite{bardeen:87a},
\cite{cheng:88} and \cite{ciuchini:90} where the effects of incomplete
GIM mechanism resulting from $m_{\rm u} \not= \mc$ for $\mu > \mc$ has
been calculated.  These analyses resulted in a scheme independent
constant ``$\frac{5}{3}$'' in place of ``1'' in  \eqn{eq:4.26} and
\eqn{eq:4.27}. Yet, the proper matching must involve a scheme dependent
constant. Our interpretation of this difference is as follows. It is
intuitively clear that an incomplete GIM cancellation in the
$Q_2^u-Q_2^c$ sector must take place for $\mu > \mc$. Yet, this effect
is at the next-to-leading level at which other scheme dependent
contributions are present. From our analysis it turns out that in the
HV and $\ndrb$ schemes the latter effects cancel exactly the term
``$\frac{5}{3}$'' found in \cite{bardeen:87a}, \cite{cheng:88} so that
$z_i(\mu) = 0$ for $\mu \ge \mc$. In the $\ndr$ scheme  however $z_i(\mc)
\not= 0$, although as we will see also in this schemes the penguin
coefficients are smaller than found in ref.~\cite{bardeen:87a}.

\subsection{The Matching Matrix $\hM$}
In order to show that a non-trivial matching matrix $\hM$ has to be
included in the evolution\footnote{This has been pointed out by
J.-M.~Schwarz and has been briefly discussed in an addendum to
\cite{burasetal:92a}.}, when going from an $f$-flavour effective
theory to an $(f-1)$-flavour effective theory, we write an amplitude
$A$ for some process as follows
\begin{equation}
A = \langle\vec Q_f(m)\rangle^T\,\vec C_f(m) =
\langle\vec Q_{f-1}(m)\rangle^T\,\vec C_{f-1}(m) \, ,
\label{eq:4.27a1}
\end{equation}
where $m$ is the threshold scale between the two effective theories in
question. $\vec Q_f$, $\vec C_f$, $\vec Q_{f-1}$, and $\vec
C_{f-1}$ are the operators and coefficients in these two theories,
respectively.

In analogy to (\ref{eq:2.31}), one finds from fig.~\ref{fig:1b}
\begin{equation}
\langle\vec Q_f(m)\rangle = \Big(\hat 1+\frac{\as(m)}{4\pi}\vardrs+
\frac{\aem}{4\pi}\vardre\Big)\,\langle\vec Q_{f-1}(m)\rangle \, ,
\label{eq:4.27a2}
\end{equation}
where
\begin{equation}
\delta \hat r_i = \hat r_i(f) - \hat r_i(f-1) \, .
\label{eq:4.27a3}
\end{equation}
The matrices $\delta \hat r$ receive only contributions from penguin
diagrams which for certain operator insertions depend on the number
of effective flavours. Inserting (\ref{eq:4.27a2}) in (\ref{eq:4.27a1}),
we find
\begin{equation}
\vec C_{f-1}(m) = \hM(m)\,\vec C_f(m) \, ,
\label{eq:4.27a4}
\end{equation}
where
\begin{equation}
\hM(m) = \hat 1 + \frac{\as(m)}{4\pi}\vardrs^T +
\frac{\aem}{4\pi}\vardre^T \, .
\label{eq:4.27a5}
\end{equation}

A simple consequence of the presence of a non-trivial matching matrix
at the next-to-leading level is the appearance of small discontinuities
in the Wilson coefficient functions, and consequently, in the matrix
elements of certain penguin operators at the thresholds between
effective theories. This is not surprising because after all for penguin
operators $\vec Q_f\neq\vec Q_{f-1}$, as explicitly seen in
(\ref{eq:2.1n}). An example of such a discontinuity is given also by
the initial conditions in (\ref{eq:4.25}), since for $\mu>\mc$ one has
$z_i(\mu)=0$ $(i\neq1,2)$. The matrices $\hM(m)$ are renormalization
scheme independent. They are given in appendix~A.

\subsection{Master Formulae for Wilson Coefficients $(\mc < \mu < \mb)$}
If $\mu$ is chosen above $\mc$ only operators $Q_{1,2}$ and $Q_{1,2}^c$
are present in the $u-c$ sector. For $Q_i$, $i=1,\ldots,10$, we simply
find the Hamiltonian in \eqn{eq:4.1} with
\begin{equation}
\left( \begin{array}{ll} z_1(\mu) \\ z_2(\mu) \end{array} \right) =
\hU_4(\mu,\mb,\aem) \; \hM(\mb) \; \hU_5(\mb,\mw,\aem) \;
\left( \begin{array}{ll} z_1(\mw) \\ z_2(\mw) \end{array} \right) \, ,
\label{eq:4.27b}
\end{equation}
\begin{equation}
\vec{v}(\mu) =
\hU_4(\mu,\mb,\aem) \; \hM(\mb) \; \hU_5(\mb,\mw,\aem) \;
\vec{v}(\mw) \, ,
\label{eq:4.27c}
\end{equation}
and $z_i=0$ for $i\neq1,2$. Since $Q_{1,2}^c$ enter the effective
Hamiltonian with the opposite sign to $Q_{1,2}^u$ and are only present
in the $(1-\tau)$ term in \eqn{eq:4.1a}, we find
\begin{equation}
z_1^c(\mu) = - \,z_1(\mu) \,, \qquad \qquad z_2^c(\mu) = - \,z_2(\mu) \,,
\label{eq:4.27d}
\end{equation}
and
\begin{equation}
y_1^c(\mu) = z_1(\mu) \,, \qquad \qquad y_2^c(\mu) = z_2(\mu) \, .
\label{eq:4.27e}
\end{equation}
On the other hand it is evident from the formulae above that
\begin{equation}
y_1(\mu) = y_2(\mu) = 0 \, ,
\label{eq:4.27f}
\end{equation}
for arbitrary $\mu$. We note that the coefficients $y_1^c$ and $y_2^c$
are large.

\subsection{Relations Between Operators}
The operators $Q_i$ given in \eqn{eq:2.1n} are linearly dependent for
$f=4$ or equivalently $\mc < \mu < \mb$. One has the relations
\begin{eqnarray}
Q_4    &=&   Q_3 + Q_2 - Q_1 + Q_2^c - Q_1^c \; , \nn \\
Q_9    &=& \frac{1}{2} ( 3 Q_1 - Q_3 ) + \frac{3}{2} Q_1^c \; ,
\label{eq:4.28} \\
Q_{10} &=&   Q_9 + Q_4 - Q_3 \; . \nn
\end{eqnarray}
For $f=3$, i.e.,~$\mu < \mc$ when the charm quark is integrated out these
relations reduce to
\begin{eqnarray}
Q_4    &=& Q_3 + Q_2 - Q_1 \, ,\nn \\
Q_9    &=& \frac{1}{2} \left(3 Q_1 - Q_3\right) \, , \\
Q_{10} &=& Q_2 + \frac{1}{2} \left(Q_1 - Q_3\right) \; .  \nn
\label{eq:4.29}
\end{eqnarray}

It should be however emphasized that these relations have been obtained
by performing Fierz transformations. Consequently, they cannot be used
a priori in regularization or renormalization schemes in which the
operators $Q_i$ and their Fierz conjugates $\widetilde{Q}_i$ are not
equivalent in $D\not=4$ dimensions. Thus, although they can be used in
the HV scheme, they receive additional contributions in the $\ndr$ and
$\ndrb$ schemes.  An inspection of the relation \eqn{eq:3.13} reveals
that in the limit $\aem=0$ only the relation for $Q_4$ receives additional
$\as$ corrections so that e.g.~in the $\ndr$ scheme we find
\begin{equation}
Q_4 = Q_3 + Q_2 - Q_1 - \frac{\as}{4\pi}
\left( Q_6 + Q_4 - \frac{1}{3} Q_3 - \frac{1}{3} Q_5 \right) \, ,
\label{eq:4.29b}
\end{equation}
which of course then has to be solved for $Q_4$.  Yet, we have decided
to work with all operators and not to use the relations
\eqn{eq:4.28}--\eqn{eq:4.29} in evaluating the coefficients $C_i(\mu)$.
As discussed at length in \cite{buchallaetal:90} one can without any
problems work with linearly dependent operators. The advantage of such
a strategy is a much clearer picture of contributions coming from
different sources.  On the other hand the relations above will turn out
to be useful in the analysis of hadronic matrix elements.

\subsection{Numerical Results for Wilson Coefficients}
In tabs.~\ref{tab:11}, \ref{tab:12}, \ref{tab:13} we give the numerical
values for the coefficients $z_i$ and $y_i$ for $\Lms^{(4)} = 200, 300,
400\mev$, in the $\ndr$ and HV schemes for $\mu=1\gev$,
$\mu=\mc=1.4\gev$, and $\mu=2\gev$, at a fixed value of $\mt=130\gev$.
In tabs.~\ref{tab:1}, \ref{tab:2}, \ref{tab:3}, we give the
$\mt$-dependence of the coefficients $y_7$ -- $y_{10}$ of the
electroweak penguin operators for $\mu=1.0\gev$, different values of
$\Lms$ and the two schemes in question. The coefficients $z_i$ do not
depend on $\mt$. The $\mt$-dependence of $y_3$ -- $y_6$ is so weak,
that it has not been shown explicitly.  The corresponding results for
$\mu=\mc=1.4\gev$ and $\mu=2.0\gev$ are given in tabs.~\ref{tab:4},
\ref{tab:5}, \ref{tab:7} and \ref{tab:8}, \ref{tab:9}, \ref{tab:10}
respectively.  The reason for giving many tables for several values of
$\mu$ is as follows: For $\mu\approx1\gev$ the calculation of hadronic
matrix elements in an approach like the $1/N$ approach can in principle
be performed as discussed in the next section. The scale $\mu=2.0\gev$
is very suitable for present lattice calculations. Finally, $\mu=\mc$
is a very convenient scale for the semi-phenomenological approach
developed in section~6. Simultaneously, $\mu=\mc$ can be used in the
case of $\dC$ decays.

For comparison, we give in the tables mentioned above the corresponding
``leading order'' results. These have been obtained using the strategy
of ref.~\cite{buchallaetal:90}. Thus we have set all the
$\mt$-independent terms in (\ref{eq:4.6}) -- (\ref{eq:4.16a}) to
zero\footnote{We should warn the reader that the authors of
ref.~\cite{ciuchini:92} keep these scheme dependent terms in their
leading order evaluation.} and we have used the leading order evolution
matrix, i.e., one--loop anomalous dimension matrices and one--loop
$\beta$-function. Clearly, the scale $\Lambda_{LO}$ in the leading
order expressions cannot be identified with $\Lms$ which enters the
next-to-leading formulae.  In addition, $\as$ at the next-to-leading
level is smaller than $\as$ in the leading order when
$\Lambda_{LO}=\Lms$ is taken. In spite of this we have given the
leading order results for $\Lambda_{LO}=\Lms$, because the range of
values for $\Lms$ used in these tables overlaps considerably with the
range used for the QCD scale in the leading order analyses. In any case
what is really relevant here are the results for $z_i$ and $y_i$ with
next-to-leading order corrections taken into account, because in this
case $\Lms$ is exactly the scale as extracted from next-to-leading
order analyses in other processes \cite{altarelli:92}.

We make the following observations:
\begin{itemize}
\item[i)] The coefficients $z_1$ and $z_2$ are suppressed through
 next-to-leading order corrections in both schemes considered, with
 the effect being stronger in the $\ndr$ scheme. Consequently, the
 ratio $z_-/z_+$ relevant for the $\dI$ rule is smaller than in the
 leading order, in contrast to the statements made in
 refs.~\cite{burasweisz:90,altarelli:81}. We will clarify this
 separately below. The $\mu$-dependence of $z_\pm$ is shown for
 $\Lms=300\mev$ in fig.~\ref{fig:1}.
\item[ii)] For $\mu>\mc$ the coefficients $z_i$ ($i\neq1,2$) are zero.
 For $\mu=\mc$ they vanish in the HV scheme and in the $\ndr$ scheme
 they are so small that we do not show them in tab.~\ref{tab:12}. They
 remain small in both schemes
 for $\mu=1\gev$ although considerable enhancements are observed
 in the case of $z_3$ -- $z_6$ for the $\ndr$ scheme. The coefficients
 $z_7$ -- $z_{10}$ being ${\cal O}(\aem)$, can be neglected for all
 practical purposes.
\item[iii)] The coefficients $y_3$ -- $y_6$ are very weakly dependent on
 $\mt$ and only their values for $\mt=130\gev$ are shown. The coefficients
 $y_6$ and $y_4$ are larger than $y_3$ and $y_5$. We notice considerable
 dependence of $y_6$ on $\mu$, $\Lms$ and the scheme considered.
 Whereas in the HV scheme $y_6$ is {\em suppressed} by $\approx$15\%
 relative to the leading order result, in the case of $\ndr$ a
 $\approx$10\% {\em enhancement} of $y_6$ for $\Lms\geq300\mev$ and $\mu
 > \mc$ is observed. For $\mu \le \mc$ however, $y_6$ in the $\ndr$
scheme is slightly suppressed relatively to the leading order result
although its absolute value is always larger than in the HV scheme as
shown in fig.~\ref{fig:2}. The small discontinuity in $y_6$ for $\mu=\mc$ in
the HV scheme is related to the matching discussed in section~4.3. In
the $\ndr$ scheme this discontinuity is larger because in addition
$z_6$ becoming non-zero at $\mu=\mc$ suppresses visibly $|y_6|$ in view
of the formula \eqn{eq:4.3}. All these changes are compensated by the
corresponding changes in the hadronic matrix elements and the
contributions of other operators such as $Q^c_{1,2}$ so that the
physical amplitudes remain $\mu$-independent.  We note that these
effects are essentially invisible in the case of $y_8$ as shown in
fig.~\ref{fig:2}.

\item[iv)] The coefficients $y_7$ and $y_8$ show strong
$\mt$-dependence as illustrated in fig.~\ref{fig:3} and in the tables.
Also as shown in fig.~\ref{fig:4} a sizable $\mt$-dependence is seen
 in the case of $y_9$ and $y_{10}$.  As seen in the tables $y_7$ and
 $y_9$ show only a weak dependence on $\mu$ and $\Lms$.  The
 corresponding dependences of $y_8$ and $y_{10}$ are much stronger.
\item[v)] We note that $y_9$ and $y_{10}$ are substantially larger than
 $y_7$ and $y_8$ and essentially do not change when going from $\ndr$
 to HV. $y_9$ is unaffected by next-to-leading order corrections, but
 $y_{10}$ is considerably suppressed. In ref.~\cite{buchallaetal:90},
 where $Q_9$ and $Q_{10}$ have been eliminated, the effect of $y_9$ and
 $y_{10}$ has been transferred to $y_1$ and $y_2$ which vanish in our
 case.  In tabs.~\ref{tab:3}, \ref{tab:7} and \ref{tab:10} we give only
the $\mt$ dependence of the sum $y_9+y_{10}$ which is relevant for
$\epe$.  We observe that for $\mt < 150\gev$ the sum $|y_9+y_{10}|$ is
enhanced by roughly 20\% over its leading order value but for higher
values of $\mt$ this enhancement is smaller.
\item[vi)] The coefficient $y_7$ is essentially the same in the schemes
considered. As noticed in \cite{flynn:89,buchallaetal:90}, $y_7$
vanishes
 for $\mt\approx145\gev$ in the leading order, and becomes positive for
 larger $\mt$-values. The inclusion of next-to-leading order
 corrections shifts this zero-point to higher values of $\mt$, such as
 $\mt\approx190\gev$ for both schemes. Consequently, $y_7$ remains
 negative for $\mt\approx{\cal O}(150\gev)$, and is roughly of the same
 order of magnitude as $y_8$. Hence for these values of $\mt$ it is
 substantially enhanced over its leading order value. However, for
 $\mt\approx{\cal O}(200\gev)$ the coefficient $y_7$ is so small in
 both schemes that the contribution of $Q_7$ to $\epe$ can be safely
 neglected for such high values of $\mt$.
\item[vii)] Most interesting is however the {\em enhancement} of $y_8$
through next-to-leading corrections as noticed already in
 ref.~\cite{ciuchini:92}, where only the HV scheme has been
 considered.  We find that this enhancement is smaller in the $\ndr$
 scheme. We also note that with increasing $\mt$ the enhancement of
$y_8$ over its leading order values becomes smaller so that for $\mt
\approx \ord(200\gev)$ the $\ndr$ result is very close to its leading
order value.
\item[viii)] As a result of different impacts of next-to-leading order
 corrections on $y_6$ and $y_8$ in the schemes considered, there is a
 strong scheme dependence in the ratio $(y_7/3+y_8)/y_6$, which roughly
 measures the relative importance of the dominant electroweak and
 QCD-penguins in the evaluation of $\epe$. We show this in
 table~\ref{tab:6} for $\mu=\mc$. We observe a large enhancement of this
ratio in the HV scheme which however decreases with increasing $\mt$. In
the $\ndr$ scheme this enhancement is substantially smaller and for $\mt
\approx 190\gev$ the LO and $\ndr$ results are very close to each
other.
\item[ix)] The coefficients in the $\ndrb$ scheme are either equal to
the $\ndr$ coefficients or only slightly different from them. The
largest differences are found for $z_6$ at $\mu < \mc$ and for $y_6$ at
$\mu > \mc$. This has to be kept in mind.
\end{itemize}

\subsection{Comparison with refs.~[12], [13] and [16]}
In refs.~\cite{gaillard:74,altarelli:74}, the enhancement of $z_-/z_+$
from 1 to roughly 3 for $\mu\approx1\gev$ has been found by calculating
QCD effects in the leading logarithmic approximation. Subsequently, it
has been stated in refs.~\cite{altarelli:81} and \cite{burasweisz:90}
that $z_-/z_+$ is further enhanced by next-to-leading logarithmic corrections.
Although the calculations in \cite{burasweisz:90} and \cite{altarelli:81}
are certainly correct, the coefficients on which this conclusion was based,
are really not the coefficients $z_\pm$ of the operators $O_\pm$, but
the scheme independent coefficients given by
\begin{equation}
\tilde z_\pm(\mu) = z_\pm(\mu)\,\Big[ 1 - \frac{\as(\mu)}{4\pi}B_\pm\Big] \, ,
\label{eq:4.43}
\end{equation}
which correspond to
\begin{equation}
\langle\widetilde O_\pm(\mu)\rangle \equiv \langle O_\pm(\mu)\rangle \,
\Big[ 1 + \frac{\as(\mu)}{4\pi}B_\pm\Big] \, .
\label{eq:4.44}
\end{equation}

The coefficients $B_\pm$ for the schemes considered in \cite{altarelli:81}
(DRED) and \cite{burasweisz:90} ($\ndr$,HV), are as follows
\begin{equation}
\begin{array}{lclclcl}
B_+ &=&  \frac{2}{3} \, , &\qquad& B_- &=& -\,\frac{16}{3}\,, \qquad
\hbox{DRED} \\ \svs
B_+ &=& \frac{11}{3} \, , &\qquad& B_- &=& -\,\frac{22}{3}\,, \qquad
\hbox{\ndr} \\ \svs
B_+ &=&  \frac{7}{3} \, , &\qquad& B_- &=& -\,\frac{14}{3}\,, \qquad
\hbox{HV}
\end{array}
\label{eq:4.45}
\end{equation}

The removal of the ${\cal O}(\as)$ corrections in (\ref{eq:4.43})
and (\ref{eq:4.44}) changes the conclusions reached in
\cite{burasweisz:90,altarelli:81}, so that agreement with our paper is
obtained. As stressed already in \cite{burasweisz:90}, the procedure of
\cite{burasweisz:90,altarelli:81} can be used, but then also the calculation
of hadronic matrix elements should include the compensating factors
given in (\ref{eq:4.44}) so that in both treatments the same amplitude is
obtained.

At this stage it should be recalled that similar correction factors are
included in the case of the parameter $B_K$ in the QCD analysis of
the $\varepsilon_K$-parameter \cite{burasjaminweisz:90}. However, there
these corrections are so small that it is irrelevant how they are treated.
In the case at hand the corrections are larger, and different conclusions
about the signs of next-to-leading order corrections can be reached,
dependently whether $z_\pm$ or $\tilde z_\pm$ are considered. Similar
comments apply to our paper of ref.~\cite{burasetal:92a}, where scheme
independent coefficients $\tilde z_1$ -- $\tilde z_6$ and
$\tilde y_1$ -- $\tilde y_6$ have been used.

Although the introduction of these additional corrections may in
principle identify the full physical effect, as discussed in
\cite{burasetal:92a}, it seems to us at present that for the future
treatment of hadronic matrix elements it is more elegant, more
convenient, and more transparent not to perform additional
``rotations'' to scheme independent bases. Indeed, on one hand such a
procedure is not unique and on the other hand it may in certain cases
cause misleading conclusions. The observed suppression of $z_-/z_+$
through next-to-leading order corrections in the scheme considered must
of course be compensated by the corresponding effects in the hadronic
matrix elements, in order to explain the $\dI$ rule as we will discuss
in more detail in section~6.

\newsection{Hadronic Matrix Elements}
\subsection{General Remarks}
An important ingredient of any analysis of non-leptonic $K$-decays are the
hadronic matrix elements of operators $Q_i$ which we denote by
\begin{equation}
\langle Q_i \rangle_I \equiv
\langle \left(\pi\pi\right)_I \left| Q_i \right| K \rangle \, ,
\qquad
I = 0,2 \, .
\label{eq:5.1}
\end{equation}
These matrix elements depend generally on the scale $\mu$ and on the
renormalization scheme used for the operators. These two dependences
are cancelled by those present in the coefficients $C_i(\mu)$ so that
the effective Hamiltonian and the resulting amplitudes do not depend on
$\mu$ and on the scheme used to renormalize the operators.

At this point, it should be emphasized that eq.~\eqn{eq:1.2} is valid
beyond perturbation theory so that in principle $\mu$ can be chosen
completely arbitrary. It can even be set to $\mu=0$. In the spirit of
ref.~\cite{bardeen:87b}, we can then divide the renormalization group
evolution into a {\em short distance evolution} from $\mw$ to $\mu$
described by
\begin{equation}
\vC(\mu) = \hU(\mu,\mw,\aem) \; \vC(\mw) \, ,
\label{eq:5.2}
\end{equation}
and a {\em long distance evolution} from scale 0 {\em up to} $\mu$,
described by
\begin{equation}
\langle \vQ^T(\mu) \rangle = \langle \vQ^T(0) \rangle \; \hU(0,\mu,\aem) \, ,
\label{eq:5.3}
\end{equation}
where $\hU(0,\mu,\aem)$ is the evolution matrix in the long distance
regime. Although this evolution cannot be evaluated at the lower end in
a perturbative framework, it has general features similar to the
evolution in the short distance regime as demonstrated in
\cite{bardeen:87b}. In particular, the $\mu$ and scheme dependences at
the {\em upper} end of the evolution (around $\mu$) must match the ones
present in the coefficients $C_i(\mu)$ in such a way that these
dependences are not present in the physical amplitudes. In other words
it is a matter of choice what belongs to the matrix element and what to
the coefficient function.

{}From this discussion it is evident that the scheme dependence of
$\langle Q_i(\mu) \rangle$ is only present at the upper end of the long
distance evolution. Consequently, for sufficiently high $\mu$ this
dependence can be calculated together with the $\mu$-dependence in a
perturbative framework. The relation between the matrix elements
calculated in two different schemes is given in \eqn{eq:3.13}. The
relation between matrix elements evaluated at two scales $m_1 < m_2$ is
simply given by
\begin{equation}
\langle \vec{Q}^T(m_2) \rangle =
\langle \vec{Q}^T(m_1) \rangle \; \hU(m_1,m_2,\aem) \, ,
\label{eq:5.3b}
\end{equation}
where $\hU$ is the evolution matrix of section 2.3. Thus, if there is a
method to calculate $\langle \vec{Q}(\mu) \rangle$ at a single scale
$\mu = m_1$, the renormalization group will give us these matrix
elements at any other scale $\mu \not= m_1$.  Unfortunately, the full
long distance evolution involves all scales down to $\mu=0$ and
consequently, the actual evaluation of $\langle Q_i(m_1) \rangle$ can
only be done in a non-perturbative framework. In spite of this, it turns
out to be very instructive to analyze the evolution of $\langle \vQ(\mu)
\rangle$ given in \eqn{eq:5.3b} in a range of $\mu$ for which
$\hU(m_1,m_2,\aem)$ can be calculated as in section~4. Since this has
not been studied so far in the literature let us recall that whereas
the evolution of $\vC(\mu)$ in \eqn{eq:5.2} in governed by $\hg^T$, the
evolution of $\langle \vQ(\mu) \rangle$ is determined by $\hg$. Now the
matrix $\hg$ is rather asymmetric
\cite{burasweisz:90}--\nocite{burasetal:92a,burasetal:92b}\cite{burasetal:92c}.
In particular the elements $\hg_{i1}=\hg_{i2}$ for $i=3,\ldots,10$
vanish. Consequently, the structure of the evolution in \eqn{eq:5.3b} is
rather different from the evolution in \eqn{eq:5.2}. Thus, whereas the
evolution of $C_{1,2}(\mu)$ is unaffected by the presence of penguin
contributions, the evolution of $\langle Q_{1,2}(\mu) \rangle$ depends
on the size of $\langle Q_i(\mu) \rangle$, $i\not=1,2$. Conversely,
whereas the evolution of $C_i(\mu)$, $i\not=1,2$ depends on the size of
$C_{1,2}(\mu)$, the evolution of the matrix elements of penguin
operators $\langle Q_i(\mu) \rangle$, $i\not=1,2$ is a sole penguin
affair. This different structure brings certain surprises as we will
see in section~6.

During the last years there have been many attempts to calculate the
matrix elements $\langle Q_i \rangle$ by using the lattice approach,
$1/N$-approach, QCD sum rules, hadronic sum rules, effective QCD action
\cite{pichderafael:91}, chiral perturbation theory \cite{kamboretal:91}
and vacuum insertion.  Yet, it is fair to say that only in the
$1/N$-expansion of refs.~\cite{bardeen:87b,bardeen:87}  and in the
vacuum insertion approach have all matrix elements $\langle Q_i
\rangle$ been calculated. Some subset of the $\langle Q_i \rangle$ is
also known from the lattice method, sum rules and effective QCD action
approach of ref.~\cite{pichderafael:91}.

At this stage it should be stressed that the $1/N$-approach, although
bearing some similarities to the vacuum insertion (factorization)
method, is really a systematic non-perturbative method for QCD
calculations, whereas this cannot be said about the vacuum insertion
approach.  Moreover, there are important quantitative differences
between these two methods in the $(Q_1,Q_2)$ sector. Whereas the $1/N$
approach offers a plausible description of $CP$-conserving $\Kpipi$
amplitudes such a description is not possible in the vacuum insertion
approach. In particular, the understanding of the $\dI$ rule is
completely missing in the factorization approach. Next, one should
mention a complete failure of the factorization method in a large class
of exclusive non-leptonic $B$- and $D$-decays, whereas a reasonable
description of these decays can be achieved in the $1/N$-approach
\cite{bauerstech:85,burasetal:86}.  Finally, the value for $B_{\rm K}
\approx 0.70 \pm 0.10$ obtained in the $1/N$-approach is rather close
to the present ``world average'' value given in \eqn{eq:7.7} which
should be contrasted with $B_{\rm K}=1$ obtained in the vacuum
insertion approach.

Concerning the matrix elements of the dominant penguin operators the
structure of the $1/N$ and the vacuum insertion formulae is similar.
However, it should be stressed that whereas the scale $\Lambda_\chi$
entering these formulae can in the $1/N$ method be related to the ratio
$F_{\rm K}/F_\pi$ \cite{georgietal:86,bardeen:87}, no such relation
exists in the vacuum insertion approach. Consequently, a quantitative
estimate of these matrix elements in the latter approach is essentially
impossible.

Concerning the $\mu$-dependence it is well known that the matrix
elements of $(Q_1,Q_2)$ calculated in the vacuum insertion method are
$\mu$-independent in contradiction with one of the general properties
of the hadronic matrix elements of operators carrying anomalous
dimensions.  In the $1/N$-approach of ref.~\cite{bardeen:87b}, the
$\mu$-dependence in $\langle Q_i(\mu) \rangle$, $i=1,2$ arises through
loop effects in a meson theory and at the semi-quantitative level the
matching in $\mu$ between long and short distance calculations is
indeed possible. A problem exists however because in the approach of
ref.~\cite{bardeen:87b} the long distance evolution in \eqn{eq:5.3} can
only be extended to scales $\mu \approx 0.6\gev$ at which perturbative
evaluation of the Wilson coefficients $C_i(\mu)$ is questionable. There
is a hope that by including vector meson contributions to the long
distance evolution the scale $\mu$ could be increased to $\mu \approx
1\gev$ \cite{gerard:88}. However, such a calculation is very difficult
and in any case not available for $\Kpipi$ at present.

On the other hand, the $\mu$-dependence of the for $\Kpipi$ dominant
penguin operators $\langle Q_6(\mu) \rangle$ and $\langle Q_8(\mu)
\rangle$ is to a large extend under control in both approaches. It is
to a very good approximation given by the $\mu$-dependence of the
running strange quark mass $1/\ms^2(\mu)$ as we will demonstrate below.
This $\mu$-dependence cancels to a very good approximation the
$\mu$-dependence of $C_6(\mu)$ and $C_8(\mu)$ so that in quantities in
which $Q_6$ and $Q_8$ dominate, the left-over $\mu$-dependence is
rather weak. The actual issue then however is what the value of $\ms$
at a fixed value of $\mu$, say $\mu\approx 1\gev$ is. We will return to
this question as well as to the $\mu$-dependence of other penguin
operators below.

Next, the renormalization scheme dependence of $\langle Q_i(\mu) \rangle$
should be addressed. The vacuum insertion method is certainly
insensitive to this dependence. It appears at present that this is also
the case of the $1/N$ approach. Evidently, in these two approaches the
scheme dependence of $C_i(\mu)$ cannot be cancelled and this feature
introduces an additional theoretical uncertainty. We will investigate
the size of this uncertainty below.

At this stage a few remarks about the lattice approach are in order
\cite{sharpe:90}--
\nocite{bernardetal:90,gavelaetal:88,kilcupetal:90,kilcup:91}\cite{sharpe:91}.
In this approach the $\mu$-dependence and the scheme dependence of
$\langle Q_i(\mu) \rangle$ is in principle calculable
\cite{martinelli:84}--
\nocite{bernardetal:87,gloterman:84,daniel:88,sheard:89}\cite{curcietal:88}.
However, the accuracy of the lattice methods is insufficient at present
to contribute very much to the issue of the $\mu$- and scheme
dependence at a quantitative level. This is evident from the strategy
of ref.~\cite{ciuchini:92} which we will discuss in more detail below.
Moreover, the matrix elements $\langle Q_1 \rangle_0$ and $\langle Q_2
\rangle_0$ have not been calculated on the lattice yet.

It will be useful in what follows to recall the matrix elements used in
refs.~\cite{buchallaetal:90,froehlich:91,heinrichetal:92,ciuchini:92},
and to compare them with each other. In section 6, we will go beyond
these approaches to find matrix elements which we believe are closer to
the true QCD matrix elements than these given by the methods just
discussed.

\subsection{Explicit Formulae for the Matrix Elements}
Here, we will give explicit formulae for $\langle Q_i\rangle_I$ which
are general enough that the matrix elements used in
refs.~\cite{buchallaetal:90,froehlich:91,heinrichetal:92,ciuchini:92}
could be compared with each other. In order to simplify the
presentation, we will use the relations between operators given in
section~4.5. As discussed there care must be taken when using these
relations in the $\ndr$ and $\ndrb$ schemes. In section~6, we will
outline our strategy in using the formulae given below.

Taking first $\mu < \mc$, we have
\begin{eqnarray}
\langle Q_1 \rangle_0 &=& -\,\frac{1}{9} X B_1^{(1/2)} \, ,
\label{eq:5.4} \\
\langle Q_2 \rangle_0 &=&  \frac{5}{9} X B_2^{(1/2)} \, ,
\label{eq:5.5} \\
\langle Q_3 \rangle_0 &=&  \frac{1}{3} X B_3^{(1/2)} \, ,
\label{eq:5.6} \\
\langle Q_4 \rangle_0 &=&  \langle Q_3 \rangle_0 + \langle Q_2 \rangle_0
                          -\langle Q_1 \rangle_0 \, ,
\label{eq:5.7} \\
\langle Q_5 \rangle_0 &=&  \frac{1}{3} B_5^{(1/2)}
                           \langle \overline{Q_6} \rangle_0 \, ,
\label{eq:5.8} \\
\langle Q_6 \rangle_0 &=&  -\,4 \sqrt{\frac{3}{2}}
\left[ \frac{m_{\rm K}^2}{\ms(\mu) + \md(\mu)}\right]^2
\frac{F_\pi}{\kappa} \,B_6^{(1/2)} \, ,
\label{eq:5.9} \\
\langle Q_7 \rangle_0 &=&
- \left[ \frac{1}{6} \langle \overline{Q_6} \rangle_0 (\kappa + 1)
         - \frac{X}{2} \right] B_7^{(1/2)} \, ,
\label{eq:5.10} \\
\langle Q_8 \rangle_0 &=&
- \left[ \frac{1}{2} \langle \overline{Q_6} \rangle_0 (\kappa + 1)
         - \frac{X}{6} \right] B_8^{(1/2)} \, ,
\label{eq:5.11} \\
\langle Q_9 \rangle_0 &=&
\frac{3}{2} \langle Q_1 \rangle_0 - \frac{1}{2} \langle Q_3 \rangle_0 \, ,
\label{eq:5.12} \\
\langle Q_{10} \rangle_0 &=&
    \langle Q_2 \rangle_0 + \frac{1}{2} \langle Q_1 \rangle_0
  - \frac{1}{2} \langle Q_3 \rangle_0 \, ,
\label{eq:5.13}
\end{eqnarray}
\begin{eqnarray}
\langle Q_1 \rangle_2 &=&
\langle Q_2 \rangle_2 = \frac{4 \sqrt{2}}{9} X B_1^{(3/2)} \, ,
\label{eq:5.14} \\
\langle Q_i \rangle_2 &=&  0 \, , \qquad i=3,\ldots,6 \, ,
\label{eq:5.15} \\
\langle Q_7 \rangle_2 &=&
  -\left[ \frac{\kappa}{6 \sqrt{2}} \langle \overline{Q_6} \rangle_0
          + \frac{X}{\sqrt{2}}
   \right] B_7^{(3/2)} \, ,
\label{eq:5.16} \\
\langle Q_8 \rangle_2 &=&
  -\left[ \frac{\kappa}{2 \sqrt{2}} \langle \overline{Q_6} \rangle_0
          + \frac{\sqrt{2}}{6} X
   \right] B_8^{(3/2)} \, ,
\label{eq:5.17} \\
\langle Q_9 \rangle_2 &=&
   \langle Q_{10} \rangle_2 = \frac{3}{2} \langle Q_1 \rangle_2 \, ,
\label{eq:5.18}
\end{eqnarray}
where
\begin{equation}
\kappa = \frac{\Lambda_\chi^2}{m_{\rm K}^2 - m_\pi^2} =
         \frac{F_\pi}{F_{\rm K} - F_\pi} = 4.55 \, ,
\label{eq:5.19}
\end{equation}
\begin{equation}
X = \sqrt{\frac{3}{2}} F_\pi \left( m_{\rm K}^2 - m_\pi^2 \right)
  = 3.71 \cdot 10^{-2}\gev^3 \, ,
\label{eq:5.20}
\end{equation}
and
\begin{equation}
\langle \overline{Q_6} \rangle_0 =
   \frac{\langle Q_6 \rangle_0}{B_6^{(1/2)}} \, .
\label{eq:5.21}
\end{equation}
For the ease of the reader, we give all the actual numerical
values for $\Lambda_\chi$, $m_{\rm K}$, $m_\pi$, $F_{\rm K}$, $F_\pi$ that
produce these ``magic numbers'' above in appendix~C.

The overall normalization in eqs.~\eqn{eq:5.4}--\eqn{eq:5.21} agrees
with \cite{buchallaetal:90}, but differs by a factor $\sqrt{3/2}$
relative to the normalization used in \cite{ciuchini:92}. This
difference cancels however in the final result for $\epe$ and other
physical quantities.

For $\mc < \mu < \mb$ the relations \eqn{eq:4.28} are valid and the
formulae for $\langle Q_4 \rangle_0$, $\langle Q_9 \rangle_0$ and
$\langle Q_{10} \rangle_0$ change:
\begin{eqnarray}
\langle Q_4 \rangle_0 &=&
\langle Q_3 \rangle_0 + \langle Q_2 \rangle_0 - \langle Q_1 \rangle_0 +
\langle Q_2^c \rangle_0 - \langle Q_1^c \rangle_0 \, ,
\label{eq:5.22} \\
\langle Q_9 \rangle_0 &=&
\frac{3}{2} \langle Q_1 \rangle_0 - \frac{1}{2} \langle Q_3 \rangle_0 +
\frac{3}{2} \langle Q_1^c \rangle_0 \, ,
\label{eq:5.23} \\
\langle Q_{10} \rangle_0 &=&
\langle Q_4 \rangle_0 + \langle Q_9 \rangle_0 - \langle Q_3 \rangle_0 \, .
\label{eq:5.24}
\end{eqnarray}
The matrix elements $\langle Q_{1,2}^c \rangle_{0,2}$ although
vanishing in the vacuum insertion method may be non-zero in general as
discussed below. Also eq.~\eqn{eq:5.15} is modified by $\ord(\aem)$
corrections.  In the vacuum insertion method $B_i \equiv 1$.

\subsection{Critical Comparison of Various Matrix Elements Estimates}
Having formulae \eqn{eq:5.4}--\eqn{eq:5.24} at hand, we can compare
these matrix elements with the ones used in
refs.~\cite{buchallaetal:90,froehlich:91,heinrichetal:92,ciuchini:92}.

In ref.~\cite{buchallaetal:90}, eqs.~\eqn{eq:5.4}--\eqn{eq:5.21} have
been used with the terms ``1'' and $X$ in \eqn{eq:5.10}, \eqn{eq:5.11}
and $X$ in \eqn{eq:5.16}, \eqn{eq:5.17} omitted and $B_i^{(\Delta I)}
\equiv 1$ except for
\begin{equation}
B_1^{(1/2)} = 5.2  \, , \qquad
B_2^{(1/2)} = 2.2  \, , \qquad
B_1^{(3/2)} = 0.55 \, ,
\label{eq:5.27}
\end{equation}
in accordance with the $1/N$-approach of ref.~\cite{bardeen:87b}. The
fact that $B_i^{(1/2)} > 1$ for $i=1,2$ and $B_1^{(3/2)} < 1$ allows to
come closer to the $\dI$ rule than it is possible in the vacuum
insertion method. Furthermore, $\mu = 1\gev$ and $125\mev \le
\ms(1\gev) \le 200\mev$ have been used. At this point we should mention
that the approach of Pich and de~Rafael \cite{pichderafael:91} gives
similar results to \cite{bardeen:87b} for $\Kpipi$ amplitudes.

The Dortmund group \cite{froehlich:91,heinrichetal:92} uses essentially
the above formulae with $0.6\gev \le \mu \le 1\gev$. The main
difference with respect to \cite{buchallaetal:90} is the modified
factor $B_6^{(1/2)}$ for the operator $Q_6$. Using the tables of
ref.~\cite{froehlich:91,heinrichetal:92}, we find
\begin{equation}
B_6^{(1/2)} \approx 2.4 \pm 0.2 \, .
\label{eq:5.28}
\end{equation}
The authors of \cite{froehlich:91,heinrichetal:92} also use $100\mev
\le \ms(1\gev) \le 175\mev$. The difference of $B_6^{(1/2)}$ from unity
follows according to ref.~\cite{froehlich:91,heinrichetal:92} from
next-to-leading $1/N$ corrections for which the details are
unfortunately not available.  $B_8^{(3/2)}$ extracted from
\cite{froehlich:91,heinrichetal:92} remains very close to unity.

The Rome group \cite{ciuchini:92} uses the set of matrix elements
relevant for $\mu = 2\gev$ with
\begin{equation}
\begin{array}{lclclcl}
B_{3,4}^{(1/2)} & = & 1\;-\;6^{(*)} \,, &\qquad&
B_{5,6}^{(1/2)} & = & 1.0 \pm 0.2 \,, \\
B_{7,8,9}^{(1/2)} & = & 1^{(*)} \,, &\qquad&
B_{7,8}^{(3/2)}&=&1.0\pm0.2\,, \\
B_{1,2}^c & = & 0\;-\;0.15^{(*)} \,, &\qquad&
B_9^{(3/2)}&=&0.8 \pm 0.2 \, ,
\end{array}
\label{eq:5.29}
\end{equation}
where entries with ${}^{(*)}$ are educated guesses and the remaining
values are taken from lattice calculations
\cite{sharpe:90}--
\nocite{bernardetal:90,gavelaetal:88,kilcupetal:90,kilcup:91}\cite{sharpe:91}.
Here, $B_{1,2}^c$ enters the matrix elements $\langle
Q_{1,2}^c \rangle_0$ as follows
\begin{equation}
\langle Q_1^c \rangle_0 = -\frac{1}{3} \; X \; B_1^c \, ,
\qquad \qquad
\langle Q_2^c \rangle_0 = X \; B_2^c \, ,
\label{eq:5.29b}
\end{equation}
where we have introduced the minus sign in order to make $B_1^c$
positive.  In the vacuum insertion approximation $B_{1,2}^c = 0$
holds.  However, in QCD it could certainly be non-zero. The range of
values for $B_{1,2}^c$ in \eqn{eq:5.29} is an educated guess by the
authors of ref.~\cite{ciuchini:92}. As we will demonstrate in section~6,
the size of $\langle Q_{1,2}^c \rangle_0$ is related to the size of
$\langle Q_i \rangle_0$, $i=3,\ldots,6$, and it is possible to go beyond
educated guesses and calculate $B_{1,2}^c$. Since as seen in
\eqn{eq:4.27e} and stressed in ref.~\cite{ciuchini:92} the coefficients
$y_{1,2}^c$ are large the operators $Q_{1,2}^c$ could have in principle
some impact on $\epe$. Our calculations in section~6 show that the
educated guess of \cite{ciuchini:92} overestimates $B_{1,2}^c$ by a
factor of four, and consequently, $Q_{1,2}^c$ play only a minor role for
$\mu \le 2\gev$. A detailed comparison with ref.~\cite{ciuchini:92}
will be given later on.

Furthermore, the Rome group uses
\begin{equation}
\ms(2\gev) = (170 \pm 30)\mev \, .
\label{eq:5.30}
\end{equation}
which corresponds to
\begin{equation}
\ms(1\gev) = (225 \pm 30)\mev \, ,
\label{eq:5.31}
\end{equation}
and is considerably higher than the values used in the other two
papers.  Consequently, the matrix elements $\langle Q_6 \rangle$ and
$\langle Q_8 \rangle$ evaluated at $\mu = 1\gev$ are smaller than in
ref.~\cite{buchallaetal:90} and substantially smaller than in
\cite{froehlich:91,heinrichetal:92}. Equivalently, the matrix elements
used by the Rome group at $\mu = 2\gev$ are close to the matrix
elements used in \cite{buchallaetal:90} where $\mu = 1\gev$ has been
taken. But since the coefficients $C_i(\mu)$ of penguin operators
decrease with increasing $\mu$ the penguin contributions to $\epe$ in
ref.~\cite{ciuchini:92} are bound to be smaller than in the remaining
two papers. We will return to this point below.

It should also be remarked that whether the terms ``1'' and $X$ are
kept or dropped in the matrix elements of $Q_7$ and $Q_8$ is
essentially immaterial for $\epe$. Although $\langle Q_7 \rangle_0$,
$\langle Q_7 \rangle_2$ and $\langle Q_8 \rangle_0$ are modified by
roughly 15\%, these matrix elements do not play a considerable role in
$\epe$. On the other hand $\langle Q_8 \rangle_2$ which is important
is modified by only 2\%.

Next, we would like to point out one additional problem with the matrix
elements given above. They have been calculated in QCD without taking
order $\ord(\aem)$ corrections into account. It is a simple matter to
convince oneself that all the matrix elements listed in
\eqn{eq:5.4}--\eqn{eq:5.18} and \eqn{eq:5.29b} receive $\ord(\aem)$
contributions because at $\ord(\aem)$ they mix under renormalization
with the operators $Q_7$ and $Q_9$ which have $\ord(1)$ matrix
elements. In particular one has
\begin{equation}
\langle Q_{3-6}   \rangle_2 = \ord(\aem) \qquad
\langle Q^c_{1,2} \rangle_2 = \ord(\aem)
\label{eq:5.32}
\end{equation}
to be compared with \eqn{eq:5.15} and $\langle Q^c_{1,2} \rangle_2=0$
used in ref.~\cite{ciuchini:92}. The $\ord(\aem)$ corrections to these
matrix elements are necessary in order to cancel the scheme dependence
present in the Wilson coefficients as discussed in detail in
section~3.  The $\ord(\aem)$ corrections to penguin matrix elements
should be dropped because they contribute $\ord(\aem\;\as)$ or
$\ord(\aem^2)$ effects to the amplitudes. On the other hand the
$\ord(\aem)$ corrections to the matrix elements of $Q_1$ and $Q_2$
should be taken into account because they are exactly at the same level
as the $\ord(\aem)$ effects which have been taken into account when
evaluating the initial conditions at $\mu \approx \mw$ in section~4.
Similarly $\ord(\aem)$ effects to $\langle Q^c_{1,2} \rangle_0$ and
$\langle Q^c_{1,2} \rangle_2$ have to be included.

After this rather critical presentation of the existing estimates of
hadronic matrix elements, we will now turn to an approach which overcomes
several difficulties of the methods just discussed.

\newsection{A Phenomenological Approach to Hadronic \ \ Matrix Elements}
\subsection{General Remarks}
We have seen that the present estimates of hadronic matrix elements by
means of existing non-perturbative methods suffer from many
deficiencies, in particular:
\begin{itemize}
\item
The hadronic matrix elements of the dominant penguin operators $Q_6$ and
$Q_8$ are poorly known due to the poor knowledge of $\ms(1\gev)$.
\item
The size of hadronic matrix elements of $Q_1$ and $Q_2$ is not fully in
accordance with the known data for $\Kpipi$.
\item
The renormalization and $\mu$-dependences of the hadronic matrix
elements are not under control.
\end{itemize}
These deficiencies imply another weak point of the $\epe$ analyses of
refs.~\cite{buchallaetal:90,froehlich:91,heinrichetal:92}.
The usual practice is to vary $\ms(1\gev)$ in eq.~\eqn{eq:5.9}, while
keeping the matrix elements of current-current operators fixed. Yet,
such a procedure fails to fit the data on $CP$-conserving $\Kpipi$
amplitudes since for a given value of $\ms$ only certain matrix
elements of current-current operators can fit these data.

We would like to propose here a phenomenological approach to obtain the
hadronic matrix elements which incorporates properly the
renormalization scheme and $\mu$-dependences of the hadronic matrix
elements as given by QCD, and at the same time assures that the
theoretical $CP$-conserving $\Kpipi$ amplitudes are in agreement with
the data. The basic ingredients in this approach are
\begin{itemize}
\item
the assumption that $\Heff$ of \eqn{eq:4.1} properly describes the
existing data on $CP$-conserving $\Kpipi$ amplitudes.
\item
the Wilson coefficient functions calculated reliably in QCD and QED in
section~4.
\item
some relations between matrix elements which are valid in QCD and are
consistent with the renormalization group properties.
\end{itemize}

Our phenomenological approach to the hadronic matrix elements involves
four basic parameters:
\begin{itemize}
\item
the QCD scale parameter $\Lms$ in the $\overline{MS}$ renormalization
scheme,
\item
the $B$-factors $B_{2}^{(1/2)}(\mc)$, $B_{6}^{(1/2)}(\mc)$,
and $B_{8}^{(3/2)}(\mc)$, parameterizing the matrix elements
$\langle Q_2(\mc)\rangle_0$, $\langle Q_6(\mc)\rangle_0$, and
$\langle Q_8(\mc)\rangle_2$ respectively, as defined in
eqs.~(\ref{eq:5.5}), (\ref{eq:5.9}), and (\ref{eq:5.17}).
\end{itemize}
The choice $\mu=\mc$ turns out to be very convenient as we will see
soon, but is not necessary.

For given values of these four parameters and the chosen
renormalization scheme for operators our semi-phenomenological approach
gives all the hadronic matrix elements of the dominant operators while
being consistent with the measured $CP$-conserving $\Kpipi$
amplitudes.  In this way an interesting correlation between matrix
elements of current-current and penguin operators absent in the
analyses of
refs.~\cite{buchallaetal:90,froehlich:91,heinrichetal:92,ciuchini:92}
arises.

Having calculated all matrix elements this way, we will be able to give
predictions for $\epe$ as functions of the four basic parameters listed
above and of $\mt$.

In what follows, we will first give the basic formulae of our
phenomenological approach. Subsequently, we will calculate the hadronic
matrix elements $\langle Q_i \rangle_{0,2}$ and we will compare our
results with those discussed in the previous section.

In our numerical analysis we use the following experimental data
\cite{particledata:92}
\begin{equation}
\RE A_0 = 33.3 \cdot 10^{-8}\gev \, ,
\qquad \qquad
\RE A_2 = 1.50 \cdot 10^{-8}\gev \, ,
\label{eq:6.3}
\end{equation}
which express the $\dI$ rule
\begin{equation}
\frac{\RE A_0}{\RE A_2} \equiv \frac{1}{\omega} = 22.2 \, .
\label{eq:6.4}
\end{equation}

\subsection{$\langle Q_i(\mu) \rangle_2$ for $(V-A) \otimes (V-A)$ Operators}
It turns out that to a very good approximation all the matrix elements
$\langle Q_i(\mu) \rangle_2$ can be determined from $\RE A_2$ in
\eqn{eq:6.3} as functions of $\Lms$, $\mu$ and the renormalization
scheme considered.

We first introduce the $Q_\pm$ operators
\begin{equation}
Q_\pm = \frac{1}{2} \big(Q_2 \pm Q_1\big) \, ,
\qquad \qquad
z_\pm = z_2 \pm z_1 \, ,
\label{eq:6.5}
\end{equation}
and using \eqn{eq:4.1a}, we find
\begin{equation}
\RE A_2 = c z_+(\mu) \langle Q_+(\mu) \rangle_2 \, ,
\label{eq:6.6}
\end{equation}
where
\begin{equation}
c \equiv \frac{G_{\rm F}}{\sqrt{2}} \; \V{ud} \V{us}^* =
         1.77 \cdot 10^{-6} \gev^{-2} \, .
\label{eq:6.7}
\end{equation}

In obtaining \eqn{eq:6.6}, we have set $\tau \; y_i = 0$ in view of
$z_i >> \tau \; y_i$, and we have neglected the contributions of the
electroweak penguin operators which having $z_i \approx \ord(\aem)$
play only a secondary role in $CP$-conserving amplitudes. Recall also
that $\langle Q_i \rangle_2 = \ord(\aem)$ for $i=3,\ldots,6$ and also
$\langle Q_{1,2}^c \rangle_2 = \ord(\aem)$. Consequently, \eqn{eq:6.6}
is valid for any scale $\mu$ in the limit $\aem=0$. Moreover for
$\mu=\mc$ the contributions of penguin operators and of $Q^c_{1,2}$ (see
below) is either zero (HV,LO) or very small ($\ndr$) and at this scale
\eqn{eq:6.6} is an excellent approximation even for $\aem \not= 0$.

We note next that (see \eqn{eq:5.14})
\begin{equation}
\langle Q_1(\mu) \rangle_2 = \langle Q_2(\mu) \rangle_2 =
\langle Q_+(\mu) \rangle_2 \; .
\label{eq:6.8}
\end{equation}
This relation is valid in pure QCD in which isospin is conserved. It is
consistent with the QCD renormalization group evolution. This follows
from the fact that the anomalous dimension matrix $\hg_{\rm s}$ in the
$(Q_1,Q_2)$ sector is symmetric  with equal diagonal entries and the
fact that $\langle Q_i \rangle_2 = 0$ for $i=3,\ldots,6$. Here, we neglect
the mixing with electroweak penguin operators which is $\ord(\aem)$. The
effect of this mixing will be analyzed below.

Using the experimental value for $\RE A_2$, together with \eqn{eq:6.6}
and \eqn{eq:6.8}, we find the matrix elements $\langle Q_{1,2} \rangle_2$,
\begin{equation}
\langle Q_1(\mu) \rangle_2 = \langle Q_2(\mu) \rangle_2 =
\frac{\RE A_2}{c\,z_+(\mu)} \, .
\label{eq:6.9}
\end{equation}
Since $z_+(\mu)$ depends on $\mu$ and the renormalization scheme used,
\eqn{eq:6.9} gives us automatically the scheme and $\mu$-dependence of
the matrix elements in question. It also gives the $\Lms$ dependence of
$\langle Q_{1,2} \rangle_2$.

Using the relation \eqn{eq:5.18} and \eqn{eq:6.9}, we find the matrix
elements $\langle Q_{9,10} \rangle_2$
\begin{equation}
\langle Q_9(\mu) \rangle_2 = \langle Q_{10}(\mu) \rangle_2 =
\frac{3\RE A_2}{2c\,z_+(\mu)} \, .
\label{eq:6.10}
\end{equation}

As discussed in section~4.5, eq.~\eqn{eq:5.18} is valid in all schemes
for $\aem=0$, and so is \eqn{eq:6.10} in this limit.

A numerical analysis shows that the $\mu$-dependence
in \eqn{eq:6.10} does not cancel the $\mu$-dependence of $z_9(\mu)$ and
$z_{10}(\mu)$. This is not surprising, because due to the mixing of $Q_9$
and $Q_{10}$ with other operators the cancellation of the $\mu$-dependence
involves more operators. It should be stressed that the matrix elements
in \eqn{eq:6.9} and \eqn{eq:6.10} do not involve any parameters except
for $\Lms$. Although the impact of $\ord(\aem)$ corrections to the
determination of these matrix elements is very small we would like to
discuss here the matrix elements $\langle Q^c_{1,2} \rangle_2$ which
as stated in \eqn{eq:5.32} are $\ord(\aem)$. The point is that these
matrix elements can be calculated ! First they can only be non-zero for
$\mu \ge \mc$ because for $\mu \le \mc$  the operators $Q^c_{1,2}$
disappear from the effective theory.  Requiring the $\mu$-independence
of $\RE A_2$ and noting that in the HV and $\ndr$ schemes $z_i(\mc)=0$
for $i=3,\ldots,10$ we find the following ``initial conditions''
\begin{equation}
\langle Q^c_1(\mc) \rangle_2 = \langle Q^c_2(\mc) \rangle_2 = 0
\label{eq:6.10a}
\end{equation}
valid in these schemes. The evolution \eqn{eq:5.3b} of these matrix
elements to scales $\mu > \mc$ generates calculable non-vanishing
values of $\ord(\aem)$. Using \eqn{eq:3.13} one can also find the
corresponding initial conditions in the $\ndr$ scheme. The inclusion of
these operators in necessary if one want to keep $\RE A_2$
$\mu$-independent.  These points will be illustrated more explicitly in
the case of $\langle Q^c_i(\mu) \rangle_0$.

Comparing \eqn{eq:6.9} and \eqn{eq:6.10} with the general formulae
\eqn{eq:5.14} and \eqn{eq:5.18}, respectively, we can extract
$B_1^{(3/2)}(\mu)$ which with the approximation used is common for $Q_1$,
$Q_2$, $Q_9$ and $Q_{10}$. The result is represented by the solid lines
in figs.~\ref{fig:12} and \ref{fig:13} for $Q_{1,2}$ and $Q_{9,10}$,
respectively. The remaining lines in these figures represent exact RG
evolution. They are obtained by evaluating first the matrix elements at
$\mu=\mc$ with the help of \eqn{eq:6.9} and \eqn{eq:6.10}, and evolving
subsequently to other values of $\mu$ by means of the evolution
equation \eqn{eq:5.3b}. In the evolution we have used $\Lms=300\mev$.

We observe the following points:
\begin{itemize}
\item
Our formulae \eqn{eq:5.6} and \eqn{eq:5.7} are good approximations
to the exact evolution including $\ord(\aem)$ effects for $1\gev \le \mu
\le 2\gev$, considered mainly in this paper. For $\langle Q_1 \rangle_2$
and $\langle Q_9 \rangle_2$ the approximations become slightly worse
when $\mu$ is increased beyond this range.
\item
We observe a sizable $\mu$-dependence of $B_1^{(3/2)}(\mu)$.
\item
The next-to-leading order corrections to $z_{1,2}(\mu)$ decrease
$B_1^{(3/2)}(\mu)$ relative to the leading order determination with the
effect being stronger in the case of the $\ndr$ scheme.
\item
Most importantly however, we find $B_{\rm HV}^{(3/2)}(\mu\approx 2\gev)
\approx 0.44$, to be compared with $B_9^{(3/2)} \approx 0.8 \pm 0.2$ in
\eqn{eq:5.29}. We conclude therefore that the lattice estimate for
$\langle Q_{9,10}(2\gev) \rangle$ of ref.~\cite{ciuchini:92}
overestimates these matrix elements by almost a factor of two. We will
investigate the impact of this on $\epe$ in section~8. The $1/N$
approach used in \cite{buchallaetal:90} is in a much better shape here
because for low $\mu$, $B_1^{(3/2)} \approx 0.55$ in \eqn{eq:5.27} is
quite consistent with figs.~\ref{fig:12} and \ref{fig:13} for $\mu
\approx 1\gev$.
\end{itemize}

\subsection{$\langle Q_i(\mu) \rangle_0$ for $(V-A)\otimes (V-A)$ Operators}
The determination of $\langle Q_i(\mu) \rangle_0$ operators is more
involved because several operators may contribute to $\RE A_0$ and a
relation like \eqn{eq:6.8} does not exist in this case. Yet, some
progress beyond the approaches of section~5 can also be made in this
case. As we will see the choice $\mu=\mc$ is again very helpful in this
respect.

Let us begin with $\mu > \mc$. In this case the Wilson coefficients
$z_i$ of penguin operators are zero and $\RE A_0$ is given as follows
\begin{equation}
\RE A_0 = c \left[
z_-(\mu) \; \left( \langle Q_- \rangle_0 - \langle Q_-^c \rangle_0 \right) +
z_+(\mu) \; \left( \langle Q_+ \rangle_0 - \langle Q_+^c \rangle_0 \right)
\right] \, ,
\label{eq:6.11}
\end{equation}
where similarly to \eqn{eq:6.6}, $\tau \; y_i$ have been set to zero. On
the other hand, for $\mu<\mc$ the operators $Q_\pm^c$ disappear from the
effective theory but instead the QCD penguin operators contribute
\begin{equation}
\RE A_0 = c \left[
z_-(\mu) \; \langle Q_-(\mu) \rangle_0 + z_+(\mu) \langle Q_+(\mu) \rangle_0
\right] + \RE A_0^P \, ,
\label{eq:6.12}
\end{equation}
where
\begin{equation}
\RE A_0^P = c \sum_{i=3}^6 z_i(\mu) \; \langle Q_i(\mu) \rangle_0 \; .
\label{eq:6.13}
\end{equation}
We again neglect the contributions of electroweak penguin operators
$Q_i$, $i=7,\ldots,10$.

Since $\RE A_0$ cannot depend on $\mu$, expressions \eqn{eq:6.11} and
\eqn{eq:6.12} have to be matched properly at $\mu=\mc$, the transition
point at which one goes from $f=4$ to $f=3$ effective theory. This
matching gives the following relation between $\langle Q_\pm^c(\mc)
\rangle_0$ and $\langle Q_i(\mc) \rangle_0$, $i=3,\ldots,6$ operators
\begin{equation}
\sum_{i=3}^6 z_i(\mc) \; \langle Q_i(\mc) \rangle =
- \left[ z_-(\mc) \; \langle Q_-^c(\mc) \rangle +
         z_+(\mc) \; \langle Q_+^c(\mc) \rangle   \right] \, ,
\label{eq:6.14}
\end{equation}
which clearly shows that the role of penguin operators in the $\dI$ rule
is for $\mu > \mc$ played by the operators $Q_{1,2}^c$. There is a very
immediate consequence of this observation. The size of the matrix
elements $\langle Q_\pm^c \rangle_0$ is correlated with the size of
$\langle Q_i \rangle_0$, $i=3,\ldots,6$.

Now, as discussed in section~4, in the HV and $\ndrb$ schemes $z_i(\mc)=0$
holds for $i=3,\ldots,6$. Consequently, in these schemes we have the
relation
\begin{equation}
z_-(\mc) \; \langle Q_-^c(\mc) \rangle_0 +
z_+(\mc) \; \langle Q_+^c(\mc) \rangle_0 = 0 \, ,
\label{eq:6.15}
\end{equation}
which is also valid for $\aem \not=0$.

If we next make a very plausible assumption consistent with all existing
non-perturbative methods that $\langle Q_\pm^c \rangle \ge 0$, we
find the following ``initial'' conditions for the matrix elements of
$Q_{1,2}^c$:
\begin{equation}
\langle Q_1^c(\mc) \rangle_0 =
\langle Q_2^c(\mc) \rangle_0 = 0 \, ,
\label{eq:6.16}
\end{equation}
valid in the HV and $\ndrb$ schemes. Using \eqn{eq:3.13} with $\aem=0$,
we then find the corresponding result for the $\ndr$ scheme
\begin{eqnarray}
\langle Q_1^c(\mc) \rangle_0 & = & 0 \, ,
\label{eq:6.17} \\
\langle Q_2^c(\mc) \rangle_0 & = & \frac{\as(\mc)}{12\pi} \,
\left[
\langle Q_6(\mc) \rangle_0 +
\langle Q_4(\mc) \rangle_0 -
\frac{1}{3} \langle Q_3(\mc) \rangle_0 -
\frac{1}{3} \langle Q_5(\mc) \rangle_0
\right] \, . \nn
\end{eqnarray}
One can easily check that this result is consistent with \eqn{eq:6.14}
and with the initial conditions for $z_i(\mc)$, $i=3,\ldots,6$, given for
the $\ndr$ scheme in \eqn{eq:4.25}. The effect of $\ord(\aem)$
corrections can easily be taken into account by using \eqn{eq:3.13}. It
has been included in our numerical analysis. With the initial conditions
\eqn{eq:6.16} and \eqn{eq:6.17} at hand we can calculate $\langle
Q_i^c(\mu) \rangle_0$ for any $\mu > \mc$. The result is given in
section~6.5.

We now want to constrain the matrix elements by the experimental value
of $\RE A_0$. To this end it is useful to set $\mu=\mc$ in
\eqn{eq:6.12}.  Since in the HV scheme for $\mu=\mc$ only $Q_1$ and
$Q_2$ operators contribute to $\RE A_0$, we find the relation
\begin{equation}
z_1(\mc) \; \langle Q_1(\mc) \rangle_0 +
z_2(\mc) \; \langle Q_2(\mc) \rangle_0 \; = \; \frac{\RE A_0}{c} \, ,
\label{eq:6.18}
\end{equation}
from which the following expression for the matrix element $\langle
Q_{1}(\mc) \rangle_0$ as a function of $\langle Q_2(\mc) \rangle_0$
can be found
\begin{equation}
\langle Q_1(\mc) \rangle_0 \; = \; \frac{\RE A_0}{c\,z_1(\mc)}
- \frac{z_2(\mc)}{z_1(\mc)} \; \langle Q_2(\mc) \rangle_0 \, ,
\label{eq:6.19}
\end{equation}
Using next the relations \eqn{eq:5.7}, \eqn{eq:5.12}, and \eqn{eq:5.13}, we
are able to obtain $\langle Q_4(\mc) \rangle_0$, $\langle Q_9(\mc) \rangle_0$,
and $\langle Q_{10}(\mc) \rangle_0$.
Because $\langle Q_3(\mc) \rangle_0$ is colour suppressed, we could set it
to zero or just use the expression of eq.~\eqn{eq:5.6}. Since $Q_3$ has
small Wilson coefficients it does not play any role in our analysis.
On the other hand $\langle Q_4(\mc) \rangle_0$ plays a substantial role
in the analysis of $\epe$ and the assumptions about $\langle Q_3(\mc)
\rangle_0$ entering \eqn{eq:5.7} matter to some extent. In the
following, we will therefore take $\langle Q_3(\mc) \rangle_0$
according to eq.~\eqn{eq:5.6}. $\langle Q_9(\mc) \rangle_0$ and
$\langle Q_{10}(\mc) \rangle_0$ are much less important for $\epe$ than
$\langle Q_4(\mc) \rangle_0$.

In the case of the $\ndr$ scheme there is a small contribution of the
penguin operators to \eqn{eq:6.16}. Moreover, the determination of
$\langle Q_4(\mc) \rangle_0$ requires the use of the relation
\eqn{eq:4.29b}. Thus, we need some assumptions about $(V-A)\otimes (V+A)$
operators which are discussed below. For $\mu=\mc$ however the issue of
$(V-A)\otimes (V+A)$ operators has only very small impact on
\eqn{eq:6.19} even in the $\ndr$ scheme. Equivalently, the matrix
elements of $(V-A)\otimes (V-A)$ operators in the $\ndr$ scheme can be
obtained from the HV matrix elements by means of \eqn{eq:3.13}.

With all this formalism at hand, we can now calculate the parameters
$B_1^{(1/2)}(\mc)$ and $B_4^{(1/2)}(\mc)$ as functions of
$B_2^{(1/2)}(\mc)$. If we in addition make the very plausible assumption
valid in all known non-perturbative approaches that $\langle Q_-(\mc)
\rangle \ge \langle Q_+(\mc) \rangle \ge 0$ the experimental value of
$\RE A_0$ implies
\begin{equation}
B_{2,LO}^{(1/2)}(\mc) = 5.8 \pm 1.1 \, , \qquad
B_{2,\ndr}^{(1/2)}(\mc) = 6.7  \pm 0.9  \, , \qquad
B_{2,HV}^{(1/2)}(\mc) = 6.3  \pm 1.0  \, . \qquad
\label{eq:6.20}
\end{equation}
This should be compared with the vacuum insertion result $B_2^{(1/2)}=1$,
and $B_2^{(1/2)} \approx 2.2$ in the $1/N$ approach. The latter value
corresponds to $\mu \approx 0.6\gev$, and when extrapolated to
$\mu=\mc$ would give
$B_2^{(1/2)}(\mc) \approx 2.8$. Yet, it is clear that the imposition of
the $\dI$ rule enhances the role of $(V-A)\otimes (V-A)$ operators.
Using \eqn{eq:6.19}, we plot $B_1^{(1/2)}(\mc)$ in fig.~\ref{fig:6} as a
function of $B_2^{(1/2)}(\mc)$. We note that if the $\dI$ rule is imposed,
the parameters $B_1^{(1/2)}(\mc)$ and $B_2^{(1/2)}(\mc)$ are strongly
correlated. A similar plot for $B_4^{(1/2)}(\mc)$ is given in
fig.~\ref{fig:14}. Here, the dependence on $B_3^{(1/2)}(\mc)$ has been
shown. These values are consistent with the upper end of the range for
$B_4^{(1/2)}$ used by the Rome group \cite{ciuchini:92}. The strong
deviation of $B_4^{(1/2)}$ from unity will have interesting consequences
for $\epe$ as already pointed out in \cite{buchallaetal:90}, and even
stronger emphasized in \cite{ciuchini:92}.

\subsection{$\langle Q_i(\mu) \rangle_{0,2}$ for $(V-A)\otimes (V+A)$
Operators}
We have seen that by setting $\mu=\mc$, we have decoupled the penguin
operators from the question of the $\dI$ rule. The matrix elements of
$(V-A)\otimes (V-A)$ penguin operators could still be determined due to
the operator relations of section~4.5. On the other hand the matrix
elements of $(V-A)\otimes (V+A)$ operators cannot be constrained by
$CP$-conserving data unless new relations between $(V-A)\otimes (V-A)$
and $(V-A)\otimes (V+A)$ operators are found. Some relations of this
type have been suggested in ref.~\cite{froehlich:91,heinrichetal:92,wu:92}.
However, in our opinion these relations are suspect as we will see
below and therefore, we will not use them here.

For $\langle Q_6(\mc) \rangle_0$, we simply use the formula
\eqn{eq:5.9}, and for $\langle Q_5(\mc) \rangle_0$ the relation
\eqn{eq:5.8}, however setting $B_5^{(1/2)}\equiv B_6^{(1/2)}$. The
assumption about $\langle Q_5(\mc) \rangle_0$ is of little importance
because this matrix element, similarly to $\langle Q_3(\mc) \rangle_0$,
is colour suppressed and has very small Wilson coefficients. Some
support for this strategy will be given soon.

For the matrix elements of the $(V-A)\otimes (V+A)$ electroweak
penguin operators $Q_7$ and $Q_8$, we will simply use the relations
\eqn{eq:5.10}, \eqn{eq:5.11}, as well as \eqn{eq:5.16} and \eqn{eq:5.17},
modified by setting $B_7^{(1/2)}\equiv B_8^{(1/2)}$ and
$B_7^{(3/2)}\equiv B_8^{(3/2)}$. Since mainly $\langle Q_8(\mc)
\rangle_2$ is of importance for $\epe$, the simplification of
identifying these $B$-parameters is immaterial. Again our analysis presented
below supports this strategy.

At this point, a remark on the strange quark mass is in order. Throughout
our analysis, we have used the value $\ms(\mc)=150\mev$, which corresponds
to $\ms(1\gev)=175\mev$. This value is somewhat lower than the central
value cited by the Particle Data Group \cite{particledata:92}. However,
the values there have
been obtained with lower values of $\Lms$ than are used in this work,
and since $\ms(1\gev)$ decreases for larger values of $\Lms$, we expect
the above value of $\ms$ to be in the right range \cite{jaminmuenz:93}.
Nevertheless, due to the existing uncertainty in the value of $\ms$, we
have allowed for a larger variation of $B_6^{(1/2)}$ and $B_8^{(1/2)}$,
which in this way include a possible variation in the strange quark mass.

As seen in \eqn{eq:5.29} the parameters $B_{5,6}^{(1/2)}$ and
$B_{7,8}^{(3/2)}$ calculated in the lattice approach \cite{sharpe:90}--
\nocite{bernardetal:90,gavelaetal:88,kilcupetal:90,kilcup:91}\cite{sharpe:91},
agree very well with the vacuum insertion value and in the case of
$B_6^{(1/2)}$ and $B_8^{(3/2)}$ with the $1/N$ approach of
ref.~\cite{burasgerard:87,bardeen:87a}. The question arises however,
whether this result is valid for a large range of $\mu$. It has been
stressed in ref.~\cite{buchallaetal:90}, that the $\mu$-dependence of
$\langle Q_6(\mu) \rangle_0$ and $\langle Q_8(\mu) \rangle_2$ is
governed in the $1/N$ approach by the $\mu$-dependence of $\ms$.
However, to our knowledge it has never been checked whether
$B_6^{(1/2)}$ and $B_8^{(3/2)}$ are indeed $\mu$-independent when $\mu$
is varied say in the range $1\gev \le \mu \le 4\gev$. With the evolution
equation \eqn{eq:5.3b} at hand, we can answer this question. Taking
$B_6^{(1/2)}(\mc)=1$ and $B_8^{(3/2)}(\mc)=1$, we have calculated the
$\mu$-dependence of these parameters in the case of LO, $\ndr$ and HV.
As seen in figs.~\ref{fig:9} and \ref{fig:11}, $B_6^{(1/2)}$ and
$B_8^{(3/2)}$ are only weakly dependent on $\mu$, demonstrating that at
least for these matrix elements the results of present non-perturbative
approaches are consistent with the renormalization group evolution. On
the other hand, one would naively expect that the $\mu$-dependence of
the colour suppressed matrix elements $\langle Q_5(\mu) \rangle_0$ and
$\langle Q_7(\mu) \rangle_2$ should be quite different from the other
two matrix elements implying strong $\mu$-dependence of
$B_{5,7}^{(1/2)}$.  After all, the Wilson coefficients $y_5$ and $y_7$
have a $\mu$-dependence very different from $y_6$ and $y_8$. This naive
expectation turns out to be completely wrong ! As shown in
figs.~\ref{fig:8} and \ref{fig:10} also $B_5^{(1/2)}(\mu)$ and
$B_7^{(1/2)}(\mu)$ are only weakly dependent on $\mu$. Moreover the
ratios $B_6^{(1/2)}(\mu)/B_5^{(1/2)}(\mu)$ and
$B_8^{(3/2)}(\mu)/B_7^{(3/2)}(\mu)$ are almost independent of $\mu$. In
our opinion this is a non-trivial result which gives some support to
the lattice results of
\cite{sharpe:90}--
\nocite{bernardetal:90,gavelaetal:88,kilcupetal:90,kilcup:91}\cite{sharpe:91}.

In this connection it is interesting to investigate whether the
relations between various operators proposed in
refs.~\cite{froehlich:91,heinrichetal:92} are consistent with the
renormalization group evolution. Taking the relations of Wu
\cite{wu:92} at face value and using the general expressions
\eqn{eq:5.4}--\eqn{eq:5.18}, one can derive relations between the
$B_i$-parameters. The most interesting among these relations are
\begin{equation}
B_{1}^{(1/2)} = \frac{5}{2} \; B_{2}^{(1/2)} \, , \qquad
B_{4}^{(1/2)} = B_{6}^{(1/2)} = \frac{5}{6} \; B_{2}^{(1/2)} \, , \qquad
B_{8}^{(3/2)} = \frac{4}{3} \; B_{1}^{(3/2)} \, .
\label{eq:6.21}
\end{equation}
It is easy to verify by using \eqn{eq:5.3b} that the relations
\eqn{eq:6.21} are incompatible with the renormalization group evolution
and can only be valid at a single value of $\mu$. The question then is
at which one ?

\subsection{The Fate of the Operators $Q_{1,2}^c$}
The authors of ref.~\cite{ciuchini:92} presented an educated guess for
the size of the matrix elements $\langle Q_{1,2}^c(2\gev) \rangle_0$,
and emphasized the possible importance of these operators. Having the
initial conditions for these matrix elements at $\mu=\mc$ given in
\eqn{eq:6.16} and \eqn{eq:6.17} for the HV and $\ndr$ scheme,
respectively, we can verify whether this is indeed true. In
fig.~\ref{fig:7}, we show $B_{1,2}^c(\mu)$ defined in \eqn{eq:5.29b} as
functions of $B_6^{(1/2)}(\mc)$ for $\mu=2$, 3, and $4\gev$ in the HV
scheme. For $\mu=2\gev$ and $B_6^{(1/2)}(\mc)=1.0 \pm 0.2$, as used in
\cite{ciuchini:92}, $B_{1,2}^c$ turn out to be at the lower end of the
range in \eqn{eq:5.29}. We conclude therefore that at least at
$\mu=2\gev$ the role of the operators $Q_{1,2}^c$ is rather minor in
spite of large coefficients $y_{1,2}^c$ given in \eqn{eq:4.27e}. They
are more important for $\mu=4\gev$, but since for $\mu<\mc$ they do not
contribute at all, the fate and the role of these operators depends
rather strongly on the choice of $\mu$ and also as seen in
fig.~\ref{fig:7} on $B_6^{(1/2)}$. Similar conclusions are reached in
the $\ndr$ scheme.

\subsection{Strategy of the Present Paper}
Making plausible approximations, using the experimental data for
$CP$-conserving amplitudes, and having calculated the Wilson coefficients
$z_i(\mu)$ in section~4, we were in a position to calculate the matrix
elements $\langle Q_i(\mu) \rangle_{0,2}$ and $\langle Q_{1,2}^c(\mu)
\rangle_{0,2}$ for any $\mu \ge 1\gev$, and
any renormalization scheme in terms of the four basic parameters
\begin{equation}
\Lms \, , \qquad
B_{2}^{(1/2)}(\mc) \, , \qquad
B_{6}^{(1/2)}(\mc) \, , \qquad
B_{8}^{(3/2)}(\mc) \, . \qquad \label{eq:6.29}
\end{equation}
The remaining two $B$-parameters $B_3^{(1/2)}$ and $B_8^{(1/2)}$, play
only a minor role in the analysis of $\epe$ and will be set to 1.

Let us summarize the basic ingredients of this approach:  Having fixed
the basic parameters in \eqn{eq:6.29} the calculation of the hadronic
matrix elements in a given scheme proceeds in several steps as
follows.
\begin{description}
\item{Step 1:}\\
$\langle Q_1(\mu) \rangle_2$, $\langle Q_2(\mu) \rangle_2$,
$\langle Q_9(\mu) \rangle_2$ and $\langle Q_{10}(\mu) \rangle_2$
are given directly by \eqn{eq:6.9} and \eqn{eq:6.10}.
\item{Step 2:}\\
The initial conditions for matrix elements $\langle Q_i(\mu) \rangle_0$
at $\mu=\mc$ can be calculated as described in section~6.3 with
the coefficients $z_i(\mc)$ calculated in section~4.
\item{Step 3:}\\
Matrix elements $\langle Q_i(\mu) \rangle_0$ for $\mu\not=\mc$ can be
found by using the evolution for the operators given in \eqn{eq:5.3b}.
This can also be done for $\langle Q_i(\mu) \rangle_2$ starting at
$\mu=\mc$ where \eqn{eq:6.9} and \eqn{eq:6.10} are most accurate.
\item{Step 4:}\\
The matrix elements $\langle Q_i(\mu) \rangle_{0,2}$ in any other
scheme can be calculated by using the relation \eqn{eq:3.13}.
\end{description}
In performing this program, care must be taken as in the case of the
Wilson coefficients, that higher order terms in $\as$ and $\aem$
generated in steps~3 and 4 are discarded. When this is done, it is an
easy matter to convince oneself that the resulting physical quantities
such as $A_0$, $A_2$ and $\epe$ are independent of $\mu$ and the scheme
considered.

Now, whereas $B_2^{(1/2)}(\mc)$ and the related $(V-A)\otimes (V-A)$
parameters taken in a given scheme can be extracted from $CP$
conserving data as seen in \eqn{eq:6.20} and in
figs.~\ref{fig:12}--\ref{fig:14} this is not the case for
$B_6^{(1/2)}(\mc)$ and $B_8^{(3/2)}(\mc)$. From \eqn{eq:3.13}, we know
that if $B_6^{(1/2)}(\mc)=B_8^{(3/2)}(\mc)=1$ in the HV scheme then for
$\Lms=300\mev$, we find $B_6^{(1/2)}(\mc)=0.84$ and
$B_8^{(3/2)}(\mc)=0.85$ in the $\ndr$ scheme. Yet, since we do not know
at present which renormalization scheme exactly gives the vacuum
insertion results, we decided to proceed in the phenomenology of $\epe$
as follows.  For all the cases considered (LO, $\ndr$, HV), we will
take universal values for the parameters $B_6^{(1/2)}(\mc)$ and
$B_8^{(3/2)}(\mc)$.  This will necessarily introduce a scheme
dependence in our results for $\epe$ because the Wilson coefficients
are scheme dependent. This left-over scheme dependence in $\epe$ will
give us some idea about the uncertainty in this ratio resulting from
the poor knowledge of the scheme dependence for $\langle Q_6(\mc)
\rangle_0$ and $\langle Q_8(\mc) \rangle_2$. On the other hand the
scheme dependence of the matrix elements of the $(V-A) \otimes (V-A)$
operators will be properly taken into account by means of our approach.

\newsection{CKM--Parameters and $\eps$}
In the analysis of $\epe$ we will need the values of various
CKM--parameters. We use the standard parameterization of the quark mixing
matrix \cite{particledata:92} in which the four basic parameters are
\begin{equation}
s_{12} = \left| \V{us} \right| \, , \qquad
s_{13} = \left| \V{ub} \right| \, , \qquad
s_{23} = \left| \V{cb} \right| \, , \qquad
\delta \, ,
\label{eq:7.1}
\end{equation}
with $\delta$ denoting the sole complex phase of this matrix.

In our numerical analysis we will set
\begin{equation}
\left| \V{us} \right| = 0.221 \, , \qquad
\left| \V{cb} \right| = 0.043 \pm 0.004 \, ,
\label{eq:7.2}
\end{equation}
\begin{equation}
\left| \V{ub}/\V{cb} \right| = 0.10 \pm 0.03 \, ,
\label{eq:7.3}
\end{equation}
and as usual we will extract $\delta$ by fitting the theoretical
expression for the parameter $\eps$ of the indirect $CP$-violation to
the data. To this end, we use
\begin{equation}
\eps = \frac{\exp(\imath\pi/4)}{\sqrt{2} \; \Delta M}
\left( \IM M_{12} + 2 \, \xi \; \RE M_{12} \right) \, ,
\label{eq:7.4}
\end{equation}
where
\begin{equation}
\xi = \frac{\IM A_0}{\RE A_0} \, , \qquad
\Delta M = 3.5 \cdot 10^{-15}\gev \, ,
\label{eq:7.5}
\end{equation}
with the amplitude $A_0 \equiv A(K \rightarrow (\pi\pi)_{\rm I=0})$
and $M_{12}$ obtained from the standard box diagrams
\begin{equation}
M_{12} =
\frac{G_{\rm F}^2}{12 \,\pi^2} \; F_{\rm K}^2 B_{\rm K} m_{\rm K} \mw^2
\left[
{\lambda^*_{\rm c}}^2 \eta_1 \; S(x_c) +
{\lambda^*_{\rm t}}^2 \eta_2 \; S(x_t) +
2 \lambda^*_{\rm c} {\lambda^*_{\rm t} \eta_3 \; S(x_c, x_t}) \right] \, .
\label{eq:7.6}
\end{equation}
Here $S(x_i)$ and $S(x_c, x_t)$ are the Inami-Lim
functions for which explicit formulae are given e.g.~in
ref.~\cite{buchallaetal:90}. Next, $F_{\rm K}= 161\mev$, $\lambda_{\rm
i} = \V{id}\V{is}^*$ and $B_{\rm K}$ is the renormalization group
invariant parameter describing the size of the matrix element $\langle
\bar{K}^0 | (\bar{s} d)_{V-A} (\bar{s} d)_{V-A} | K^0 \rangle$. In our
numerical analysis, we will choose
\begin{equation}
B_{\rm K} = 0.65 \pm 0.15 \, ,
\label{eq:7.7}
\end{equation}
which is in the ball park of most recent estimates
\cite{burasharlander:92}. For the QCD-factors
$\eta_i$, we will use
\begin{equation}
\eta_1 = 0.85 \, , \qquad
\eta_2 = 0.58 \, , \qquad
\eta_3 = 0.36 \, ,
\label{eq:7.8}
\end{equation}
where $\eta_2$ includes next-to-leading order corrections calculated in
ref.~\cite{burasjaminweisz:90}. Because the second term in
eq.~\eqn{eq:7.6} is the dominant one, we included next-to-leading order
corrections in $\eta_2$ although for the other two smaller terms they
are still unknown. Without these corrections one would have
$(\eta_2)_{\rm lo} \approx 0.62$.

For fixed values of the parameters in \eqn{eq:7.2}, \eqn{eq:7.3},
\eqn{eq:7.7} and \eqn{eq:7.8}, and for a fixed value of $\mt$ comparison
of \eqn{eq:7.4} with the experimental data for $\eps$ gives two
solutions for the phase $\delta$ in the first and second quadrant. With
the experimental constraint from $B^0-\bar{B}^0$-mixing it is
sometimes possible to exclude one of the two solutions. In particular
for $\mt > 150\gev$ and $F_{\rm B} > 200\mev$ the solution in the first
quadrant is favoured. Due to remaining substantial uncertainties in
$F_{\rm B}$ and $\mt$, we will however not use the constraint coming from
$B^0-\bar{B}^0$-mixing and we will present the results for $\epe$
corresponding to two solutions for $\delta$.

In the left half of fig.~\ref{fig:5}, we show $\IM \lambda_{\rm t}$ as
a function of $\mt$ when $|\V{cb}|$, $|\V{ub}/\V{cb}|$ and $B_{\rm K}$
are varied in the ranges given by \eqn{eq:7.2}, \eqn{eq:7.3}, and
\eqn{eq:7.7}.  In the right half of fig.~\ref{fig:5} similar plots are
given for the parameter ranges to be expected in a few years time
\begin{equation}
|\V{cb}| \; = \; 0.043 \pm 0.002 \, ,
\qquad
|\V{ub}/\V{cb}| \; = \; 0.10 \pm 0.01 \, ,
\qquad
B_{\rm K} \; = \; 0.65 \pm 0.05 \, ,
\label{eq:7.9}
\end{equation}
We observe that $\IM\lambda_{\rm t}$ decreases with $\mt$ and is larger
for the solution $0 < \delta < \pi/2$.  The uncertainty in
$\IM\lambda_{\rm t}$ is considerable at present.

\newsection{$\epe$ Beyond Leading Logarithms}
\subsection{Basic Formulae}
The parameter $\eps'$ is given in terms of the amplitudes $A_0$ and $A_2$
as follows
\begin{equation}
\eps' = - \frac{\omega}{\sqrt{2}} \,\xi \left( 1-\Omega \right)
\exp^{\imath \phi} \, ,
\label{eq:8.1}
\end{equation}
where
\begin{equation}
\xi = \frac{\IM A_0}{\RE A_0}\, ,
\qquad
\omega = \frac{\RE A_2}{\RE A_0} \, ,
\qquad
\Omega = \frac{1}{\omega} \; \frac{\IM A_2}{\IM A_0} \, ,
\label{eq:8.2}
\end{equation}
and $\phi = \pi/2 + \delta_2 - \delta_0 \approx \pi/4$.

Using $\Heff(\dS)$ and the experimental data for $\omega$, $\RE A_0$
and $\eps$, we find
\begin{equation}
\frac{\eps'}{\eps} \; = \; 10^{-4}\,\left[\frac{\IM\lambda_{\rm t}}
{1.7\cdot 10^{-4}}\right]\, \left[\,P^{(1/2)}-P^{(3/2)}\,\right] \, ,
\label{eq:8.3}
\end{equation}
where
\begin{eqnarray}
P^{(1/2)} & = & \sum P^{(1/2)}_i \; = \;
              r \sum y_i \langle Q_i
\rangle_0\, \big(1-\Omega_{\eta+\eta'}\big) \, , \label{eq:8.4} \\
P^{(3/2)} & = & \sum P^{(3/2)}_i \; = \;
\frac{r}{\omega} \sum y_i \langle Q_i \rangle_2 \, , \label{eq:8.5}
\end{eqnarray}
with
\begin{equation}
r \; = \; 1.7\,\frac{G_F\,\omega}{2\,|\eps|\,\RE A_0}
\; = \; 594\gev^{-3} \, .
\label{eq:8.6}
\end{equation}
In \eqn{eq:8.5} and \eqn{eq:8.5} the sum runs over all contributing
operators. This means for $\mu > \mc$ also the contribution from
operators $Q^c_{1,2}$ to $P^{(1/2)}$ and $P^{(3/2)}$ have to be taken
into account. This is necessary for $P^{(1/2)}$ and $P^{(3/2)}$ to be
independent of $\mu$. Next
\begin{equation}
\Omega_{\eta+\eta'} \; = \; \frac{1}{\omega}\,\frac{(\IM A_2)_{I.B.}}
{\IM A_0} \, ,
\label{eq:8.7}
\end{equation}
represents the contribution of the isospin breaking in the quark
masses ($\mup\neq \md$). For $\Omega_{\eta+\eta'}$, we will take
\begin{equation}
\Omega_{\eta+\eta'} \; = \; 0.25\,\pm\,0.05 \,
\label{eq:8.8}
\end{equation}
which is in the ball park of the values obtained in the $1/N$ approach
\cite{burasgerard:87} and in chiral perturbation theory
\cite{donoghueetal:86,lusignoli:89}. $\Omega_{\eta+\eta'}$ is
independent of $\mt$.

Now the formulation of the phenomenology of $\epe$ found in refs.
\cite{flynn:89}--\nocite{buchallaetal:90,paschos:91}\cite{lusignoli:92}
and \cite{froehlich:91,heinrichetal:92,ciuchini:92} uses instead of
$P^{(1/2)}$ and $P^{(3/2)}$ defined in \eqn{eq:8.4} and \eqn{eq:8.5}
the ratios
\begin{equation}
\Omega_i^{(1/2)} \; \equiv \; -\frac{P^{(1/2)}_i}{P^{(1/2)}_6} \, ,
\qquad
\Omega_i^{(3/2)} \; \equiv \; \frac{P^{(3/2)}_i}{P^{(1/2)}_6}\,
\big(1-\Omega_{\eta+\eta'}\big) \, ,
\label{eq:8.9}
\end{equation}
with $\Omega_{\eta+\eta'}$ being treated separately.

We would like to point out one drawback of this formulation:  The
separate contributions $\Omega_i^{(1/2)}$ and $\Omega_i^{(3/2)}$ depend
on $\mu$ and on the renormalization scheme.  In this respect the
example of the operator $Q_2^c$ is very instructive. As pointed out in
ref.~\cite{ciuchini:92}, $Q_2^c$ can have some impact on $\epe$ for
$\mu \approx \ord(2\gev)$. Yet, for $\mu \leq \mc$ the operator $Q_2^c$
is absent in the effective theory, hence $\Omega_{2,c}^{(1/2)}=0$, and
its effects are hidden in the contributions of operators present in the
effective three quark theory.

On the other hand, $P^{(1/2)}$ and $P^{(3/2)}$ in \eqn{eq:8.4} and
\eqn{eq:8.5} are independent of $\mu$ and the renormalization scheme
considered. We have verified numerically that this is always the case.
Consequently, they are more suitable for the phenomenological analysis.
$P^{(1/2)}$ is dominated by the contribution of the $Q_6$ operator but
receives also important contributions from $Q_4$. $P^{(3/2)}$ receives
only contributions from electroweak operators $Q_i$ ($i=7\ldots10$).
Although $Q_8$ dominates $P^{(3/2)}$, the operators $Q_7$, $Q_9$ and
$Q_{10}$ have also some impact on the final result for $\epe$

\subsection{$B_i$-Expansions and the Four Dominant Contributions to $\epe$}
The contributions $P^{(1/2)}$ and $P^{(3/2)}$ can be written as linear
combinations of the $B_i$-parameters introduced in section~5. In our
approach of section~6, there are only three relevant $B_i$-parameters
introduced in \eqn{eq:6.29}. We then find
\begin{eqnarray}
P^{(1/2)} & = & a_0^{(1/2)} + a_2^{(1/2)}\,B_2^{(1/2)} +
                  a_6^{(1/2)}\,B_6^{(1/2)} \, ,
\label{eq:8.10} \\
\mvs
P^{(3/2)} & = & a_0^{(3/2)} + a_8^{(3/2)}\,B_8^{(3/2)} \, ,
\label{eq:8.11}
\end{eqnarray}
where the parameters $B_i$ will be taken at $\mu=\mc$. Then the
coefficients $a_i^{(1/2)}$ and  $ a_i^{(3/2)}$ depend only on $\Lms$,
$\mt$, and the renormalization scheme considered. These dependencies
are given in tabs.~\ref{tab:21} and \ref{tab:22}.

Looking at the expression \eqn{eq:8.10} and \eqn{eq:8.11}, we identify
four contributions which govern the ratio $\epe$:
\begin{itemize}
\item[i)]
The contribution of $(V-A)\otimes (V-A)$ operators to $P^{(1/2)}$ is
represented by the first two terms in \eqn{eq:8.10}. These two terms
include also small contributions from $(V-A)\otimes (V+A)$ electroweak
penguin operators, and if we would use $\mu \ge \mc$ also operators
$Q_{1,2}^c$ would contribute here.  These two terms are dominated by
the contribution of the operator $Q_4$. We observe that the sum of the
two terms in question is {\em negative} and only weakly dependent on
$B_2^{(1/2)}$, especially if we vary this parameter in the ranges given
in \eqn{eq:6.18}. The dependence of $\Lms$ is also weak. These weak
dependences on renormalization scheme, $B_2^{(1/2)}$ and $\Lms$  result
from  the fact that in our approach of section~6 the matrix elements
entering the first two terms in $P^{(1/2)}$ are more or less fixed by
the experimental value of $A_0$.  The weak dependence on $\mt$ results
from the contributions of electroweak penguin operators. It should be
stressed that the constraint from $CP$-conserving data suppresses
considerably $\epe$. In order to see this we calculated the sum of the
first two terms in \eqn{eq:8.10} using the matrix elements with
$B_i=1$.  Taking $\Lms=300\mev$, $\mu=\mc$ and $\mt=130\gev$ we have
found $a_0^{(1/2)} + a_2^{(1/2)} B_2^{(1/2)}$ equal to -0.75, -0.72,
-0.74 for LO, $\ndr$ and HV, respectively. This should be compared with
-3.78, -3.99 and -3.92 used here.
\item[ii)]
The contribution of $(V-A)\otimes (V+A)$ QCD penguin operators to
$P^{(1/2)}$ is given by the last term in \eqn{eq:8.10}. This contribution
is large and {\em positive}. The coefficient $a_6^{(1/2)}$ depends
sensitively on $\Lms$ which is the result of the strong dependence of
$y_6$ on the QCD scale. $a_6^{(1/2)}$ is suppressed by next-to-leading
corrections at $\mu=\mc$ although due to matching effects discussed in
section~4 it is somewhat enhanced for $\mu > \mc$ in the $\ndr$ scheme.
\item[iii)]
The contribution of the $(V-A)\otimes (V-A)$ electroweak penguin operators
$Q_9$ and $Q_{10}$ to $P^{(3/2)}$ is represented by the first term in
$P^{(3/2)}$. As in the case of the contribution i), the matrix elements
contributing to $a_0^{(3/2)}$ are fixed by the $CP$-conserving data.
This time by the amplitude $A_2$. Consequently, the scheme dependence and
the $\Lms$ dependence of $a_0^{(3/2)}$ are weak. The sizable
$\mt$-dependence of $a_0^{(3/2)}$ results from the corresponding
dependence of $y_9 + y_{10}$. $a_0^{(3/2)}$ contributes {\em positively}
to $\epe$. As seen in tab.~\ref{tab:22} this contribution is slightly
enhanced through next-to-leading order corrections. On the other hand it
should be stressed that it is considerably suppressed by the constraint
from $CP$-conserving data. Indeed as seen in fig.~\ref{fig:13}
$B^{(3/2)}(\mc)$ is by a factor of 2 below the vacuum insertion value.
\item[iv)]
The contribution of the $(V-A)\otimes (V+A)$ electroweak penguin operators
$Q_7$ and $Q_{8}$ to $P^{(3/2)}$ is represented by the second term in
\eqn{eq:8.11}. This contribution depends sensitively on $\mt$ and
$\Lms$. It contributes {\em negatively} to $\epe$. The $\mt$ dependence
of this contribution is governed roughly by $y_7/3 + y_8$. For $\mt <
150\gev$ enhancement through next-to-leading corrections is observed but
for $\mt > 150\gev$  the leading and next-to-leading order results are
quite similar to each other. For $\mt < 150\gev$ the operator $Q_7$ is
important when next-to-leading corrections are taken into account but
for $\mt \approx 200\gev$ its contribution can be neglected.
\end{itemize}
The competition between these four contributions depends on $\mt$ and on
the values of $B_6^{(1/2)}$ and $B_8^{(3/2)}$. Setting first
$B_6^{(1/2)}=B_8^{(3/2)}=1$, we observe that the contribution ii) is
always largest and the contribution iii) smallest. For lower values of
$\mt$ the contribution i) is more important than iv), and it is this
contribution which represents the main suppression of $\epe$ at $\mt \le
150\gev$ together with $\Omega_{\eta+\eta'} \not= 0$. With increasing
$\mt$, the contribution iv) grows and for $\mt \ge 170\gev$, it is more
important than the contribution i). At such high values of $\mt$, the
collaboration of i) and iv) is so strong that it becomes as important as
the contribution ii) and hence, $\epe$ becomes very small. For even
higher values of $\mt$ the sum i)+iv) wins the competition, and $\epe$
becomes negative.

\subsection{The $\mu$-Dependence of Various Contributions}
It is instructive to study the $\mu$-dependence of the various
contributions in \eqn{eq:8.4} and \eqn{eq:8.5}. In fig.~\ref{fig:16}
we show $P_5^{(1/2)}$, $P_6^{(1/2)}$, $P_7^{(3/2)}$, $P_8^{(3/2)}$
and the remaining contributions to $P^{(1/2)}$ and $P^{(3/2)}$ as
functions of $\mu$ for $\Lms=300\mev$, $\mt=130\gev$ and
$B_5^{(1/2)}(\mc)=B_6^{(1/2)}(\mc)=B_7^{(3/2)}(\mc)=B_8^{(3/2)}(\mc)=1$.
We observe
\begin{itemize}
\item[i)]
The various contributions show visible $\mu$-dependence. The sum of all
contributions is however $\mu$-independent. The discontinuities in
$\ndr$ are the result of the matching at $\mc$.
\item[ii)]
The $\mu$-dependence of $P_6^{(1/2)}(\mu)$ and $P_8^{(3/2)}(\mu)$
demonstrates the importance of $1/N$ corrections because in the large-$N$
limit they should be $\mu$-independent. These $\mu$-dependence  although
visible is rather weak.
\item[iii)]
Substantially stronger $\mu$ dependence is observed for the remaining
smaller contributions in accordance with our previous discussions.
$P_5^{(1/2)}$ remains however always very small.
\end{itemize}

\subsection{Final Numerical Results for $\epe$}
We are now in the position to present our final results for $\epe$ which
are based on the Wilson coefficients of section~4, the hadronic matrix
elements of section~6 and results on $\IM \lambda_{\rm t}$ of section~7.

Let us first emphasize that with the help of the $CP$-conserving data we
were able to determine two out of four contributions to $\epe$
discussed in section~8.2. These are the contributions i) and iii). Taken
together these two contributions give a {\em negative} value for $\epe$
which is very weakly dependent on $\Lms$ and on the renormalization
scheme (LO,NDR,HV).  For instance for $\Lms = 300\mev$ and $\mt =
130\gev$ we find $\epe \approx -2.7 \times 10^{-4}$.  We observe that
with increasing $\mt$ this part of $\epe$ becomes smaller.

We should also stress that these two contributions are very different
in the vacuum insertion method. In fact in the latter method the sum of
the contributions i) and iii) gives a {\em positive} $\epe \approx 1.8
\times 10^{-4}$ for $\mt=130\gev$ instead of the negative contribution
found here.

The fate of $\epe$ in the Standard Model depends then mainly on the
following quantities:
\begin{equation}
\IM \lambda_{\rm t}, \qquad
\mt, \qquad
\Lms, \qquad
B_6^{(1/2)}(\mc), \qquad
B_8^{(3/2)}(\mc) \; .
\label{eq:8.12}
\end{equation}

This dependence is shown in tabs.~\ref{tab:15}--\ref{tab:20} and in
figs.~\ref{fig:17}--\ref{fig:19}. In obtaining these results we have set
$\ms(1\gev)=175\mev$ but we varied the parameters $B_6^{(1/2)}(\mc)$
and $B_8^{(3/2)}(\mc)$ in a broad range which effectively could take into
account variations in $\ms$. Recall that the relevant combinations are
$B_i/\ms^2$. $B_{6,8}=0.75, 1, 1.25, 1.5\ {\rm and}\ 2.0$ corresponds
then effectively to $\ms(1\gev)=202, 175, 157, 143\ {\rm and}\
124\mev$ if the matrix elements $\langle Q_{6,8} \rangle$ were kept at
the vacuum insertion values. The tabs.~\ref{tab:15}--\ref{tab:17} in
which $B_6=B_8$ show then effectively the variation of $\ms(1\gev)$ in
the range $143\mev \le \ms(1\gev) \le 202\mev$. Tabs.~\ref{tab:18},
\ref{tab:19} and \ref{tab:20}  on the other hand show examples of
$B_6\not=B_8$. We observe the following main dependencies:
\begin{itemize}
\item
$\epe$ decreases with increasing $\mt$ and $B_8^{(3/2)}$.
\item
$\epe$ increases with increasing $\Lms$ and $B_6^{(1/2)}$
\end{itemize}
\noindent
Moreover as one can deduce from the analysis of $\IM \lambda_{\rm t}$
\begin{itemize}
\item
$\epe$ increases with decreasing $B_{\rm K}$, $\V{cb}$ and $\V{ub}$.
\end{itemize}

\noindent
The main conclusion which we can draw on the basis of these tables are
as follows
\begin{itemize}
\item[i)]
If $B_6=B_8=1$ the values of $\epe$ are below $10^{-3}$ and fully
consistent with the result of the experiment E731. The next-to-leading
corrections calculated in the $\ndr$ scheme suppress $\epe$ by roughly
10--20\% relative to the leading order result. Much stronger suppression
is observed in the HV scheme, especially at higher values of $\mt$. Yet
in the full range of the parameters considered, $\epe$ remains positive.
\item[ii)]
We stress however that the increase of $\langle Q_6(\mc) \rangle_0$ by
only a factor of two with $\langle Q_8(\mc) \rangle_2$ kept fixed moves
$\epe$ as shown in tab.~\ref{tab:20} in the ball park of the NA31 result
if $\mt < 150\gev$, $\Lms \ge 300\mev$ and the $\ndr$ scheme is
considered. The values in the HV scheme are lower but still consistent
with the NA31 result if the CKM phase $\delta$ is in the first quadrant.
\item[iii)]
For the intermediate choices of the $B_i$ parameters shown in
tabs.~\ref{tab:16}, \ref{tab:18} and \ref{tab:19} the ratio $\epe$ is
found somewhere in between the values of the two experiments in question
although for higher values of $\mt$ and lower values of $\Lms$ they
certainly favour the E731 experiment.
\item[iv)]
An important point to be stressed here is the relatively strong
dependence of $\epe$ on the value of $\Lms$. It is a result of large
anomalous dimensions of penguin operators. This dependence is evident in
the tables  but is even more clearly depicted in figs.~\ref{fig:17},
\ref{fig:18} and \ref{fig:19} where $\epe$ is plotted as a function of
$\Lms$ for various choices of $\mt$ and $B_i$. We note that for the
cases $(B_6=B_8=1.5)$ and $(B_6=1.5, B_8=1)$ with $\mt=130\gev$ the
increase of $\Lms$ from $150\mev$ to $450\mev$ moves essentially $\epe$
from the E731 value to the NA31 result.
\end{itemize}

We should stress that although our different contributions to $\epe$
have been presented for $\mu=\mc$ we have checked that the final result
for $\epe$ does not depend on the actual choice of $\mu$.

In summary, our results for $\epe$ are fully compatible with both
experiments when $\mt < 150\gev$ but favour E731 when $\mt$ is higher.
The progress in predicting a more accurate value of $\epe$ can only be
achieved through the reduction of the uncertainties in the values of the
parameters listed in \eqn{eq:8.12}.

\subsection{Comparison with Other Analyses}
Our results for $\epe$ cover the range of values between those obtained
by the Dortmund \cite{froehlich:91,heinrichetal:92} and Rome
\cite{ciuchini:92} groups. Depending on the choice of the parameters
listed in \eqn{eq:8.12} one can obtain the results of these two groups.

Here we would like to make a comparison with ref.\cite{ciuchini:92}
because this is the only other paper in which full next-to-leading order
corrections to the $\dS$ Hamiltonian have been computed.

Since these authors presented their results only for the HV scheme, only
the comparison for this scheme is possible. Here are our findings:
\begin{itemize}
\item[i)]
The results for $y_6$ and $y_8$ obtained in \cite{ciuchini:92} agree
with ours to better than 3\%. Since the details on the two--loop
anomalous dimensions are not yet available from the Rome group it is
impossible for us to identify the origin of this difference\footnote{
We have just been informed by Guido Martinelli that the Rome group
confirmed our $\ndr$ two--loop results of
refs.~\cite{burasetal:92b,burasetal:92c}. The final results on two--loop
anomalous dimensions in the HV scheme also agree, although there are
apparently some differences at the intermediate stages of the
calculations. We guess then that the 3\% difference in the Wilson
coefficients originates in the numerical integration of the
renormalization group equations done in \cite{ciuchini:92} which in our
paper has been done analytically.}. Fortunately this is immaterial
for any phenomenological applications. We should however remark that
the next-to-leading corrections in the HV scheme found by us are larger
than those given in ref.~\cite{ciuchini:92} because there some of these
corrections have been included in the leading order.
\item[ii)]
{}From the smooth $\mu$-dependence of $y_6$ presented in
\cite{ciuchini:92} we conclude that the matching effects discussed in
section~4.3 have not been taken into account there. In the case of the
HV scheme as seen in fig.~\ref{fig:2} this is fortunately only a small
effect.
\end{itemize}
Next we would like to list several differences between our own analysis
and the one of ref.\cite{ciuchini:92} which are relevant when taken
individually but which cancel each other to some extent when taken as a
whole.
\begin{itemize}
\item[iii)]
Our analysis of section~6.5 shows that the contributions of $Q^c_{1,2}$
operators to $\epe$ evaluated at $\mu \approx 2\gev$ lie at the lower
end of the educated guess of ref.~\cite{ciuchini:92}. We find
$\Omega^c_1 \approx -0.01$, $\Omega^c_2 \approx -0.04$ to be compared
with $\Omega^c_1 \approx 0.02$, $\Omega^c_2 \approx -0.18$ quoted in
\cite{ciuchini:92}. This reduces the value of $\epe$ found by the Rome
group.
\item[iv)]
Our analysis of $B_1^{(3/2)}$ shows that the lattice calculations
overestimate the contributions of the operators $Q_{9,10}$ by almost a
factor of two. When this is corrected for, the value of $\epe$ of
ref.~\cite{ciuchini:92} is further reduced.
\item[v)]
The suppression of $\epe$ through the imposition of the $\dI$ rule found
by us is compatible with the one given in \cite{ciuchini:92} although
somewhat stronger.
\item[vi)]
As discussed in section~5 the values of $\ms(1\gev)$ used in
\cite{ciuchini:92} are higher than those used by us. With our values the
results for $\epe$ obtained in \cite{ciuchini:92} would be increased.
\item[vii)]
The values of $|\V{ub}/\V{cb}|$, $B_{\rm K}$ and in particular of
$|\V{cb}|$ used in \cite{ciuchini:92} are systematically higher than
those used here which results in a smaller value of $\IM \lambda_{\rm
t}$. With our choices of these parameters the authors of
\cite{ciuchini:92} would find a higher $\epe$. In view of the increased
$B$-meson life time and the decrease of $B_{\rm K}$ in the recent
lattice calculations \cite{guptaetal:92} we think that our estimate of $\IM
\lambda_{\rm t}$ is better than the one of ref.~\cite{ciuchini:92}.
\end{itemize}

As already stated above these effects cancel each other to some extent
when taken as a whole and consequently our result given in
tab.~\ref{tab:15} for $\epe$ in the HV scheme with $B_6=B_8=1$ are
certainly consistent with the ones obtained in ref.~\cite{ciuchini:92}.
We have however varied the parameters $B_6$ and $B_8$ in a larger range
and did in addition the same analysis in the $\ndr$ scheme. Consequently
we reached somewhat different conclusions than ref.~\cite{ciuchini:92}
regarding the two experimental results for $\epe$ given in \eqn{eq:1.1}.

\newsection{Wilson Coefficients for $\dB$ Decays}
This section can be viewed as the generalization of the leading order
analyses of Grinstein \cite{grinstein:89} and Kramer and Palmer
\cite{kramerpalmer:92}, beyond the leading order approximation. In
addition, the coefficients of electroweak penguin operators will be
calculated.

We will focus on the $\Delta B = 1$, $\Delta C = \Delta U = 0$ part of
the effective Hamiltonian which is of particular interest for the study
of $CP$ violation, i.e., decays to $CP$ self-conjugate final states.

At the tree level the effective Hamiltonian is simply given by

\begin{equation}
{\cal H}_{eff}(\Delta B=1) \; = \; \frac{G_{F}}{\sqrt{2}} \,
\sum_{q=u,c} \sum_{q'=d,s} V_{q b}^{*} V_{q q'} \,
\big( \bar{b} \, q \big)_{V-A}
\big( \bar{q} \, {q'} \big)_{V-A} \; .
\label{eq:9.1}
\end{equation}
The cases $q' = d$ or $s$ can be treated separately and have the same
coefficients $C_{i}(\mu)$. Specifying to $q' = d$ and using the
unitarity of the CKM matrix we find for $\mu\approx {\cal O}(\mb)$
($\xi_{i} = V_{i b}^{*} V_{i d}$)
\begin{eqnarray}
{\cal H}_{eff}(\Delta B=1) \; = \; \frac{G_{F}}{\sqrt{2}} \, \Biggl\{ \,
  & \xi_{c} & \!\!\! \biggl(\, z_{1}(\mu)\,O_{1}^{c} + z_{2}(\mu)\,O_{2}^{c}
\,\biggr) \nn \\
+ & \xi_{u} & \!\!\! \biggl(\, z_{1}(\mu)\,O_{1}^{u} + z_{2}(\mu)\,O_{2}^{u}
\,\biggr) \nn \\
- & \xi_{t} & \!\!\! \biggl(\,\, \sum_{i=3}^{10} y_{i}(\mu)\,O_{i}
\,\biggr) \, \Biggr\} \; ,
\label{eq:9.2}
\end{eqnarray}
where $O_{i}$, $i=3,\ldots,10$, are obtained from $Q_{i}$,
$i=3,\ldots,10$ in \eqn{eq:2.1n} by replacing $s$ by $b$,
\begin{equation}
O_{1}^{u} \; = \;
\big( \bar{b}_{\alpha} \, u_{\beta}  \big)_{V-A}
\big( \bar{u}_{\beta}  \, d_{\alpha} \big)_{V-A}
\qquad ; \qquad
O_{2}^{u} \; = \;
\big( \bar{b} \, u \big)_{V-A}
\big( \bar{u} \, d \big)_{V-A}
\label{5.3}
\end{equation}
and the $O_{i}^{c}$ are obtained from (\ref{5.3}) through the replacement
$u \rightarrow c$.

In spite of the presence of the CKM factors in (\ref{eq:9.2}), it is an
easy matter to convince oneself that for $\mu=m_{b}$ all ten coefficients
building the vector $\vC(\mu)$ can be found using
\begin{equation}
\vec{C}(m_{b}) \; = \; \hat{U}_{5}(m_{b}, M_{W}, \aem) \, \vec{C}(M_{W}) \; ,
\label{eq:9.4}
\end{equation}
with the evolution matrix of eq. (\ref{eq:2.11}) given in an effective
theory with 5 flavours and the initial conditions given in
eqs.~\eqn{eq:4.6}--\eqn{eq:4.15}. The numerical results for
$\mt=130\gev$ are given in tab.~\ref{tab:14}. They have a structure
similar to the results in tab.~\ref{tab:13} and therefore do not
require further discussion.

\newsection{Ten Messages from This Paper}
Using the two--loop anomalous dimension matrices of
refs.~\cite{burasweisz:90}--
\nocite{burasetal:92a,burasetal:92b}\cite{burasetal:92c},
we have constructed the effective Hamiltonian for $\dS$, $\dC$, and
$\dB$ transitions with the Wilson coefficients $C_i(\mu)$ including
leading and next-to-leading QCD corrections and leading order
corrections in the electromagnetic coupling constant $\aem$. As the
main application of the constructed $\dS$ Hamiltonian, we have
calculated the ratio $\epe$ as a function of $\mt$, $\Lms$ and of a set
of non-perturbative parameters $B_i$. This new anatomy of $\epe$
improves the leading order anatomy of 1989 \cite{buchallaetal:90} in
two respects:
\begin{itemize}
\item
The Wilson coefficients $C_i(\mu)$ include next-to-leading logarithmic
corrections and consequently, the scale $\Lms$ extracted from
deep-inelastic scattering and from $e^+e^-$-data at LEP can be
meaningfully used for the first time in the evaluation of $\epe$. Since
$\epe$, being dominated by penguin contributions with large anomalous
dimensions, is sensitive to the strength of the QCD coupling constant,
the meaningful use of $\Lms$ increases considerably the precision of
the theory provided the QCD scale can be extracted accurately from high
energy data.
\item
The hadronic matrix elements $\langle Q_i(\mu) \rangle$ used in the
analysis are at all stages consistent with the data on $CP$-conserving
$\Kpipi$ decays. In particular, $\dI$ enhancement and $\dIth$
suppression are exactly built into our analysis.
\end{itemize}

We have presented a detailed renormalization group analysis of both the
Wilson coefficient functions and of the hadronic matrix elements. In
particular, we have stressed that the Wilson coefficients are
renormalization scheme dependent and that this scheme dependence can
only be cancelled by the one present in $\langle Q_i(\mu) \rangle$.
Since the existing methods are not capable to study this scheme
and scale dependence accurately enough, we have presented a
semi-phenomenological approach to hadronic matrix elements with the help
of which we could determine several of these matrix elements from
existing $CP$-conserving data.

To summarize, the ten main messages of our paper are as follows:
\begin{itemize}
\item[i)]
The coefficients $z_-(\dI)$ and $z_+(\dIth)$ calculated in the $\ndr$
and HV schemes are contrary to the common wisdom suppressed and
enhanced by next-to-leading order corrections, respectively. These
effects must be compensated by those present in the hadronic matrix
elements so that the experimentally known $\dI$ and $\dIth$ amplitudes
are reproduced.
\item[ii)]
The contribution of QCD penguins to the $\dI$ rule depends strongly on
$\mu$ and the renormalization scheme considered, but is rather small
with respect to the whole amplitude. For $\mu > \mc$ it vanishes, but
even for $\mu \approx 1\gev$ it does not exceed $20\%$ of $A_0$ for
$\ms(1\gev) > 125\mev$. In view of these features, relating the size of
$\epe$ to the penguin contribution to $A_0$ is in our opinion not a good
idea.
\item[iii)]
The coefficient $y_6(\mc)$ is suppressed by roughly $15\%$ in the HV
scheme but it is enhanced for $\mu > \mc$ by $10\%$ in the $\ndr$
scheme relative to the leading order value. Simulaneously, for $\mt <
150\gev$ we find a strong enhancement of $y_8(\mc)$ calculated in the
HV scheme, to be compared with modest enhancement of $y_8(\mc)$
obtained in the $\ndr$ scheme. These enhancements of $y_8$ become
smaller for large $\mt$ values.
\item[iv)]
The choice $\mu=\mc$ allows to extract efficiently several hadronic
matrix elements from the data on $CP$-conserving $\Kpipi$ decays. Indeed,
at this scale the contributions of all penguin operators and furthermore
$Q_{1,2}^c$ operators to the amplitudes $A_0$ and $A_2$ vanish for the HV
scheme or are very small for the $\ndr$ scheme. This allows for obtaining
the matrix elements $\langle Q_{1,2}(\mc) \rangle_2$ directly from $A_2$
as functions of $\Lms$. The matrix elements $\langle Q_{1,2}(\mc) \rangle_0$
are given as a function of $B_2^{(1/2)}(\mc)$ which is found to be in
the range $5 \le B_2^{(1/2)}(\mc) \le 7$. The relations between
operators allow to obtain subsequently all matrix elements of
$(V-A)\otimes (V-A)$  penguin operators. This analysis shows that the
extracted matrix elements differ considerably from the ones obtained by
the vacuum insertion method and in the case of $\langle Q_i \rangle_2$,
$i=1,2,9,10$ also from the lattice results. We have also demonstrated
that $\langle Q^c_{1,2}(\mu) \rangle_{0,2}$ can be calculated and that
for $\mu \le 2\gev$ the role of these operators in $\epe$ is small.
\item[v)]
Our renormalization group analysis of $\langle Q_i(\mu) \rangle$ shows
an interesting result that the $B_i$ parameters of all $(V-A)\otimes
(V+A)$ operators are almost $\mu$-independent. All $B_i$ parameters of
$(V-A)\otimes (V-A)$ operators show a substantial $\mu$-dependence. The
$\mu$- and scheme dependences of $\langle Q_{1,2}(\mu) \rangle_2$ and
$\langle Q_{9,10}(\mu) \rangle_2$ are given to a very good
approximation by $1/z_+(\mu)$ in the range $1\gev < \mu < 2\gev$.
The $\mu$-dependences of $\langle
Q_5(\mu) \rangle_0$, $\langle Q_6(\mu) \rangle_0$, $\langle Q_7(\mu)
\rangle_2$ and $\langle Q_8(\mu) \rangle_2$ are given within $5\%$
accuracy by $1/\ms^2(\mu)$. We also point out that the contributions
$P_6^{(1/2)}(\mu)$ and $P_8^{(3/2)}(\mu)$, which are $\mu$-independent
in the large-N limit show visible $\mu$-dependences when exact
renormalization group analysis with mixing between operators is taken
into account.
\item[vi)]
As a preparation for future developments in the evaluation of hadronic
matrix elements, we have presented $\epe$ in form of an expansion in
the independent non-perturbative parameters $B_i(\mc)$. The
coefficients of this expansion are functions of renormalization scheme,
$\Lms$ and $\mt$ and the values given in tabs.~\ref{tab:21} and
\ref{tab:22} are such that the $CP$-conserving $A_0$ and $A_2$
amplitudes are always consistent with the data.
\item[vii)]
Among several contributions to $\epe$ there are four which govern this
ratio. Two of them can to a good accuracy be fixed by the
$CP$-conserving data and consequently are only weakly dependent on
$\Lms$ and the renormalization scheme. These are the negative
$\mt$-independent contribution of $Q_4$ and a smaller positive
contribution of $Q_9+Q_{10}$ which moderately increases with $\mt$. The
former contribution is substantially enhanced when the $\dI$ rule is
taken into account as pointed out already in
ref.~\cite{buchallaetal:90}, and recently reemphasized by the authors
of \cite{ciuchini:92}. On the other hand, $Q_9$ and $Q_{10}$
contributions are similarly to $A_2$ suppressed substantially
relatively to their vacuum insertion and lattice estimates.
Consequently, if only the contributions of $(V-A)\otimes (V-A)$
penguins $Q_4$, $Q_9$ and $Q_{10}$ are taken into account $\epe$ is
negative and of $\ord(10^{-4})$.
\item[viii)]
The main uncertainties in $\epe$ result from $\IM\lambda_{\rm t}$ and
from the contributions of the $(V-A)\otimes (V+A)$ penguin operators
$Q_6$ and $Q_8$ which enter $\epe$ with opposite signs and are
sensitive functions of $\Lms$. None of them can be determined from the
existing $CP$-conserving data. Their matrix elements depend on $\ms$
and the parameters $B_6^{(1/2)}$ and $B_8^{(3/2)}$. For
$\ms(\mc)=150\mev$ and $B_6^{(1/2)}(\mc)=B_8^{(3/2)}(\mc)=1$, the
$Q_6$ contribution dominates for $\mt < 200\gev$, although this
dominance is substantially suppressed in the HV scheme and moderately
in the $\ndr$ scheme (see message iii) above). For $\mt < 150\gev$ the
contribution of $Q_8$ is smaller than the one of $Q_4$ and
consequently, for $\mt \approx \ord(130\gev)$ the ratio $\epe$ is
mainly determined by the QCD penguins $Q_6$ and $Q_4$ with $Q_6$ being
stronger and giving a positive $\epe$. With increasing $\mt$ the role
of $Q_8$ increases rapidly and for $\mt \approx \ord(200\gev)$ it is
the competition of $Q_6$ and $Q_8$ which dominates $\epe$. $Q_8$ being
helped by the negative contribution of $Q_4$ wins this competition for
$\mt > 200\gev$ and $\epe$ becomes negative similarly to the results of
\cite{buchallaetal:90}.
\item[ix)]
Although the parameters $B_6$ and $B_8$ are essentially
$\mu$-independent and in accordance with the vacuum insertion method
and the $1/N$ approach, there is an important question whether their
actual values are in the neighborhood of unity. Recall that other
parameters $B_i$ show substantial deviation from $B_i=1$. In this
respect, we sympathize with the efforts made by the authors of
ref.~\cite{froehlich:91,heinrichetal:92,wu:92} who questioned the use of
$B_6=B_8=1$. Yet, in our opinion the chapter on the actual values of
$B_6$ and $B_8$ is just being opened and it may well turn out that
these parameters are indeed special with values close to unity as given
in the leading order of the $1/N$ approach, in the vacuum insertion
method, and in lattice calculations.  A closely related issue is the
actual value of $\ms$.
\item[x)]
Our numerical analysis of $\epe$ summarized in
tabs.~\ref{tab:15}--\ref{tab:20} shows that for $B_6=B_8=1$ and high
values of $\mt$, at which strong cancellation takes place, the actual
values for $\epe$ depend sensitively on whether the ratio $y_8/y_6$ is
calculated in the $\ndr$ or HV scheme. This demonstrates very clearly
the necessity of a solid evaluation of $\langle Q_6 \rangle_0$ and
$\langle Q_8 \rangle_2$ which would cancel this dependence. For higher
values of $B_6$ and lower values of $\mt$ this scheme dependence is
weaker. Generally, the values of $\epe$ obtained in the HV scheme are
below the ones obtained in the $\ndr$ scheme and roughly in agreement
with the results of ref.~\cite{ciuchini:92}. We have also stressed that
$\epe$ depends sensitively on $\Lms$.

In the $\ndr$ scheme for $B_6^{(1/2)}(\mc)=B_8^{(3/2)}(\mc)=1$,
$\ms(\mc)=150\mev$, $\Lms=300\mev$, and $\mt=130\gev$, we find
\begin{equation}
\epe = \left\{
\begin{array}{ll}
(6.7 \pm 2.6) \times 10^{-4} & \qquad 0 < \delta \le \frac{\pi}{2} \, , \\
(4.8 \pm 2.2) \times 10^{-4} & \qquad \frac{\pi}{2} < \delta < \pi  \, ,
\end{array}
\right.
\label{eq:9.5}
\end{equation}
in perfect agreement with the E731 result. However, for
$B_6^{(1/2)}(\mc)=2$ and $B_8^{(3/2)}(\mc)=1$ we find
\begin{equation}
\epe = \left\{
\begin{array}{ll}
(20.0 \pm 6.5) \times 10^{-4} & \qquad 0 < \delta \le \frac{\pi}{2} \, ,\\
(14.4 \pm 5.6) \times 10^{-4} & \qquad \frac{\pi}{2} < \delta < \pi \, ,
\end{array}
\right.
\label{eq:9.6}
\end{equation}
in agreement with the findings of NA31. The LO results are slightly
higher than these values. In the HV scheme somewhat lower values are
found.
\end{itemize}

The fate of theoretical estimates of $\epe$ in the coming years depends
crucially on whether it will be possible to reduce the uncertainties in
the values of $\mt$, $\Lms$, $\ms$, $B_6^{(1/2)}$, $B_8^{(3/2)}$,
$B_{\rm K}$, $|\V{cb}|$ and $|\V{ub}/\V{cb}|$. If the standard $CP$
violating interactions parametrized by the CKM matrix are the main
origin of $\eps' \not=0$ also the fate of experimental searchers for
$\epe' \not=0$ depends crucially on these parameters. They should pray
that the top quark will be found soon or that $B_6 > 1.5$, $\ms(1\gev)
< 175\mev$ and $\Lms > 250\mev$.


\vskip 1cm
\begin{center}
{\large\bf Acknowledgement}
\end{center}
\noindent
We are grateful to Gerhard Buchalla, Robert Fleischer and Peter Weisz
for several stimulating and illuminating discussions.  M.~J. would like
to thank Toni Pich for discussions and A.J.~B. would like to thank
Guido Martinelli for e-mail discussions.


\newpage
\appendix{\LARGE\bf\noindent Appendices}
\renewcommand{\baselinestretch}{1.0}

\newsection{Quark Threshold Matching Matrices}
The matrix $\vardrs$ in \eqn{eq:4.27a5} has the following non-vanishing
rows.

$\mu = \mb$:
\begin{equation}
\vardrs(Q_4) = \vardrs(Q_6) = -2 \; \vardrs(Q_8) = -2 \; \vardrs(Q_{10}) =
-\frac{5}{9} \, P
\label{eq:a.1}
\end{equation}

$\mu = \mc$:
\begin{equation}
\vardrs(Q_4) = \vardrs(Q_6) = \vardrs(Q_8) = \vardrs(Q_{10}) =
-\frac{5}{9} \, P
\label{eq:a.2}
\end{equation}
where
\begin{equation}
P = (0,0,-\frac{1}{3},1,-\frac{1}{3},1,0,0,0,0) \, .
\label{eq:a.3}
\end{equation}
The matrix $\vardre$ has the following non-vanishing rows.

$\mu = \mb$:
\begin{equation}
\vardre(Q_3) =
3 \; \vardre(Q_4) =
\vardre(Q_5) =
3 \; \vardre(Q_6) =
\frac{20}{27} \, \bar{P}
\label{eq:a.4}
\end{equation}
\begin{equation}
\vardre(Q_7) =
3 \; \vardre(Q_8) =
\vardre(Q_9) =
3 \; \vardre(Q_{10}) =
-\frac{10}{27} \, \bar{P}
\label{eq:a.5}
\end{equation}

$\mu = \mc$:
\begin{equation}
\vardre(Q_3) = \vardre(Q_5) = \vardre(Q_7) = \vardre(Q_9) =
-\frac{40}{27} \, \bar{P}
\label{eq:a.6}
\end{equation}
\begin{equation}
\vardre(Q_4) = \vardre(Q_6) = \vardre(Q_8) = \vardre(Q_{10}) =
-\frac{40}{81} \, \bar{P}
\label{eq:a.7}
\end{equation}
where
\begin{equation}
P = (0,0,0,0,0,0,1,0,1,0) \, .
\label{eq:a.8}
\end{equation}

\newpage

\newsection{Numerical Results for Wilson Coefficients, $\epe$
and $B_i$-Expansion}
{\small
\begin{table}[h]
\caption{$\dS$ Wilson coefficients at $\mu=1\gev$ for
$\mt=130\gev$. In our approach $y_1 = y_2 \equiv 0$ holds.
\label{tab:11}}
\begin{center}
\begin{tabular}{|c|c|c|c||c|c|c||c|c|c|}
\hline
& \multicolumn{3}{c||}{$\Lms=0.2\gev$} &
  \multicolumn{3}{c||}{$\Lms=0.3\gev$} &
  \multicolumn{3}{c| }{$\Lms=0.4\gev$} \\
\hline
Scheme & LO & $\ndr$ & HV & LO &
$\ndr$ & HV & LO & $\ndr$ & HV \\
\hline
\hline
$z_1$ & -0.587 & -0.397 & -0.477 & -0.715 &
-0.486 & -0.606 & -0.854 & -0.588 & -0.774 \\
\hline
$z_2$ & 1.319 & 1.204 & 1.256 & 1.409 &
1.262 & 1.345 & 1.511 & 1.333 & 1.472 \\
\hline
\hline
$z_3$ & 0.004 & 0.008 & 0.004 & 0.005 &
0.013 & 0.008 & 0.007 & 0.021 & 0.015 \\
\hline
$z_4$ & -0.010 & -0.023 & -0.011 & -0.014 &
-0.035 & -0.018 & -0.018 & -0.054 & -0.030 \\
\hline
$z_5$ & 0.003 & 0.006 & 0.003 & 0.004 &
0.007 & 0.004 & 0.006 & 0.009 & 0.006 \\
\hline
$z_6$ & -0.011 & -0.023 & -0.010 & -0.015 &
-0.035 & -0.016 & -0.020 & -0.056 & -0.026 \\
\hline
\hline
$z_7/\aem$ & 0.005 & 0.000 & -0.005 & 0.008 &
0.008 & -0.003 & 0.012 & 0.017 & -0.002 \\
\hline
$z_8/\aem$ & 0.001 & 0.009 & 0.007 & 0.001 &
0.016 & 0.011 & 0.002 & 0.028 & 0.018 \\
\hline
$z_9/\aem$ & 0.005 & 0.005 & -0.001 & 0.009 &
0.016 & 0.004 & 0.013 & 0.030 & 0.010 \\
\hline
$z_{10}/\aem$ & -0.001 & -0.006 & -0.007 & -0.001 &
-0.009 & -0.011 & -0.001 & -0.015 & -0.017 \\
\hline
\hline
$z_-$ & 1.907 & 1.600 & 1.733 & 2.123 &
1.748 & 1.951 & 2.365 & 1.921 & 2.245 \\
\hline
$z_+$ & 0.732 & 0.807 & 0.779 & 0.694 &
0.776 & 0.740 & 0.657 & 0.746 & 0.698 \\
\hline
$z_-/z_+$ & 2.604 & 1.982 & 2.226 & 3.060 &
2.252 & 2.638 & 3.597 & 2.576 & 3.217 \\
\hline
\multicolumn{9}{c}{} \\
\hline
$y_3$ & 0.027 & 0.022 & 0.025 & 0.034 &
0.029 & 0.033 & 0.042 & 0.036 & 0.043 \\
\hline
$y_4$ & -0.048 & -0.044 & -0.046 & -0.056 &
-0.052 & -0.055 & -0.064 & -0.060 & -0.065 \\
\hline
$y_5$ & 0.011 & 0.005 & 0.013 & 0.012 &
0.001 & 0.015 & 0.013 & -0.008 & 0.018 \\
\hline
$y_6$ & -0.078 & -0.071 & -0.065 & -0.102 &
-0.099 & -0.087 & -0.130 & -0.142 & -0.118 \\
\hline
\hline
$y_7/\aem$ & -0.025 & -0.083 & -0.082 & -0.017 &
-0.080 & -0.079 & -0.009 & -0.078 & -0.076 \\
\hline
$y_8/\aem$ & 0.053 & 0.069 & 0.081 & 0.074 &
0.095 & 0.110 & 0.101 & 0.133 & 0.154 \\
\hline
$y_9/\aem$ & -1.160 & -1.161 & -1.161 & -1.222 &
-1.225 & -1.226 & -1.297 & -1.314 & -1.318 \\
\hline
$y_{10}/\aem$ & 0.488 & 0.400 & 0.407 & 0.592 &
0.502 & 0.512 & 0.706 & 0.634 & 0.647 \\
\hline
\end{tabular}
\end{center}
\end{table}

\begin{table}[h]
\caption{$\dS$ Wilson coefficients at $\mu=1.4\gev$ for
$\mt=130\gev$. In our approach $y_1 = y_2 \equiv 0$ holds.
\label{tab:12}}
\begin{center}
\begin{tabular}{|c|c|c|c||c|c|c||c|c|c|}
\hline
& \multicolumn{3}{c||}{$\Lms=0.2\gev$} &
  \multicolumn{3}{c||}{$\Lms=0.3\gev$} &
  \multicolumn{3}{c| }{$\Lms=0.4\gev$} \\
\hline
Scheme & LO & $\ndr$ & HV & LO &
$\ndr$ & HV & LO & $\ndr$ & HV \\
\hline
\hline
$z_1$ & -0.486 & -0.321 & -0.383 & -0.573 &
-0.379 & -0.459 & -0.660 & -0.438 & -0.543 \\
\hline
$z_2$ & 1.252 & 1.158 & 1.195 & 1.310 &
1.193 & 1.244 & 1.369 & 1.231 & 1.301 \\
\hline
\hline
$z_-$ & 1.738 & 1.479 & 1.578 & 1.883 &
1.571 & 1.704 & 2.030 & 1.669 & 1.844 \\
\hline
$z_+$ & 0.766 & 0.836 & 0.812 & 0.736 &
0.814 & 0.785 & 0.709 & 0.793 & 0.758 \\
\hline
$z_-/z_+$ & 2.269 & 1.769 & 1.943 & 2.558 &
1.930 & 2.170 & 2.862 & 2.105 & 2.432 \\
\hline
\multicolumn{9}{c}{} \\
\hline
$y_3$ & 0.024 & 0.020 & 0.022 & 0.029 &
0.025 & 0.028 & 0.035 & 0.030 & 0.035 \\
\hline
$y_4$ & -0.047 & -0.044 & -0.045 & -0.055 &
-0.052 & -0.054 & -0.063 & -0.060 & -0.063 \\
\hline
$y_5$ & 0.013 & 0.007 & 0.013 & 0.014 &
0.006 & 0.015 & 0.015 & 0.004 & 0.017 \\
\hline
$y_6$ & -0.068 & -0.061 & -0.056 & -0.085 &
-0.077 & -0.070 & -0.102 & -0.097 & -0.087 \\
\hline
\hline
$y_7/\aem$ & -0.023 & -0.082 & -0.081 & -0.014 &
-0.080 & -0.079 & -0.006 & -0.078 & -0.076 \\
\hline
$y_8/\aem$ & 0.041 & 0.059 & 0.067 & 0.054 &
0.076 & 0.085 & 0.068 & 0.097 & 0.106 \\
\hline
$y_9/\aem$ & -1.103 & -1.110 & -1.110 & -1.139 &
-1.144 & -1.143 & -1.179 & -1.182 & -1.182 \\
\hline
$y_{10}/\aem$ & 0.403 & 0.320 & 0.325 & 0.474 &
0.379 & 0.385 & 0.545 & 0.443 & 0.450 \\
\hline
\end{tabular}
\end{center}
\end{table}

\begin{table}[h]
\caption{$\dS$ Wilson coefficients at $\mu=2\gev$ for
$\mt=130\gev$. In our approach $y_1 = y_2 \equiv 0$ holds.
\label{tab:13}}
\begin{center}
\begin{tabular}{|c|c|c|c||c|c|c||c|c|c|}
\hline
& \multicolumn{3}{c||}{$\Lms=0.2\gev$} &
  \multicolumn{3}{c||}{$\Lms=0.3\gev$} &
  \multicolumn{3}{c| }{$\Lms=0.4\gev$} \\
\hline
Scheme & LO & $\ndr$ & HV & LO &
$\ndr$ & HV & LO & $\ndr$ & HV \\
\hline
\hline
$z_1$ & -0.403 & -0.261 & -0.311 & -0.465 &
-0.300 & -0.362 & -0.523 & -0.338 & -0.413 \\
\hline
$z_2$ & 1.200 & 1.123 & 1.152 & 1.238 &
1.145 & 1.182 & 1.276 & 1.168 & 1.214 \\
\hline
\hline
$z_-$ & 1.603 & 1.384 & 1.463 & 1.703 &
1.445 & 1.545 & 1.798 & 1.505 & 1.627 \\
\hline
$z_+$ & 0.797 & 0.862 & 0.840 & 0.773 &
0.845 & 0.820 & 0.752 & 0.830 & 0.801 \\
\hline
$z_-/z_+$ & 2.010 & 1.606 & 1.741 & 2.202 &
1.710 & 1.884 & 2.390 & 1.814 & 2.030 \\
\hline
\multicolumn{9}{c}{} \\
\hline
$y_3$ & 0.019 & 0.019 & 0.018 & 0.023 &
0.022 & 0.021 & 0.026 & 0.026 & 0.025 \\
\hline
$y_4$ & -0.039 & -0.045 & -0.038 & -0.045 &
-0.052 & -0.044 & -0.050 & -0.060 & -0.050 \\
\hline
$y_5$ & 0.011 & 0.010 & 0.011 & 0.012 &
0.010 & 0.013 & 0.013 & 0.011 & 0.014 \\
\hline
$y_6$ & -0.053 & -0.056 & -0.045 & -0.064 &
-0.067 & -0.054 & -0.074 & -0.080 & -0.063 \\
\hline
\hline
$y_7/\aem$ & -0.023 & -0.077 & -0.073 & -0.016 &
-0.073 & -0.072 & -0.010 & -0.069 & -0.071 \\
\hline
$y_8/\aem$ & 0.032 & 0.047 & 0.054 & 0.040 &
0.057 & 0.065 & 0.048 & 0.068 & 0.076 \\
\hline
$y_9/\aem$ & -1.060 & -1.073 & -1.070 & -1.084 &
-1.092 & -1.091 & -1.108 & -1.111 & -1.114 \\
\hline
$y_{10}/\aem$ & 0.333 & 0.264 & 0.268 & 0.384 &
0.304 & 0.309 & 0.432 & 0.344 & 0.349 \\
\hline
\end{tabular}
\end{center}
\end{table}

\begin{table}[thb]
\caption{$\dS$ Wilson coefficient $y_7/\aem$ at $\mu=1\gev$ for various
$\mt$ values.
\label{tab:1}}
\begin{center}
\begin{tabular}{|c|c|c|c||c|c|c||c|c|c|}
\hline
& \multicolumn{3}{c||}{$\Lms=0.2\gev$} &
  \multicolumn{3}{c||}{$\Lms=0.3\gev$} &
  \multicolumn{3}{c| }{$\Lms=0.4\gev$} \\
\hline
$\mt [\gev]$ & LO & $\ndr$ & HV & LO &
$\ndr$ & HV & LO & $\ndr$ & HV \\
\hline
\hline
110 & -0.045 & -0.102 & -0.102 & -0.036 &
-0.100 & -0.099 & -0.028 & -0.097 & -0.094 \\
\hline
130 & -0.025 & -0.083 & -0.082 & -0.017 &
-0.080 & -0.079 & -0.009 & -0.078 & -0.076 \\
\hline
150 & -0.002 & -0.058 & -0.058 & 0.006 &
-0.057 & -0.056 & 0.014 & -0.056 & -0.053 \\
\hline
170 & 0.025 & -0.031 & -0.031 & 0.033 &
-0.030 & -0.029 & 0.039 & -0.030 & -0.027 \\
\hline
190 & 0.056 & -0.000 & -0.000 & 0.062 &
-0.000 & 0.001 & 0.068 & -0.001 & 0.002 \\
\hline
\end{tabular}
\end{center}
\end{table}

\begin{table}[thb]
\caption{$\dS$ Wilson coefficient $y_8/\aem$ at $\mu=1\gev$ for various
$\mt$ values.
\label{tab:2}}
\begin{center}
\begin{tabular}{|c|c|c|c||c|c|c||c|c|c|}
\hline
& \multicolumn{3}{c||}{$\Lms=0.2\gev$} &
  \multicolumn{3}{c||}{$\Lms=0.3\gev$} &
  \multicolumn{3}{c| }{$\Lms=0.4\gev$} \\
\hline
$\mt [\gev]$ & LO & $\ndr$ & HV & LO &
$\ndr$ & HV & LO & $\ndr$ & HV \\
\hline
\hline
110 & 0.031 & 0.050 & 0.062 & 0.045 &
0.070 & 0.085 & 0.062 & 0.098 & 0.119 \\
\hline
130 & 0.053 & 0.069 & 0.081 & 0.074 &
0.095 & 0.110 & 0.101 & 0.133 & 0.154 \\
\hline
150 & 0.079 & 0.090 & 0.102 & 0.109 &
0.124 & 0.140 & 0.145 & 0.174 & 0.195 \\
\hline
170 & 0.109 & 0.115 & 0.126 & 0.148 &
0.158 & 0.173 & 0.196 & 0.221 & 0.242 \\
\hline
190 & 0.142 & 0.142 & 0.153 & 0.191 &
0.195 & 0.210 & 0.252 & 0.273 & 0.293 \\
\hline
\end{tabular}
\end{center}
\end{table}

\begin{table}[thb]
\caption{$\dS$ Wilson coefficient sum $(y_9+y_{10})/\aem$ at $\mu=1\gev$
 for various $\mt$ values.
\label{tab:3}}
\begin{center}
\begin{tabular}{|c|c|c|c||c|c|c||c|c|c|}
\hline
& \multicolumn{3}{c||}{$\Lms=0.2\gev$} &
  \multicolumn{3}{c||}{$\Lms=0.3\gev$} &
  \multicolumn{3}{c| }{$\Lms=0.4\gev$} \\
\hline
$\mt [\gev]$ & LO & $\ndr$ & HV & LO &
$\ndr$ & HV & LO & $\ndr$ & HV \\
\hline
\hline
110 & -0.567 & -0.652 & -0.646 & -0.531 &
-0.620 & -0.613 & -0.497 & -0.585 & -0.576 \\
\hline
130 & -0.672 & -0.760 & -0.754 & -0.630 &
-0.722 & -0.715 & -0.591 & -0.680 & -0.671 \\
\hline
150 & -0.779 & -0.872 & -0.866 & -0.732 &
-0.828 & -0.820 & -0.688 & -0.779 & -0.770 \\
\hline
170 & -0.891 & -0.987 & -0.981 & -0.838 &
-0.937 & -0.929 & -0.788 & -0.881 & -0.872 \\
\hline
190 & -1.008 & -1.108 & -1.102 & -0.948 &
-1.051 & -1.043 & -0.893 & -0.988 & -0.978 \\
\hline
\end{tabular}
\end{center}
\end{table}

\begin{table}[thb]
\caption{$\dS$ Wilson coefficient $y_7/\aem$ at $\mu=1.4\gev$ for various
$\mt$ values.
\label{tab:4}}
\begin{center}
\begin{tabular}{|c|c|c|c||c|c|c||c|c|c|}
\hline
& \multicolumn{3}{c||}{$\Lms=0.2\gev$} &
  \multicolumn{3}{c||}{$\Lms=0.3\gev$} &
  \multicolumn{3}{c| }{$\Lms=0.4\gev$} \\
\hline
$\mt [\gev]$ & LO & $\ndr$ & HV & LO &
$\ndr$ & HV & LO & $\ndr$ & HV \\
\hline
\hline
110 & -0.043 & -0.102 & -0.102 & -0.034 &
-0.100 & -0.099 & -0.025 & -0.098 & -0.096 \\
\hline
130 & -0.023 & -0.082 & -0.081 & -0.014 &
-0.080 & -0.079 & -0.006 & -0.078 & -0.076 \\
\hline
150 & 0.001 & -0.057 & -0.057 & 0.010 &
-0.056 & -0.054 & 0.018 & -0.054 & -0.052 \\
\hline
170 & 0.029 & -0.029 & -0.029 & 0.037 &
-0.028 & -0.027 & 0.044 & -0.027 & -0.025 \\
\hline
190 & 0.060 & 0.002 & 0.003 & 0.067 &
0.003 & 0.004 & 0.074 & 0.003 & 0.005 \\
\hline
\end{tabular}
\end{center}
\end{table}

\begin{table}[thb]
\caption{$\dS$ Wilson coefficient $y_8/\aem$ at $\mu=1.4\gev$ for various
$\mt$ values.
\label{tab:5}}
\begin{center}
\begin{tabular}{|c|c|c|c||c|c|c||c|c|c|}
\hline
& \multicolumn{3}{c||}{$\Lms=0.2\gev$} &
  \multicolumn{3}{c||}{$\Lms=0.3\gev$} &
  \multicolumn{3}{c| }{$\Lms=0.4\gev$} \\
\hline
$\mt [\gev]$ & LO & $\ndr$ & HV & LO &
$\ndr$ & HV & LO & $\ndr$ & HV \\
\hline
\hline
110 & 0.024 & 0.045 & 0.053 & 0.032 &
0.059 & 0.068 & 0.042 & 0.075 & 0.085 \\
\hline
130 & 0.041 & 0.059 & 0.067 & 0.054 &
0.076 & 0.085 & 0.068 & 0.097 & 0.106 \\
\hline
150 & 0.061 & 0.075 & 0.084 & 0.079 &
0.097 & 0.106 & 0.099 & 0.122 & 0.132 \\
\hline
170 & 0.084 & 0.094 & 0.102 & 0.107 &
0.120 & 0.129 & 0.133 & 0.151 & 0.161 \\
\hline
190 & 0.109 & 0.115 & 0.123 & 0.139 &
0.146 & 0.155 & 0.172 & 0.183 & 0.193 \\
\hline
\end{tabular}
\end{center}
\end{table}

\begin{table}[thb]
\caption{$\dS$ Wilson coefficient sum $(y_9+y_{10})/\aem$ at $\mu=1.4\gev$
 for various $\mt$ values.
\label{tab:7}}
\begin{center}
\begin{tabular}{|c|c|c|c||c|c|c||c|c|c|}
\hline
& \multicolumn{3}{c||}{$\Lms=0.2\gev$} &
  \multicolumn{3}{c||}{$\Lms=0.3\gev$} &
  \multicolumn{3}{c| }{$\Lms=0.4\gev$} \\
\hline
$\mt [\gev]$ & LO & $\ndr$ & HV & LO &
$\ndr$ & HV & LO & $\ndr$ & HV \\
\hline
\hline
110 & -0.591 & -0.677 & -0.672 & -0.560 &
-0.655 & -0.649 & -0.532 & -0.634 & -0.627 \\
\hline
130 & -0.700 & -0.791 & -0.785 & -0.665 &
-0.764 & -0.758 & -0.633 & -0.739 & -0.732 \\
\hline
150 & -0.812 & -0.907 & -0.902 & -0.773 &
-0.877 & -0.871 & -0.738 & -0.847 & -0.840 \\
\hline
170 & -0.929 & -1.028 & -1.023 & -0.886 &
-0.993 & -0.987 & -0.846 & -0.959 & -0.952 \\
\hline
190 & -1.052 & -1.155 & -1.149 & -1.003 &
-1.115 & -1.109 & -0.959 & -1.076 & -1.069 \\
\hline
\end{tabular}
\end{center}
\end{table}

\begin{table}[thb]
\caption{$\dS$ Wilson coefficient $y_7/\aem$ at $\mu=2\gev$ for various
$\mt$ values.
\label{tab:8}}
\begin{center}
\begin{tabular}{|c|c|c|c||c|c|c||c|c|c|}
\hline
& \multicolumn{3}{c||}{$\Lms=0.2\gev$} &
  \multicolumn{3}{c||}{$\Lms=0.3\gev$} &
  \multicolumn{3}{c| }{$\Lms=0.4\gev$} \\
\hline
$\mt [\gev]$ & LO & $\ndr$ & HV & LO &
$\ndr$ & HV & LO & $\ndr$ & HV \\
\hline
\hline
110 & -0.043 & -0.098 & -0.094 & -0.036 &
-0.093 & -0.093 & -0.030 & -0.089 & -0.091 \\
\hline
130 & -0.023 & -0.077 & -0.073 & -0.016 &
-0.073 & -0.072 & -0.010 & -0.069 & -0.071 \\
\hline
150 & 0.002 & -0.051 & -0.048 & 0.008 &
-0.048 & -0.047 & 0.014 & -0.044 & -0.046 \\
\hline
170 & 0.030 & -0.023 & -0.020 & 0.036 &
-0.020 & -0.019 & 0.041 & -0.017 & -0.018 \\
\hline
190 & 0.062 & 0.009 & 0.013 & 0.067 &
0.012 & 0.013 & 0.072 & 0.015 & 0.013 \\
\hline
\end{tabular}
\end{center}
\end{table}

\begin{table}[thb]
\caption{$\dS$ Wilson coefficient $y_8/\aem$ at $\mu=2\gev$ for various
$\mt$ values.
\label{tab:9}}
\begin{center}
\begin{tabular}{|c|c|c|c||c|c|c||c|c|c|}
\hline
& \multicolumn{3}{c||}{$\Lms=0.2\gev$} &
  \multicolumn{3}{c||}{$\Lms=0.3\gev$} &
  \multicolumn{3}{c| }{$\Lms=0.4\gev$} \\
\hline
$\mt [\gev]$ & LO & $\ndr$ & HV & LO &
$\ndr$ & HV & LO & $\ndr$ & HV \\
\hline
\hline
110 & 0.019 & 0.036 & 0.043 & 0.024 &
0.044 & 0.052 & 0.029 & 0.052 & 0.061 \\
\hline
130 & 0.032 & 0.047 & 0.054 & 0.040 &
0.057 & 0.065 & 0.048 & 0.068 & 0.076 \\
\hline
150 & 0.048 & 0.059 & 0.067 & 0.059 &
0.072 & 0.080 & 0.070 & 0.086 & 0.095 \\
\hline
170 & 0.066 & 0.074 & 0.081 & 0.081 &
0.090 & 0.098 & 0.096 & 0.106 & 0.115 \\
\hline
190 & 0.086 & 0.090 & 0.098 & 0.104 &
0.109 & 0.117 & 0.124 & 0.129 & 0.138 \\
\hline
\end{tabular}
\end{center}
\end{table}

\begin{table}[thb]
\caption{$\dS$ Wilson coefficient sum $(y_9+y_{10})/\aem$ at $\mu=2\gev$
 for various $\mt$ values.
\label{tab:10}}
\begin{center}
\begin{tabular}{|c|c|c|c||c|c|c||c|c|c|}
\hline
& \multicolumn{3}{c||}{$\Lms=0.2\gev$} &
  \multicolumn{3}{c||}{$\Lms=0.3\gev$} &
  \multicolumn{3}{c| }{$\Lms=0.4\gev$} \\
\hline
$\mt [\gev]$ & LO & $\ndr$ & HV & LO &
$\ndr$ & HV & LO & $\ndr$ & HV \\
\hline
\hline
110 & -0.613 & -0.693 & -0.685 & -0.589 &
-0.674 & -0.669 & -0.569 & -0.656 & -0.653 \\
\hline
130 & -0.727 & -0.810 & -0.802 & -0.700 &
-0.788 & -0.782 & -0.676 & -0.767 & -0.764 \\
\hline
150 & -0.844 & -0.931 & -0.923 & -0.814 &
-0.905 & -0.900 & -0.787 & -0.882 & -0.879 \\
\hline
170 & -0.966 & -1.056 & -1.048 & -0.932 &
-1.028 & -1.022 & -0.902 & -1.001 & -0.998 \\
\hline
190 & -1.093 & -1.187 & -1.179 & -1.055 &
-1.155 & -1.150 & -1.022 & -1.125 & -1.123 \\
\hline
\end{tabular}
\end{center}
\end{table}

\begin{table}[thb]
\caption{$\dS$ Wilson coefficient ratio $\frac{-(y_7/3+y_8)/\aem}{y_6}$
at $\mu=1.4\gev$ for various $\mt$ values.
\label{tab:6}}
\begin{center}
\begin{tabular}{|c|c|c|c||c|c|c||c|c|c|}
\hline
& \multicolumn{3}{c||}{$\Lms=0.2\gev$} &
  \multicolumn{3}{c||}{$\Lms=0.3\gev$} &
  \multicolumn{3}{c| }{$\Lms=0.4\gev$} \\
\hline
$\mt [\gev]$ & LO & $\ndr$ & HV & LO &
$\ndr$ & HV & LO & $\ndr$ & HV \\
\hline
\hline
110 & 0.144 & 0.183 & 0.349 & 0.252 &
0.331 & 0.499 & 0.330 & 0.443 & 0.615 \\
\hline
130 & 0.492 & 0.525 & 0.714 & 0.582 &
0.644 & 0.838 & 0.646 & 0.730 & 0.935 \\
\hline
150 & 0.898 & 0.923 & 1.138 & 0.964 &
1.007 & 1.233 & 1.013 & 1.066 & 1.308 \\
\hline
170 & 1.352 & 1.370 & 1.619 & 1.395 &
1.418 & 1.680 & 1.426 & 1.445 & 1.729 \\
\hline
190 & 1.855 & 1.867 & 2.146 & 1.870 &
1.871 & 2.173 & 1.883 & 1.864 & 2.195 \\
\hline
\end{tabular}
\end{center}
\end{table}
}

{\small
\begin{table}[thb]
\caption{Coefficients in the expansion of $P^{(1/2)}$. \label{tab:21}}
\begin{center}
\begin{tabular}{|c|c|c|c|c||c|c|c||c|c|c|}
\hline
& & \multicolumn{3}{c||}{LO} &
  \multicolumn{3}{c||}{NDR} &
  \multicolumn{3}{c|}{HV} \\
\hline
$\Lms$ & $\mt [\gev]$ & $a_0^{(1/2)}$ & $a_2^{(1/2)}$ & $a_6^{(1/2)}$ &
$a_0^{(1/2)}$ & $a_2^{(1/2)}$ & $a_6^{(1/2)}$ &
$a_0^{(1/2)}$ & $a_2^{(1/2)}$ & $a_6^{(1/2)}$ \\
\hline
\hline
   & 110 & -6.49 & 0.48 &  9.84 & -7.66 & 0.57 &  8.97 & -7.59 & 0.60 &  8.03\\
\hline
   & 130 & -6.18 & 0.45 &  9.97 & -7.29 & 0.53 &  9.07 & -7.22 & 0.56 &  8.13\\
\hline
200& 150 & -5.84 & 0.41 & 10.07 & -6.89 & 0.49 &  9.15 & -6.82 & 0.52 &  8.20\\
\hline
   & 170 & -5.46 & 0.38 & 10.15 & -6.45 & 0.45 &  9.21 & -6.38 & 0.48 &  8.26\\
\hline
   & 190 & -5.05 & 0.34 & 10.21 & -5.97 & 0.41 &  9.26 & -5.90 & 0.43 &  8.32\\
\hline
\hline
   & 110 & -6.62 & 0.47 & 12.29 & -7.94 & 0.57 & 11.56 & -7.84 & 0.61 & 10.07\\
\hline
   & 130 & -6.33 & 0.44 & 12.44 & -7.61 & 0.54 & 11.68 & -7.51 & 0.57 & 10.19\\
\hline
300& 150 & -6.00 & 0.42 & 12.56 & -7.24 & 0.51 & 11.77 & -7.14 & 0.54 & 10.28\\
\hline
   & 170 & -5.65 & 0.39 & 12.66 & -6.84 & 0.47 & 11.85 & -6.74 & 0.50 & 10.36\\
\hline
   & 190 & -5.26 & 0.35 & 12.74 & -6.41 & 0.43 & 11.91 & -6.31 & 0.46 & 10.42\\
\hline
\hline
   & 110 & -6.68 & 0.46 & 14.94 & -8.13 & 0.57 & 14.78 & -8.00 & 0.60 & 12.49\\
\hline
   & 130 & -6.40 & 0.44 & 15.13 & -7.82 & 0.54 & 14.92 & -7.69 & 0.57 & 12.64\\
\hline
400& 150 & -6.08 & 0.41 & 15.27 & -7.48 & 0.51 & 15.03 & -7.35 & 0.54 & 12.75\\
\hline
   & 170 & -5.73 & 0.39 & 15.39 & -7.10 & 0.48 & 15.12 & -6.97 & 0.51 & 12.84\\
\hline
   & 190 & -5.35 & 0.36 & 15.48 & -6.70 & 0.45 & 15.19 & -6.57 & 0.48 & 12.91\\
\hline
\end{tabular}
\end{center}
\end{table}

\begin{table}[thb]
\caption{Coefficients in the expansion of $P^{(3/2)}$. \label{tab:22}}
\begin{center}
\begin{tabular}{|c|c|c|c||c|c||c|c|}
\hline
& & \multicolumn{2}{c||}{LO} &
  \multicolumn{2}{c||}{NDR} &
  \multicolumn{2}{c|}{HV} \\
\hline
$\Lms$ & $\mt [\gev]$ & $a_0^{(3/2)}$ & $a_8^{(3/2)}$ & $a_0^{(3/2)}$ &
$a_8^{(3/2)}$ & $a_0^{(3/2)}$ & $a_8^{(3/2)}$ \\
\hline
\hline
    & 110 & -1.01 & 0.66 & -1.04 & 0.87 & -1.08 & 1.35 \\
\hline
    & 130 & -1.19 & 1.97 & -1.22 & 2.01 & -1.26 & 2.49 \\
\hline
200 & 150 & -1.38 & 3.52 & -1.41 & 3.36 & -1.45 & 3.84 \\
\hline
    & 170 & -1.58 & 5.28 & -1.60 & 4.89 & -1.64 & 5.37 \\
\hline
    & 190 & -1.79 & 7.24 & -1.80 & 6.59 & -1.84 & 7.07 \\
\hline
\hline
    & 110 & -0.99 & 1.29 & -1.03 & 1.68 & -1.08 & 2.22 \\
\hline
    & 130 & -1.18 & 2.84 & -1.21 & 3.02 & -1.26 & 3.55 \\
\hline
300 & 150 & -1.37 & 4.67 & -1.40 & 4.59 & -1.45 & 5.13 \\
\hline
    & 170 & -1.57 & 6.76 & -1.59 & 6.38 & -1.64 & 6.92 \\
\hline
    & 190 & -1.78 & 9.07 & -1.79 & 8.37 & -1.84 & 8.91 \\
\hline
\hline
    & 110 & -0.98 &  1.96 & -1.02 &  2.66 & -1.08 & 3.24 \\
\hline
    & 130 & -1.16 &  3.78 & -1.20 &  4.23 & -1.26 & 4.81 \\
\hline
400 & 150 & -1.36 &  5.92 & -1.39 &  6.07 & -1.44 & 6.66 \\
\hline
    & 170 & -1.56 &  8.36 & -1.58 &  8.17 & -1.64 & 8.75 \\
\hline
    & 190 & -1.76 & 11.06 & -1.78 & 10.50 & -1.84 &11.09 \\
\hline
\end{tabular}
\end{center}
\end{table}
}

\clearpage

{\small
\begin{table}[thb]
\caption{$\epe$ in units of $10^{-4}$ for $B_6^{(1/2)}\,=\,1.0$ and
$B_8^{(3/2)}\,=\,1.0$. \label{tab:15}}
\begin{center}
\begin{tabular}{|c|c|c|c||c|c||c|c|}
\hline
& & \multicolumn{2}{c||}{LO} &
  \multicolumn{2}{c||}{NDR} &
  \multicolumn{2}{c|}{HV} \\
\hline
$\Lms$ & $\mt$ & 1. Quad. & 2. Quad. & 1. Quad. & 2. Quad. &
1. Quad. & 2. Quad. \\
\hline
\hline
   & 130 & 6.4 $\pm$ 2.4 & 4.6 $\pm$ 2.0 & 5.2 $\pm$ 2.2 & 3.7 $\pm$ 1.8 & 3.7
$\pm$ 1.7 & 2.6 $\pm$ 1.4 \\
\hline
200& 150 & 4.6 $\pm$ 1.9 & 3.2 $\pm$ 1.6 & 3.7 $\pm$ 1.7 & 2.6 $\pm$ 1.4 & 2.4
$\pm$ 1.3 & 1.6 $\pm$ 1.0 \\
\hline
   & 170 & 3.0 $\pm$ 1.3 & 2.0 $\pm$ 1.1 & 2.3 $\pm$ 1.2 & 1.6 $\pm$ 1.0 & 1.1
$\pm$ 0.8 & 0.8 $\pm$ 0.6 \\
\hline
\hline
   & 130 & 8.0 $\pm$ 2.9 & 5.8 $\pm$ 2.5 & 6.7 $\pm$ 2.6 & 4.8 $\pm$ 2.2 & 4.6
$\pm$ 2.0 & 3.3 $\pm$ 1.6 \\
\hline
300& 150 & 5.8 $\pm$ 2.2 & 4.1 $\pm$ 1.9 & 4.9 $\pm$ 2.0 & 3.4 $\pm$ 1.7 & 2.9
$\pm$ 1.5 & 2.0 $\pm$ 1.2 \\
\hline
   & 170 & 3.8 $\pm$ 1.6 & 2.6 $\pm$ 1.4 & 3.1 $\pm$ 1.5 & 2.1 $\pm$ 1.2 & 1.4
$\pm$ 0.9 & 1.0 $\pm$ 0.7 \\
\hline
\hline
   & 130 & 9.9 $\pm$ 3.4 & 7.1 $\pm$ 2.9 & 8.8 $\pm$ 3.2 & 6.3 $\pm$ 2.7 & 5.7
$\pm$ 2.3 & 4.1 $\pm$ 1.9 \\
\hline
400& 150 & 7.2 $\pm$ 2.7 & 5.0 $\pm$ 2.3 & 6.5 $\pm$ 2.5 & 4.5 $\pm$ 2.2 & 3.7
$\pm$ 1.7 & 2.6 $\pm$ 1.4 \\
\hline
   & 170 & 4.7 $\pm$ 1.9 & 3.2 $\pm$ 1.6 & 4.3 $\pm$ 1.9 & 3.0 $\pm$ 1.6 & 1.8
$\pm$ 1.1 & 1.3 $\pm$ 0.9 \\
\hline
\end{tabular}
\end{center}
\end{table}

\begin{table}[thb]
\caption{$\epe$ in units of $10^{-4}$ for $B_6^{(1/2)}\,=\,1.5$ and
$B_8^{(3/2)}\,=\,1.5$. \label{tab:16}}
\begin{center}
\begin{tabular}{|c|c|c|c||c|c||c|c|}
\hline
& & \multicolumn{2}{c||}{LO} &
  \multicolumn{2}{c||}{NDR} &
  \multicolumn{2}{c|}{HV} \\
\hline
$\Lms$ & $\mt$ & 1. Quad. & 2. Quad. & 1. Quad. & 2. Quad. &
1. Quad. & 2. Quad. \\
\hline
\hline
   & 130 & 10.9 $\pm$ 3.8 & 7.9 $\pm$ 3.2 & 9.2 $\pm$ 3.3 & 6.6 $\pm$ 2.8 & 6.9
$\pm$ 2.7 & 5.0 $\pm$ 2.2 \\
\hline
200& 150 & 8.0 $\pm$ 2.9 & 5.6 $\pm$ 2.5 & 6.6 $\pm$ 2.6 & 4.6 $\pm$ 2.2 & 4.6
$\pm$ 1.9 & 3.2 $\pm$ 1.6 \\
\hline
   & 170 & 5.2 $\pm$ 2.1 & 3.6 $\pm$ 1.8 & 4.3 $\pm$ 1.8 & 2.9 $\pm$ 1.6 & 2.5
$\pm$ 1.2 & 1.7 $\pm$ 1.0 \\
\hline
\hline
   & 130 & 13.5 $\pm$ 4.5 & 9.7 $\pm$ 3.9 & 11.6 $\pm$ 4.1 & 8.3 $\pm$ 3.4 &
8.3 $\pm$ 3.1 & 6.0 $\pm$ 2.6 \\
\hline
300& 150 & 9.9 $\pm$ 3.5 & 6.9 $\pm$ 3.0 & 8.5 $\pm$ 3.2 & 5.9 $\pm$ 2.7 & 5.6
$\pm$ 2.3 & 3.9 $\pm$ 1.9 \\
\hline
   & 170 & 6.5 $\pm$ 2.5 & 4.5 $\pm$ 2.2 & 5.6 $\pm$ 2.3 & 3.8 $\pm$ 2.0 & 3.0
$\pm$ 1.4 & 2.0 $\pm$ 1.2 \\
\hline
\hline
   & 130 & 16.3 $\pm$ 5.4 & 11.8 $\pm$ 4.6 & 14.8 $\pm$ 5.0 & 10.6 $\pm$ 4.3 &
10.2 $\pm$ 3.7 & 7.3 $\pm$ 3.1 \\
\hline
400& 150 & 12.0 $\pm$ 4.1 & 8.4 $\pm$ 3.6 & 11.0 $\pm$ 3.9 & 7.7 $\pm$ 3.4 &
6.9 $\pm$ 2.7 & 4.8 $\pm$ 2.3 \\
\hline
   & 170 & 8.0 $\pm$ 3.0 & 5.5 $\pm$ 2.6 & 7.5 $\pm$ 2.9 & 5.1 $\pm$ 2.5 & 3.7
$\pm$ 1.7 & 2.5 $\pm$ 1.4 \\
\hline
\end{tabular}
\end{center}
\end{table}

\begin{table}[thb]
\caption{$\epe$ in units of $10^{-4}$ for $B_6^{(1/2)}\,=\,0.75$ and
$B_8^{(3/2)}\,=\,0.75$. \label{tab:17}}
\begin{center}
\begin{tabular}{|c|c|c|c||c|c||c|c|}
\hline
& & \multicolumn{2}{c||}{LO} &
  \multicolumn{2}{c||}{NDR} &
  \multicolumn{2}{c|}{HV} \\
\hline
$\Lms$ & $\mt$ & 1. Quad. & 2. Quad. & 1. Quad. & 2. Quad. &
1. Quad. & 2. Quad. \\
\hline
\hline
   & 130 & 4.1 $\pm$ 1.8 & 3.0 $\pm$ 1.4 & 3.2 $\pm$ 1.6 & 2.3 $\pm$ 1.3 & 2.1
$\pm$ 1.3 & 1.5 $\pm$ 1.0 \\
\hline
200& 150 & 2.9 $\pm$ 1.4 & 2.0 $\pm$ 1.1 & 2.2 $\pm$ 1.2 & 1.6 $\pm$ 1.0 & 1.2
$\pm$ 0.9 & 0.9 $\pm$ 0.7 \\
\hline
   & 170 & 1.8 $\pm$ 1.0 & 1.2 $\pm$ 0.8 & 1.3 $\pm$ 0.9 & 0.9 $\pm$ 0.7 & 0.4
$\pm$ 0.6 & 0.3 $\pm$ 0.4 \\
\hline
\hline
   & 130 & 5.3 $\pm$ 2.1 & 3.8 $\pm$ 1.7 & 4.3 $\pm$ 1.9 & 3.1 $\pm$ 1.6 & 2.7
$\pm$ 1.4 & 1.9 $\pm$ 1.2 \\
\hline
300& 150 & 3.8 $\pm$ 1.6 & 2.6 $\pm$ 1.4 & 3.1 $\pm$ 1.5 & 2.1 $\pm$ 1.2 & 1.6
$\pm$ 1.0 & 1.1 $\pm$ 0.8 \\
\hline
   & 170 & 2.4 $\pm$ 1.2 & 1.6 $\pm$ 1.0 & 1.9 $\pm$ 1.1 & 1.3 $\pm$ 0.9 & 0.6
$\pm$ 0.7 & 0.4 $\pm$ 0.5 \\
\hline
\hline
   & 130 & 6.6 $\pm$ 2.5 & 4.8 $\pm$ 2.1 & 5.8 $\pm$ 2.4 & 4.2 $\pm$ 2.0 & 3.5
$\pm$ 1.7 & 2.5 $\pm$ 1.4 \\
\hline
400& 150 & 4.8 $\pm$ 1.9 & 3.4 $\pm$ 1.6 & 4.2 $\pm$ 1.9 & 2.9 $\pm$ 1.5 & 2.2
$\pm$ 1.2 & 1.5 $\pm$ 1.0 \\
\hline
   & 170 & 3.1 $\pm$ 1.4 & 2.1 $\pm$ 1.2 & 2.8 $\pm$ 1.4 & 1.9 $\pm$ 1.1 & 0.9
$\pm$ 0.8 & 0.6 $\pm$ 0.6 \\
\hline
\end{tabular}
\end{center}
\end{table}

\begin{table}[thb]
\caption{$\epe$ in units of $10^{-4}$ for $B_6^{(1/2)}\,=\,1.25$ and
$B_8^{(3/2)}\,=\,0.75$. \label{tab:18}}
\begin{center}
\begin{tabular}{|c|c|c|c||c|c||c|c|}
\hline
& & \multicolumn{2}{c||}{LO} &
  \multicolumn{2}{c||}{NDR} &
  \multicolumn{2}{c|}{HV} \\
\hline
$\Lms$ & $\mt$ & 1. Quad. & 2. Quad. & 1. Quad. & 2. Quad. &
1. Quad. & 2. Quad. \\
\hline
\hline
   & 130 & 9.8 $\pm$ 3.4 & 7.0 $\pm$ 2.9 & 8.3 $\pm$ 3.1 & 6.0 $\pm$ 2.6 & 6.7
$\pm$ 2.6 & 4.8 $\pm$ 2.2 \\
\hline
200& 150 & 8.1 $\pm$ 2.9 & 5.6 $\pm$ 2.5 & 6.9 $\pm$ 2.7 & 4.8 $\pm$ 2.3 & 5.4
$\pm$ 2.2 & 3.8 $\pm$ 1.9 \\
\hline
   & 170 & 6.6 $\pm$ 2.5 & 4.5 $\pm$ 2.2 & 5.6 $\pm$ 2.3 & 3.8 $\pm$ 1.9 & 4.3
$\pm$ 1.8 & 2.9 $\pm$ 1.5 \\
\hline
\hline
   & 130 & 12.4 $\pm$ 4.2 & 8.9 $\pm$ 3.6 & 10.9 $\pm$ 3.9 & 7.8 $\pm$ 3.3 &
8.5 $\pm$ 3.2 & 6.1 $\pm$ 2.7 \\
\hline
300& 150 & 10.3 $\pm$ 3.6 & 7.1 $\pm$ 3.1 & 9.1 $\pm$ 3.3 & 6.3 $\pm$ 2.9 & 6.9
$\pm$ 2.7 & 4.8 $\pm$ 2.3 \\
\hline
   & 170 & 8.3 $\pm$ 3.1 & 5.7 $\pm$ 2.7 & 7.4 $\pm$ 2.9 & 5.0 $\pm$ 2.5 & 5.4
$\pm$ 2.2 & 3.7 $\pm$ 1.9 \\
\hline
\hline
   & 130 & 15.3 $\pm$ 5.0 & 11.0 $\pm$ 4.3 & 14.2 $\pm$ 4.9 & 10.2 $\pm$ 4.1 &
10.7 $\pm$ 3.8 & 7.7 $\pm$ 3.2 \\
\hline
400& 150 & 12.7 $\pm$ 4.3 & 8.8 $\pm$ 3.8 & 11.9 $\pm$ 4.2 & 8.3 $\pm$ 3.7 &
8.7 $\pm$ 3.2 & 6.1 $\pm$ 2.8 \\
\hline
   & 170 & 10.3 $\pm$ 3.7 & 7.0 $\pm$ 3.2 & 9.8 $\pm$ 3.6 & 6.6 $\pm$ 3.2 & 6.9
$\pm$ 2.7 & 4.7 $\pm$ 2.3 \\
\hline
\end{tabular}
\end{center}
\end{table}

\begin{table}[thb]
\caption{$\epe$ in units of $10^{-4}$ for $B_6^{(1/2)}\,=\,1.5$ and
$B_8^{(3/2)}\,=\,1.0$. \label{tab:19}}
\begin{center}
\begin{tabular}{|c|c|c|c||c|c||c|c|}
\hline
& & \multicolumn{2}{c||}{LO} &
  \multicolumn{2}{c||}{NDR} &
  \multicolumn{2}{c|}{HV} \\
\hline
$\Lms$ & $\mt$ & 1. Quad. & 2. Quad. & 1. Quad. & 2. Quad. &
1. Quad. & 2. Quad. \\
\hline
\hline
   & 130 & 12.1 $\pm$ 4.1 & 8.7 $\pm$ 3.5 & 10.3 $\pm$ 3.7 & 7.4 $\pm$ 3.1 &
8.3 $\pm$ 3.1 & 6.0 $\pm$ 2.6 \\
\hline
200& 150 & 9.8 $\pm$ 3.5 & 6.8 $\pm$ 3.0 & 8.4 $\pm$ 3.1 & 5.8 $\pm$ 2.7 & 6.6
$\pm$ 2.6 & 4.6 $\pm$ 2.2 \\
\hline
   & 170 & 7.7 $\pm$ 2.9 & 5.2 $\pm$ 2.5 & 6.6 $\pm$ 2.6 & 4.5 $\pm$ 2.2 & 5.0
$\pm$ 2.1 & 3.4 $\pm$ 1.7 \\
\hline
\hline
   & 130 & 15.1 $\pm$ 5.0 & 10.9 $\pm$ 4.3 & 13.3 $\pm$ 4.6 & 9.6 $\pm$ 3.9 &
10.4 $\pm$ 3.7 & 7.5 $\pm$ 3.1 \\
\hline
300& 150 & 12.3 $\pm$ 4.2 & 8.6 $\pm$ 3.7 & 10.9 $\pm$ 3.9 & 7.6 $\pm$ 3.4 &
8.2 $\pm$ 3.1 & 5.7 $\pm$ 2.6 \\
\hline
   & 170 & 9.7 $\pm$ 3.5 & 6.6 $\pm$ 3.1 & 8.6 $\pm$ 3.3 & 5.9 $\pm$ 2.8 & 6.2
$\pm$ 2.5 & 4.2 $\pm$ 2.1 \\
\hline
\hline
   & 130 & 18.5 $\pm$ 6.0 & 13.3 $\pm$ 5.1 & 17.2 $\pm$ 5.7 & 12.4 $\pm$ 4.9 &
12.9 $\pm$ 4.5 & 9.3 $\pm$ 3.8 \\
\hline
400& 150 & 15.1 $\pm$ 5.1 & 10.5 $\pm$ 4.5 & 14.2 $\pm$ 4.9 & 9.9 $\pm$ 4.3 &
10.3 $\pm$ 3.7 & 7.2 $\pm$ 3.2 \\
\hline
   & 170 & 11.9 $\pm$ 4.3 & 8.1 $\pm$ 3.7 & 11.3 $\pm$ 4.2 & 7.7 $\pm$ 3.6 &
7.8 $\pm$ 3.0 & 5.3 $\pm$ 2.6 \\
\hline
\end{tabular}
\end{center}
\end{table}

\begin{table}[thb]
\caption{$\epe$ in units of $10^{-4}$ for $B_6^{(1/2)}\,=\,2.0$ and
$B_8^{(3/2)}\,=\,1.0$. \label{tab:20}}
\begin{center}
\begin{tabular}{|c|c|c|c||c|c||c|c|}
\hline
& & \multicolumn{2}{c||}{LO} &
  \multicolumn{2}{c||}{NDR} &
  \multicolumn{2}{c|}{HV} \\
\hline
$\Lms$ & $\mt$ & 1. Quad. & 2. Quad. & 1. Quad. & 2. Quad. &
1. Quad. & 2. Quad. \\
\hline
\hline
   & 130 & 17.8 $\pm$ 5.8 & 12.8 $\pm$ 5.0 & 15.5 $\pm$ 5.2 & 11.1 $\pm$ 4.4 &
12.9 $\pm$ 4.5 & 9.3 $\pm$ 3.8 \\
\hline
200& 150 & 15.0 $\pm$ 5.0 & 10.4 $\pm$ 4.4 & 13.1 $\pm$ 4.5 & 9.1 $\pm$ 4.0 &
10.8 $\pm$ 3.8 & 7.5 $\pm$ 3.3 \\
\hline
   & 170 & 12.4 $\pm$ 4.4 & 8.5 $\pm$ 3.9 & 10.9 $\pm$ 4.0 & 7.4 $\pm$ 3.5 &
8.8 $\pm$ 3.3 & 6.0 $\pm$ 2.9 \\
\hline
\hline
   & 130 & 22.2 $\pm$ 7.1 & 16.0 $\pm$ 6.1 & 20.0 $\pm$ 6.5 & 14.4 $\pm$ 5.6 &
16.2 $\pm$ 5.4 & 11.6 $\pm$ 4.6 \\
\hline
300& 150 & 18.8 $\pm$ 6.2 & 13.1 $\pm$ 5.5 & 16.9 $\pm$ 5.7 & 11.8 $\pm$ 5.0 &
13.5 $\pm$ 4.7 & 9.4 $\pm$ 4.1 \\
\hline
   & 170 & 15.6 $\pm$ 5.4 & 10.6 $\pm$ 4.8 & 14.1 $\pm$ 5.1 & 9.6 $\pm$ 4.4 &
11.1 $\pm$ 4.1 & 7.5 $\pm$ 3.5 \\
\hline
\hline
   & 130 & 27.1 $\pm$ 8.5 & 19.5 $\pm$ 7.4 & 25.7 $\pm$ 8.2 & 18.5 $\pm$ 7.1 &
20.1 $\pm$ 6.6 & 14.5 $\pm$ 5.6 \\
\hline
400& 150 & 22.9 $\pm$ 7.5 & 16.0 $\pm$ 6.6 & 21.8 $\pm$ 7.2 & 15.2 $\pm$ 6.4 &
16.8 $\pm$ 5.7 & 11.7 $\pm$ 5.0 \\
\hline
   & 170 & 19.1 $\pm$ 6.6 & 13.0 $\pm$ 5.8 & 18.3 $\pm$ 6.4 & 12.5 $\pm$ 5.6 &
13.8 $\pm$ 5.0 & 9.4 $\pm$ 4.3 \\
\hline
\end{tabular}
\end{center}
\end{table}
}

{\small
\begin{table}[thb]
\caption{$\dB$ Wilson coefficients at $\mu=\mb$ for
$\mt=130\gev$.
\label{tab:14}}
\begin{center}
\begin{tabular}{|c|c|c|c||c|c|c||c|c|c|}
\hline
& \multicolumn{3}{c||}{$\Lms=0.2\gev$} &
  \multicolumn{3}{c||}{$\Lms=0.3\gev$} &
  \multicolumn{3}{c| }{$\Lms=0.4\gev$} \\
\hline
Scheme & LO & $\ndr$ & HV & LO &
$\ndr$ & HV & LO & $\ndr$ & HV \\
\hline
\hline
$z_1$ & -0.255 & -0.152 & -0.187 & -0.286 &
-0.170 & -0.210 & -0.312 & -0.185 & -0.230 \\
\hline
$z_2$ & 1.115 & 1.066 & 1.083 & 1.131 &
1.075 & 1.095 & 1.146 & 1.082 & 1.106 \\
\hline
\hline
$y_3$ & 0.011 & 0.011 & 0.010 & 0.012 &
0.013 & 0.011 & 0.014 & 0.014 & 0.013 \\
\hline
$y_4$ & -0.025 & -0.029 & -0.024 & -0.028 &
-0.033 & -0.027 & -0.030 & -0.036 & -0.029 \\
\hline
$y_5$ & 0.007 & 0.008 & 0.008 & 0.008 &
0.008 & 0.008 & 0.009 & 0.009 & 0.009 \\
\hline
$y_6$ & -0.030 & -0.033 & -0.026 & -0.034 &
-0.037 & -0.030 & -0.038 & -0.041 & -0.033 \\
\hline
\hline
$y_7/\aem$ & -0.011 & -0.056 & -0.045 & -0.008 &
-0.054 & -0.046 & -0.006 & -0.053 & -0.046 \\
\hline
$y_8/\aem$ & 0.019 & 0.028 & 0.033 & 0.022 &
0.032 & 0.037 & 0.025 & 0.035 & 0.041 \\
\hline
$y_9/\aem$ & -0.981 & -1.003 & -0.992 & -0.991 &
-1.010 & -1.001 & -1.000 & -1.017 & -1.009 \\
\hline
$y_{10}/\aem$ & 0.210 & 0.166 & 0.169 & 0.235 &
0.185 & 0.188 & 0.256 & 0.201 & 0.204 \\
\hline
\end{tabular}
\end{center}
\end{table}
}
\vfill

\renewcommand{\baselinestretch}{1.2}

\clearpage
\newsection{Collection of Numerical Input Parameters}
\leftline{\large\bf Quark Masses}

\begin{displaymath}
\begin{array}{lclclcl}
\mt &=& 100\;-\;200\gev &\qquad& \mt^{\rm central} &=& 130\gev \\
\mb &=& 4.8\gev    &\qquad& \mc &=& 1.4\gev \\
\md(\mc) &=& 8\mev &\qquad& \ms(\mc) &=& 150\mev
\end{array}
\end{displaymath}

\medskip

\leftline{\large\bf Scalar Meson Masses and Decay Constants}

\begin{displaymath}
\begin{array}{lclclcl}
m_{\rm K} &=& 498\mev &\qquad& F_{\rm K} &=& 161\mev \\
m_\pi     &=& 135\mev &\qquad& F_\pi     &=& 132\mev
\end{array}
\end{displaymath}

\medskip

\leftline{\large\bf QCD and Electroweak Parameters}

\begin{displaymath}
\begin{array}{lclclcl}
\Lms      &=& 200\;-\;400\mev &\qquad& \Lms^{\rm central}  &=& 300\mev \\
\aem      &=& 1/128           &\qquad& \mw                 &=& 80.0\gev
\end{array}
\end{displaymath}

\medskip

\leftline{\large\bf CKM Elements}

\begin{displaymath}
\begin{array}{lclclcl}
\left|\V{us}\right| &=& 0.221 &\qquad&
\left|\V{ud}\right| &=& 0.9753 \\
\left| \V{cb} \right| &=& 0.043 \pm 0.004 &\qquad&
\left| \V{ub}/\V{cb} \right| &=& 0.10 \pm 0.03 \quad {\rm (range~I)} \\
\left| \V{cb} \right| &=& 0.043 \pm 0.002 &\qquad&
\left| \V{ub}/\V{cb} \right| &=& 0.10 \pm 0.01 \quad {\rm (range~II)}
\end{array}
\end{displaymath}

\medskip

\leftline{\large\bf $\Kpipi$ Decays and $K^0-\bar{K}^0$ Mixing}

\begin{displaymath}
\begin{array}{lclclcl}
\RE A_0 &=& 3.33 \times 10^{-7}\gev &\qquad&
\RE A_2 &=& 1.50 \times 10^{-8}\gev \\
\omega  &=& 1/22.2              &\qquad& \Omega_{\eta\eta'} &=& 0.25 \\
\eps    &=& (2.258 \pm 0.018) \times 10^{-3} &\qquad& \Delta M_{\rm K}
&=& 3.5 \times 10^{-15}\gev \\
B_{\rm K} &=& 0.65 \pm 0.15 \quad {\rm (range~I)} &\qquad&
B_{\rm K} &=& 0.65 \pm 0.05 \quad {\rm (range~II)} \\
\eta_1  &=& 0.85 &\qquad& \eta_2 &=& 0.58 \\
\eta_3  &=& 0.36 &\qquad& && \\
\end{array}
\end{displaymath}

\medskip

\leftline{\large\bf Hadronic Matrix Elements}

\begin{displaymath}
\begin{array}{lclclclclcl}
B_{2,LO}^{(1/2)}(\mc)   &=& 5.8 \pm 1.1 &\qquad&
B_{2,\ndr}^{(1/2)}(\mc) &=& 6.7 \pm 0.9 &\qquad&
B_{2,HV}^{(1/2)}(\mc)   &=& 6.3 \pm 1.0 \\
\end{array}
\end{displaymath}

\begin{displaymath}
B_3^{(1/2)} = B_5^{(1/2)} = B_6^{(1/2)} = B_7^{(1/2)} = B_8^{(1/2)} =
B_7^{(3/2)} = B_8^{(3/2)} = 1
\qquad
\hbox{\rm (central values)}
\end{displaymath}

\clearpage
\newsection{Figures}

\begin{figure}[h]
\vspace{0.15in}
\centerline{
\epsfysize=3.1in
\epsffile{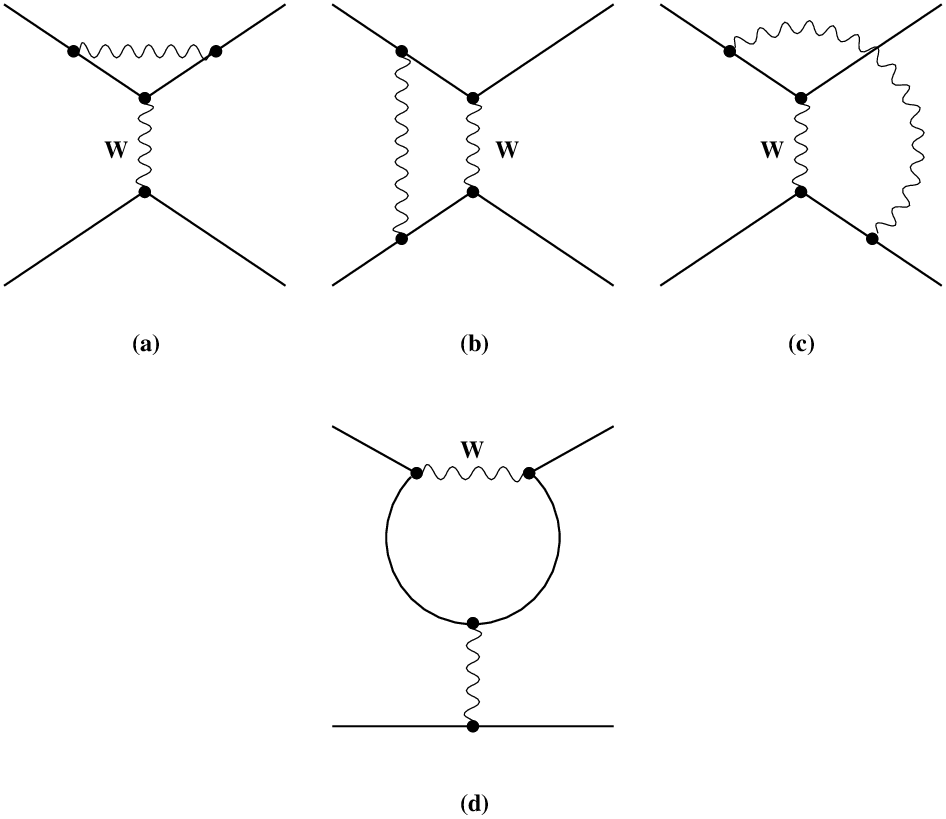}
}
\vspace{0.15in}
\caption[]{
\label{fig:1a}}
\end{figure}

\begin{figure}[h]
\vspace{0.15in}
\centerline{
\epsfysize=3.1in
\epsffile{effdiag.ps}
}
\vspace{0.15in}
\caption[]{
\label{fig:1b}}
\end{figure}

\begin{figure}[h]
\vspace{0.15in}
\centerline{
\epsfysize=7in
\rotate[r]{
\epsffile{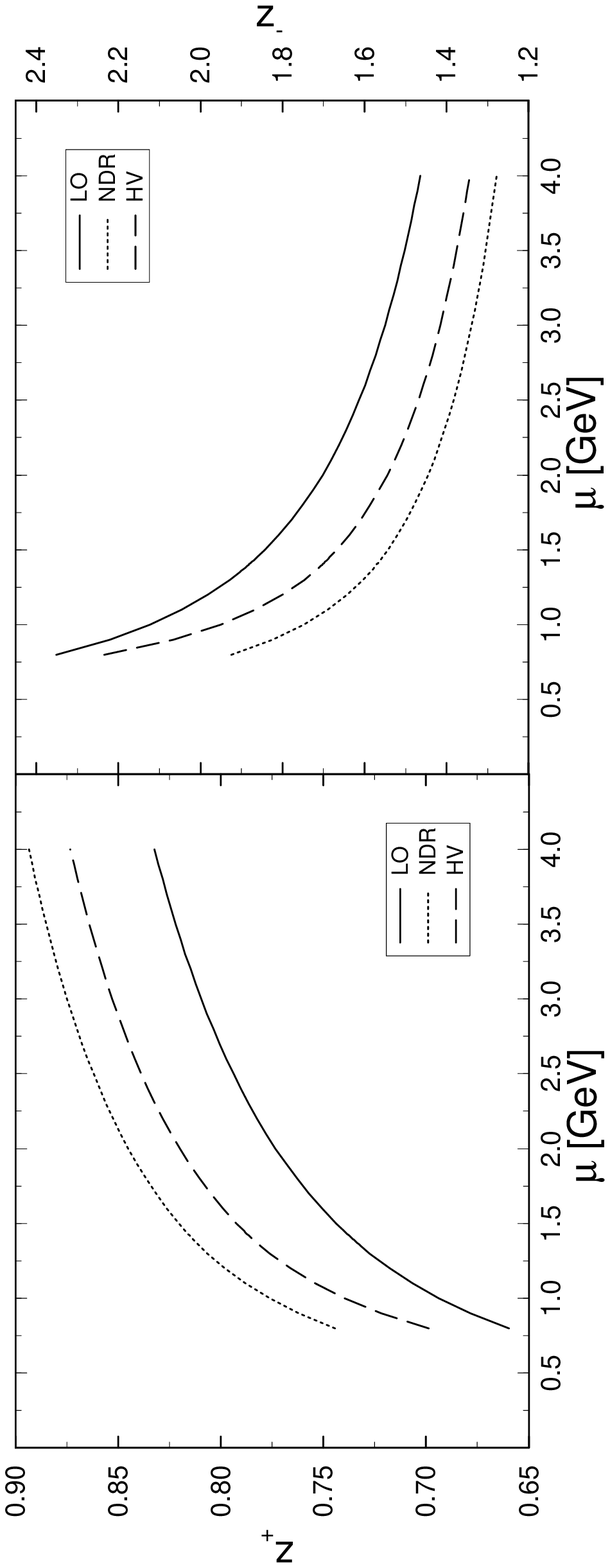}
}}
\vspace{0.15in}
\caption[]{
\label{fig:1}}
\end{figure}

\begin{figure}[h]
\vspace{0.15in}
\centerline{
\epsfysize=7in
\rotate[r]{
\epsffile{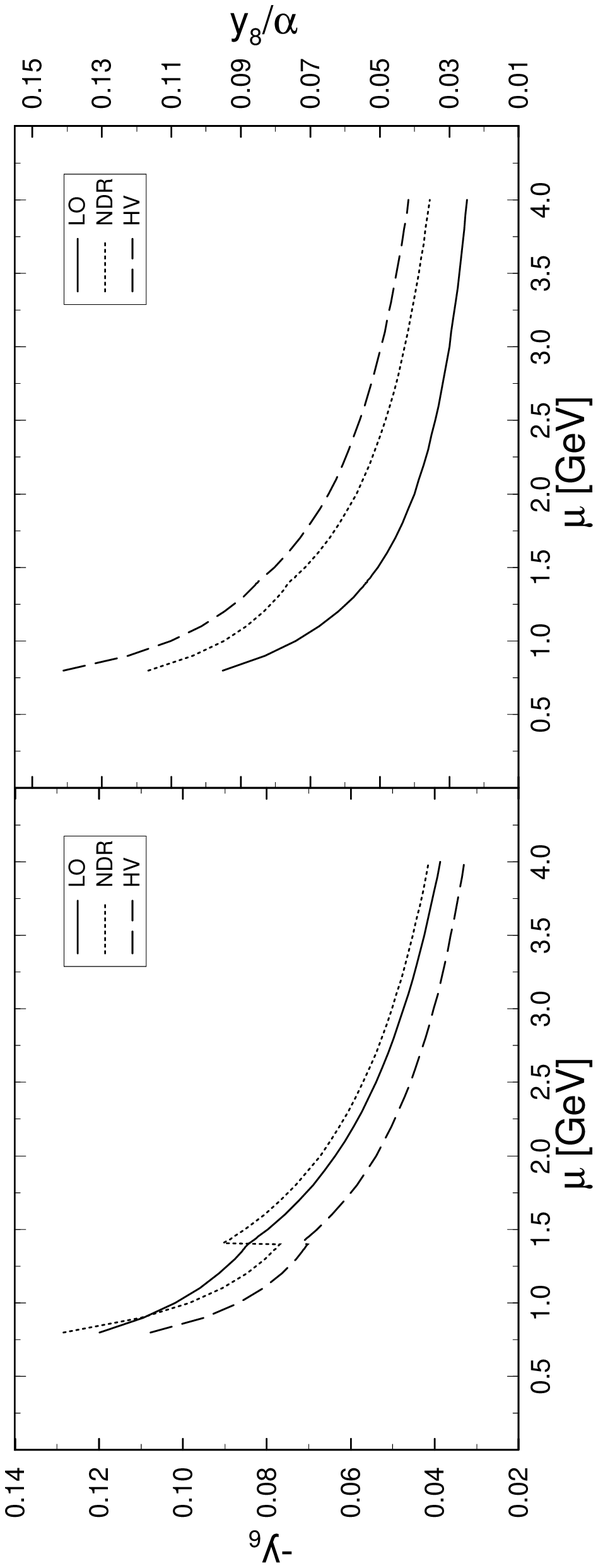}
}}
\vspace{0.15in}
\caption[]{
\label{fig:2}}
\end{figure}

\begin{figure}[h]
\vspace{0.15in}
\centerline{
\epsfysize=7in
\rotate[r]{
\epsffile{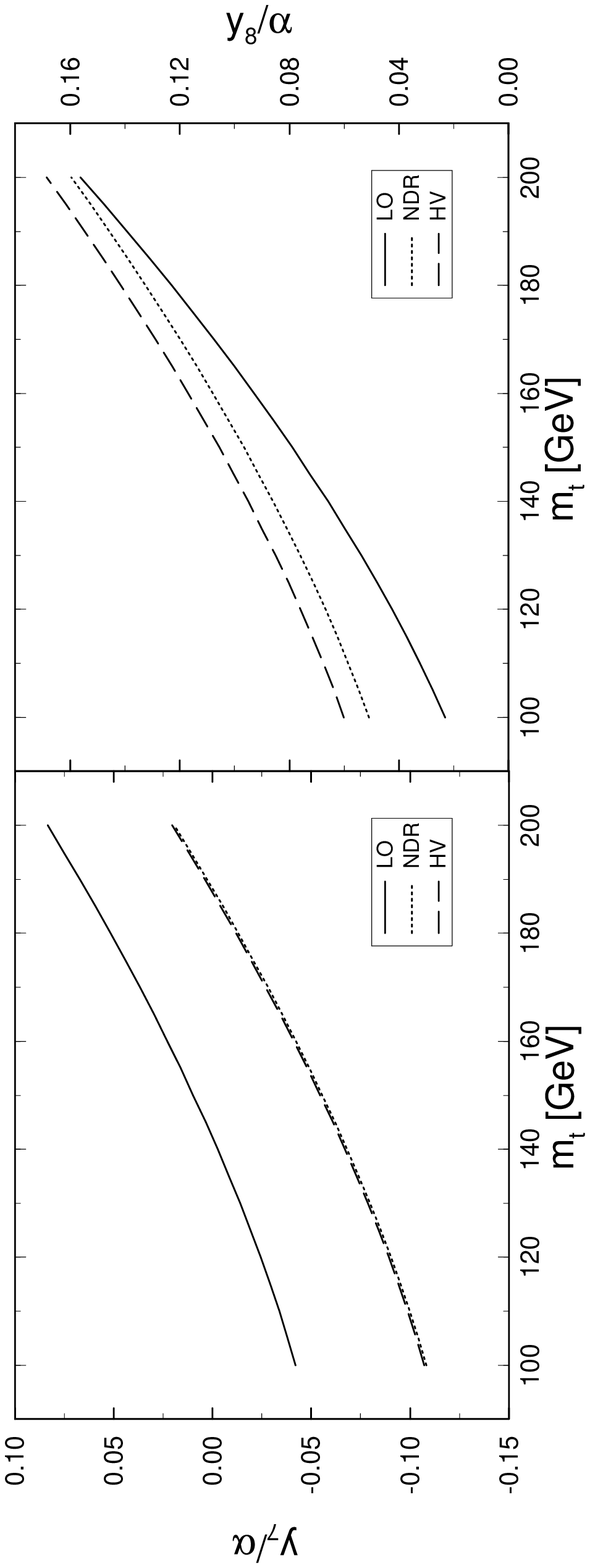}
}}
\vspace{0.15in}
\caption[]{
\label{fig:3}}
\end{figure}

\begin{figure}[h]
\vspace{0.15in}
\centerline{
\epsfysize=7in
\rotate[r]{
\epsffile{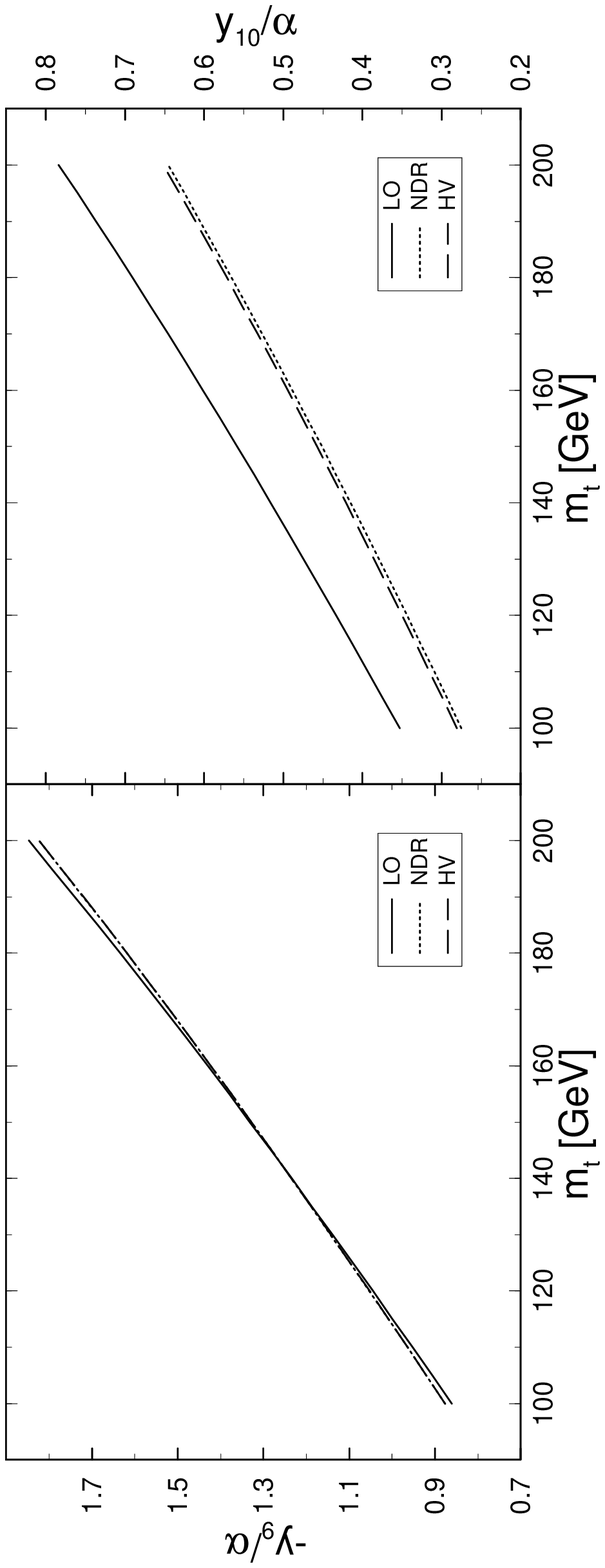}
}}
\vspace{0.15in}
\caption[]{
\label{fig:4}}
\end{figure}

\begin{figure}[h]
\vspace{0.15in}
\centerline{
\epsfysize=4.5in
\rotate[r]{
\epsffile{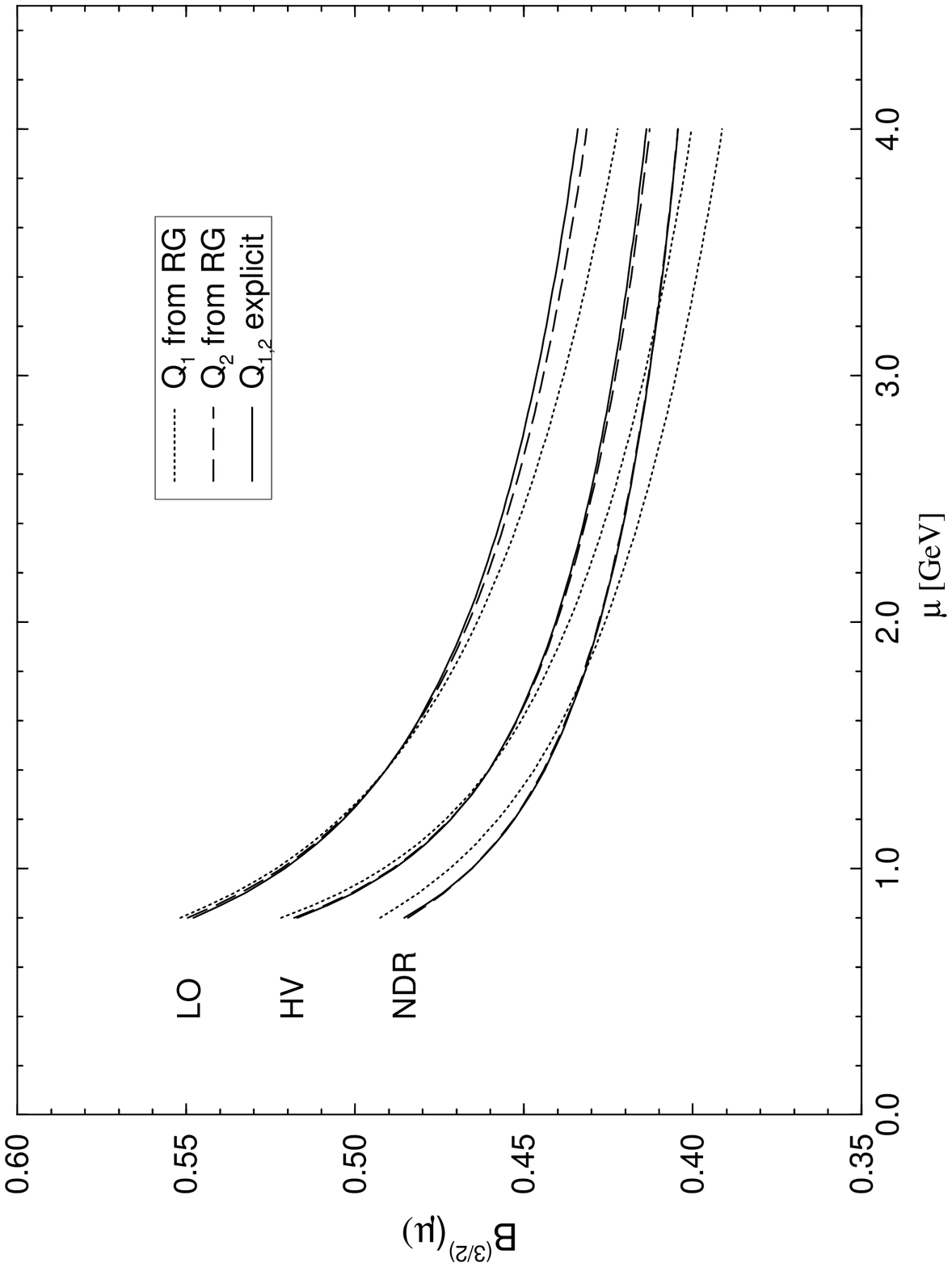}
}}
\vspace{0.15in}
\caption[]{
\label{fig:12}}
\end{figure}

\begin{figure}[h]
\vspace{0.15in}
\centerline{
\epsfysize=4.5in
\rotate[r]{
\epsffile{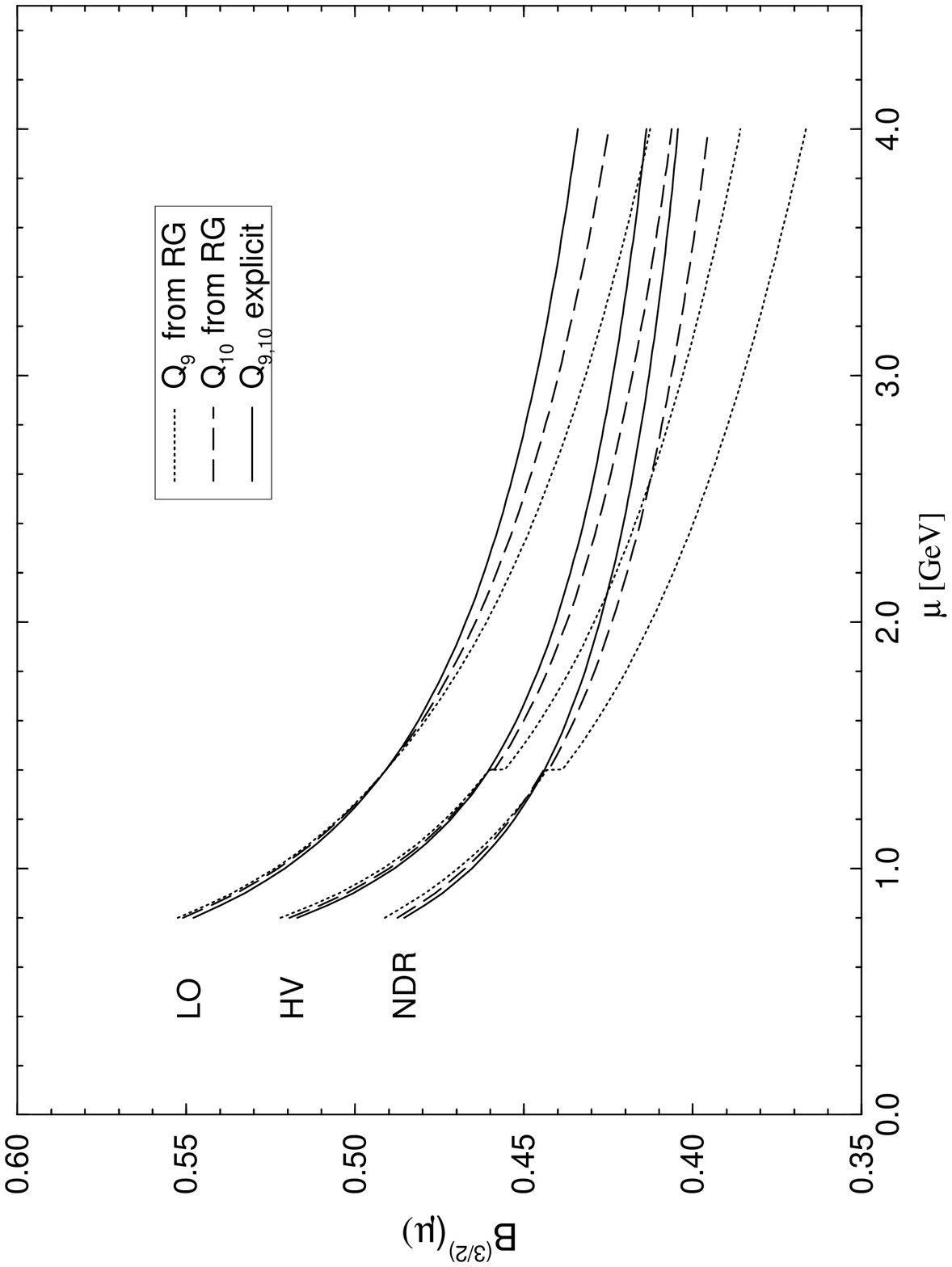}
}}
\vspace{0.15in}
\caption[]{
\label{fig:13}}
\end{figure}

\begin{figure}[h]
\vspace{0.15in}
\centerline{
\epsfysize=4.5in
\rotate[r]{
\epsffile{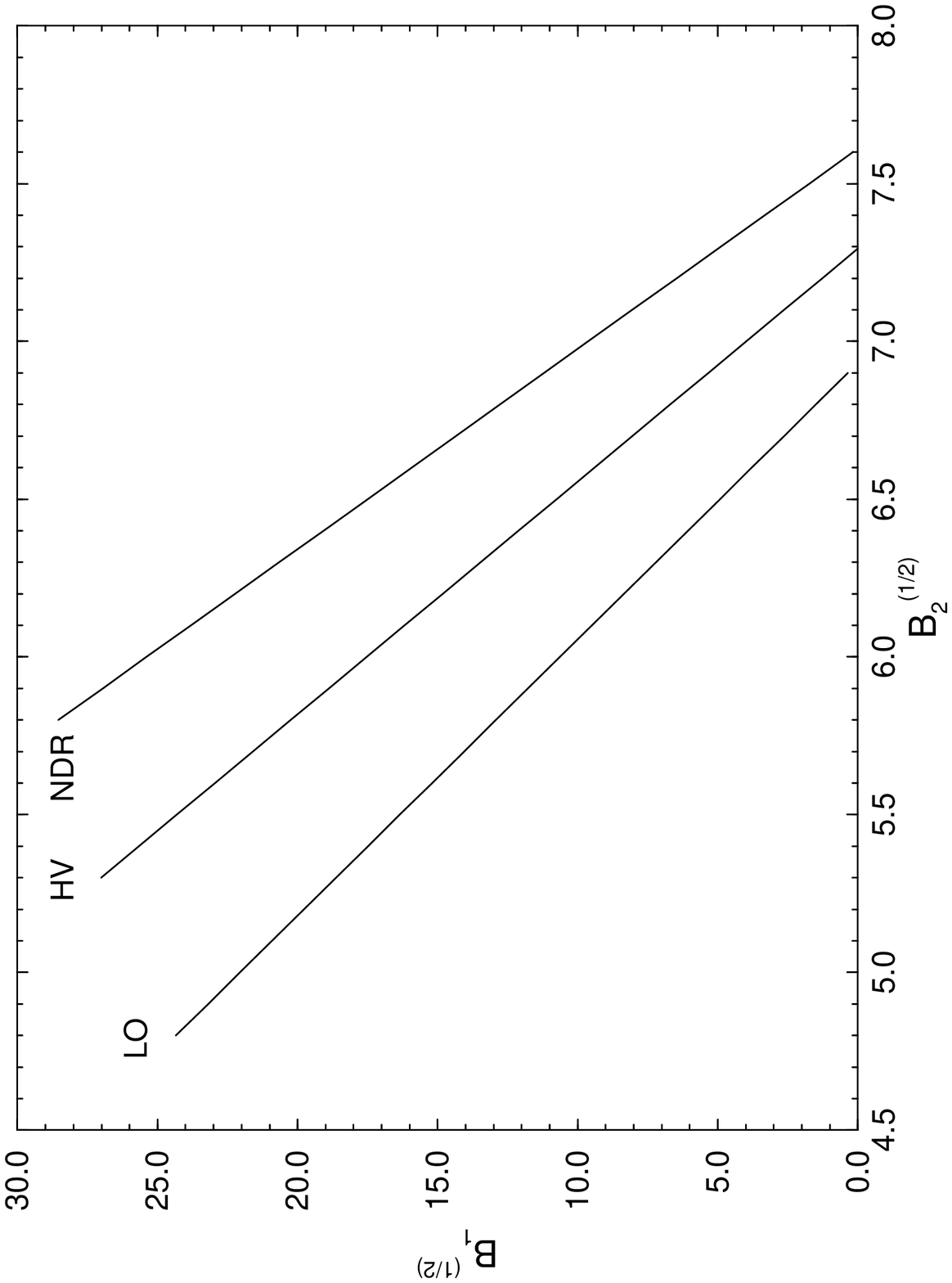}
}}
\vspace{0.15in}
\caption[]{
\label{fig:6}}
\end{figure}

\begin{figure}[h]
\vspace{0.15in}
\centerline{
\epsfysize=4.5in
\rotate[r]{
\epsffile{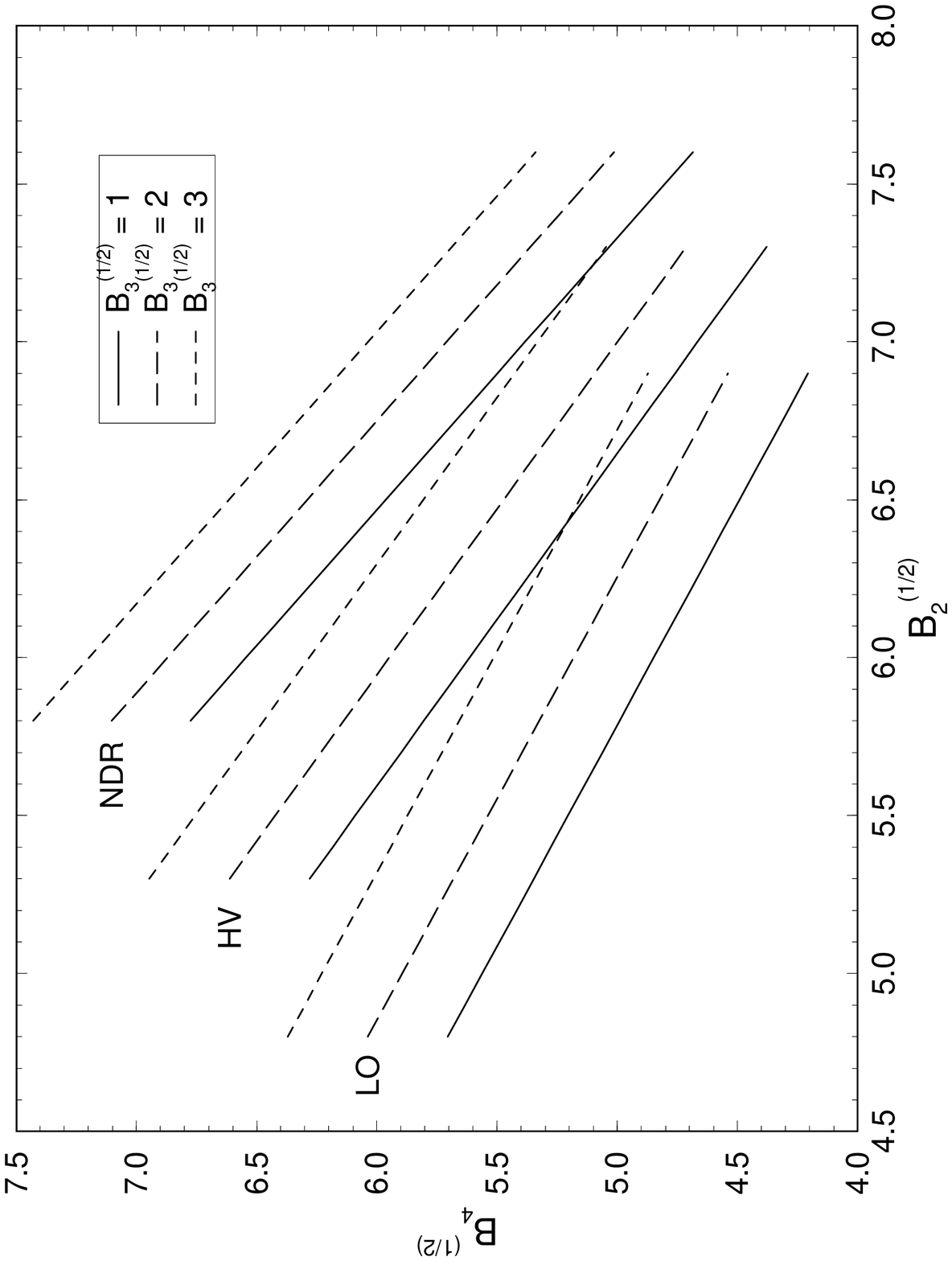}
}}
\vspace{0.15in}
\caption[]{
\label{fig:14}}
\end{figure}

\begin{figure}[h]
\vspace{0.15in}
\centerline{
\epsfysize=4.5in
\rotate[r]{
\epsffile{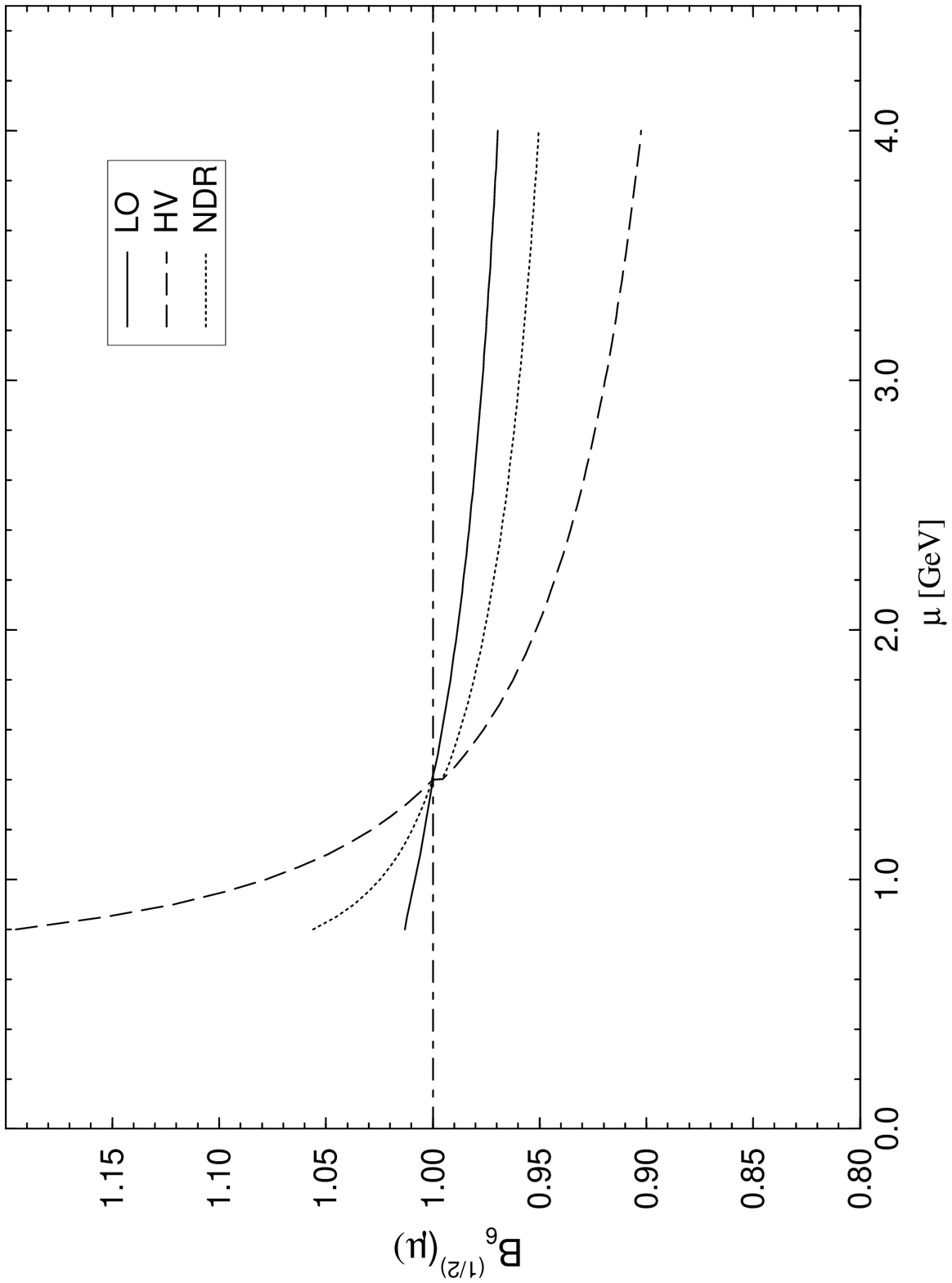}
}}
\vspace{0.15in}
\caption[]{
\label{fig:9}}
\end{figure}

\begin{figure}[h]
\vspace{0.15in}
\centerline{
\epsfysize=4.5in
\rotate[r]{
\epsffile{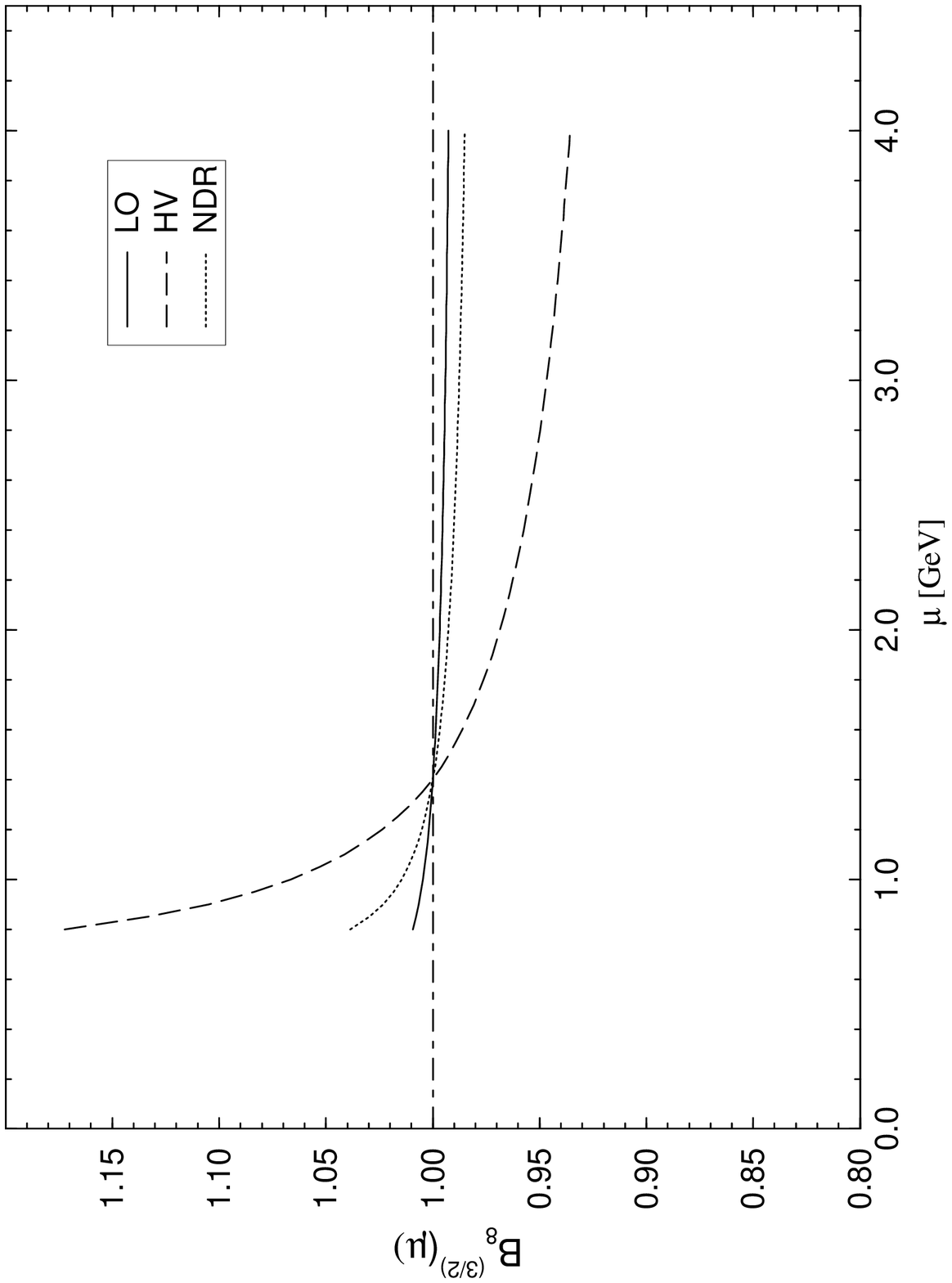}
}}
\vspace{0.15in}
\caption[]{
\label{fig:11}}
\end{figure}

\begin{figure}[h]
\vspace{0.15in}
\centerline{
\epsfysize=4.5in
\rotate[r]{
\epsffile{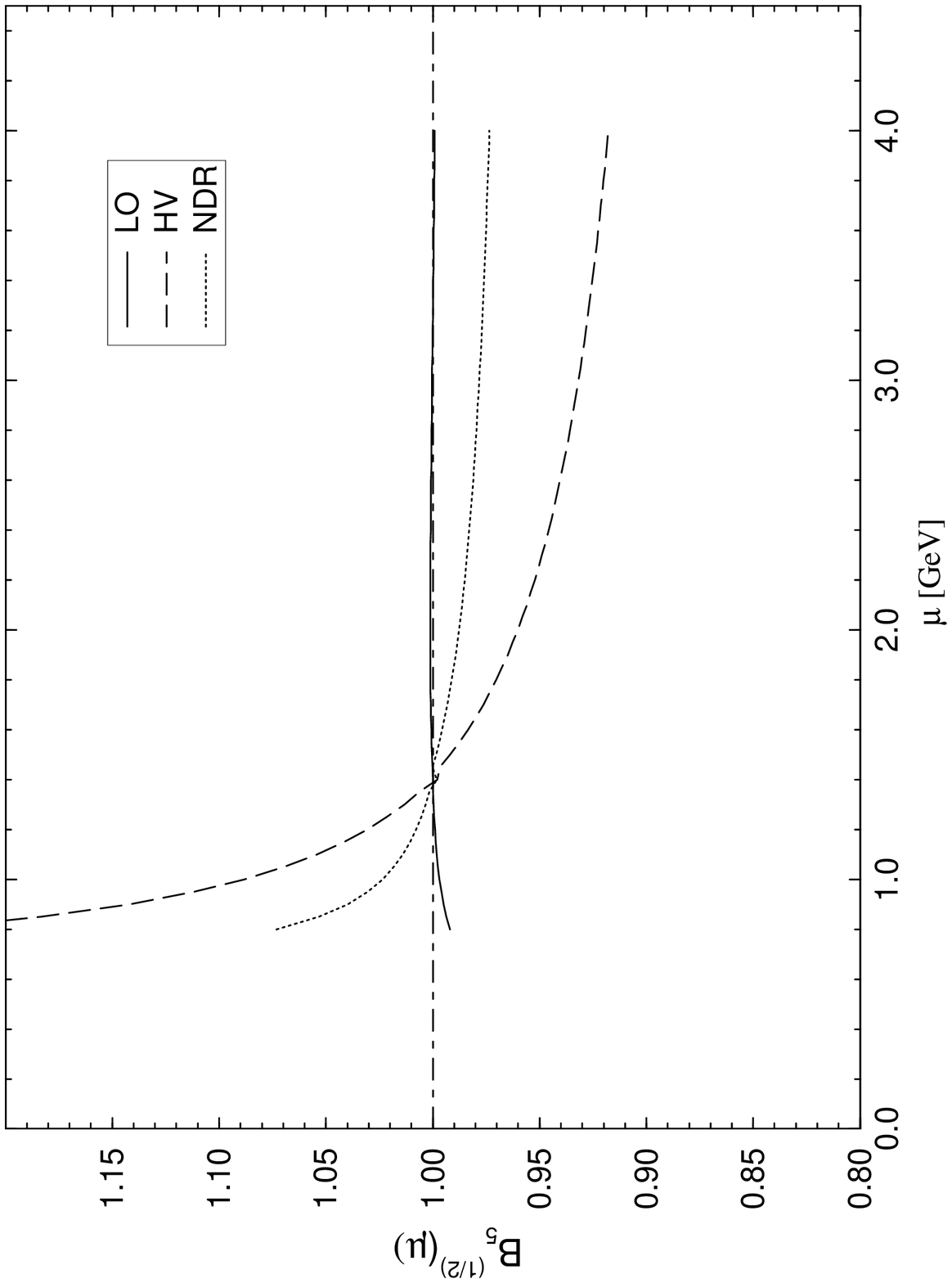}
}}
\vspace{0.15in}
\caption[]{
\label{fig:8}}
\end{figure}

\begin{figure}[h]
\vspace{0.15in}
\centerline{
\epsfysize=4.5in
\rotate[r]{
\epsffile{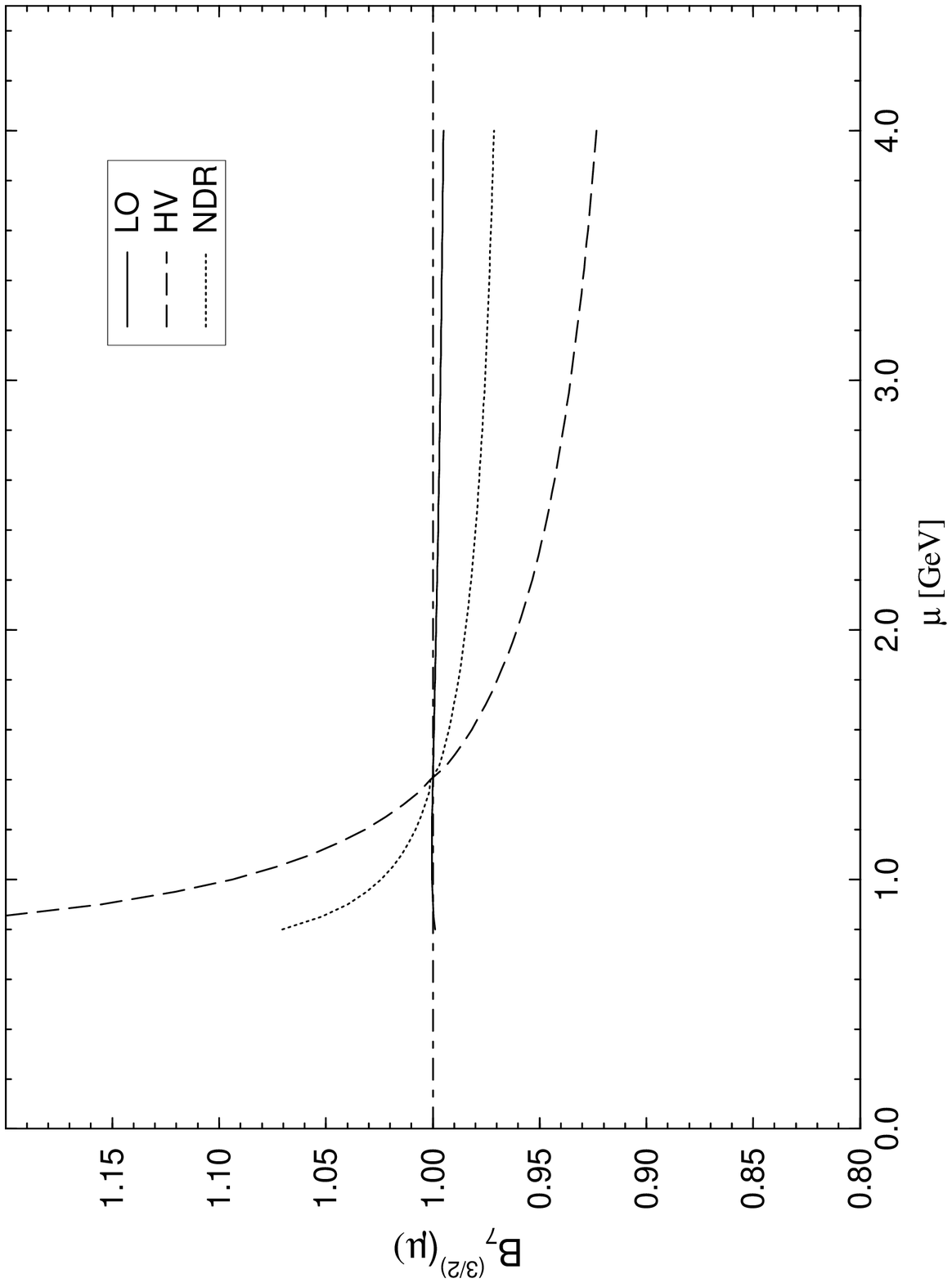}
}}
\vspace{0.15in}
\caption[]{
\label{fig:10}}
\end{figure}

\begin{figure}[h]
\vspace{0.15in}
\centerline{
\epsfysize=4.5in
\rotate[r]{
\epsffile{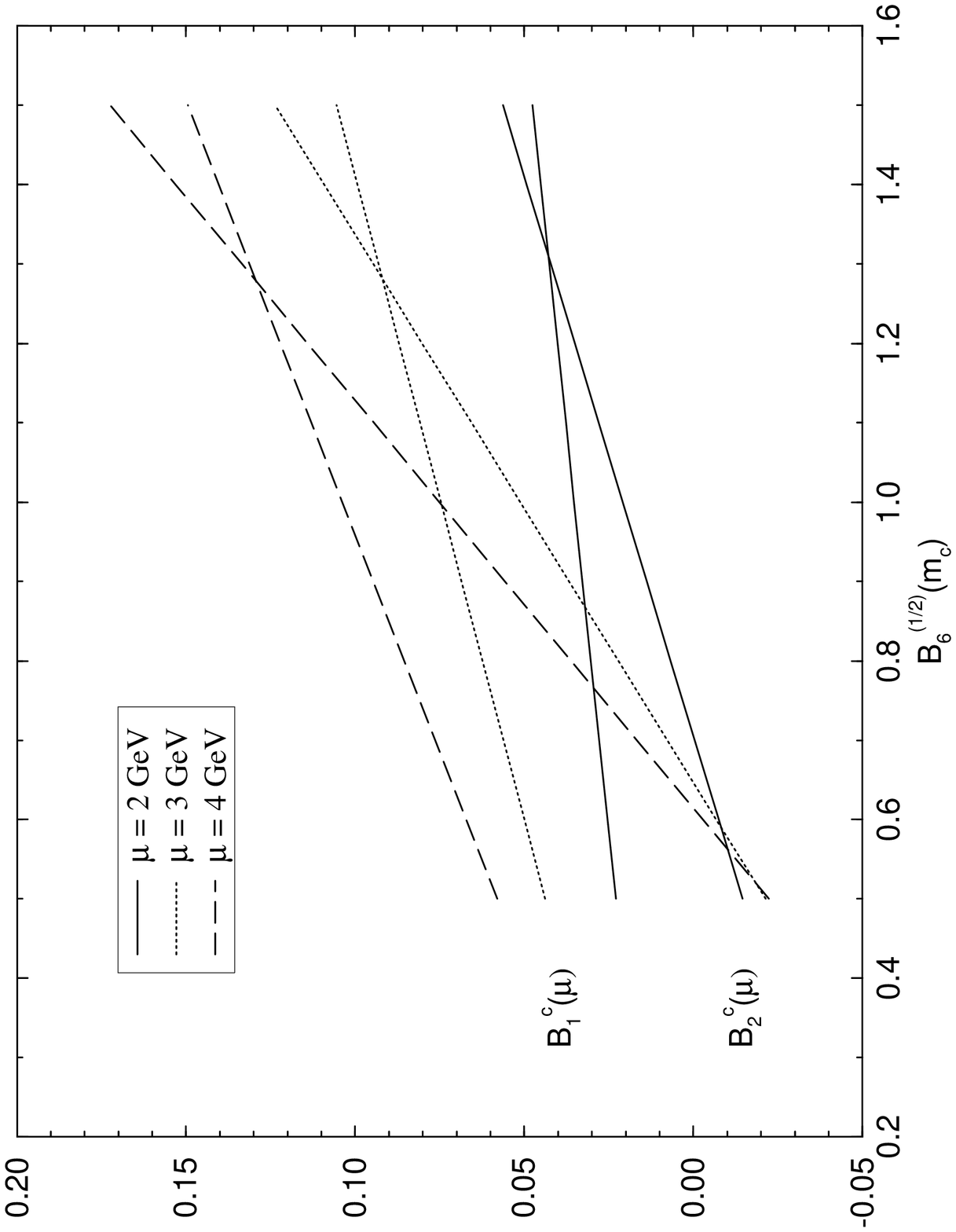}
}}
\vspace{0.15in}
\caption[]{
\label{fig:7}}
\end{figure}

\begin{figure}[h]
\vspace{0.15in}
\centerline{
\epsfysize=4.5in
\rotate[r]{
\epsffile{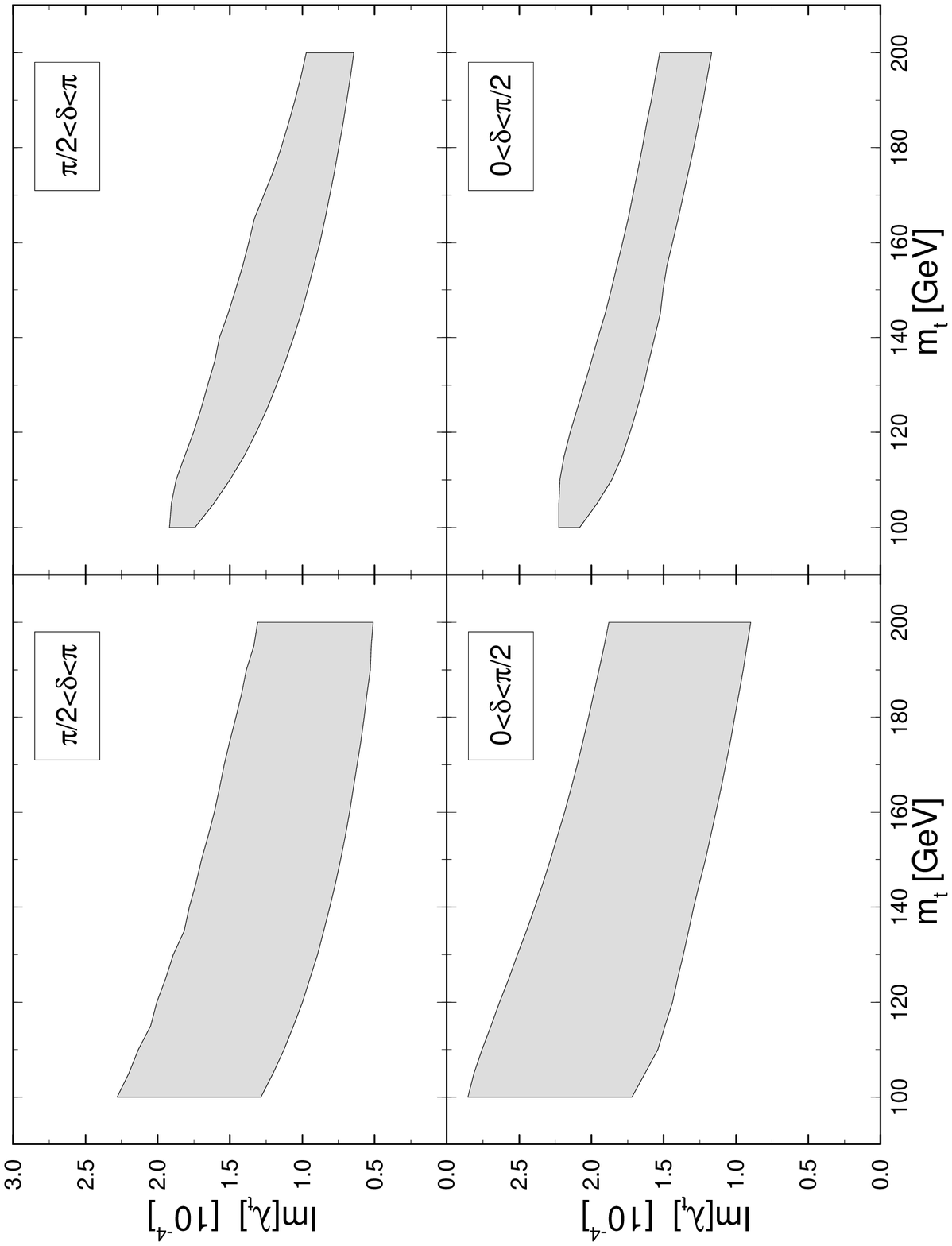}
}}
\vspace{0.15in}
\caption[]{
\label{fig:5}}
\end{figure}

\begin{figure}[h]
\vspace{0.15in}
\centerline{
\epsfysize=7in
\epsffile{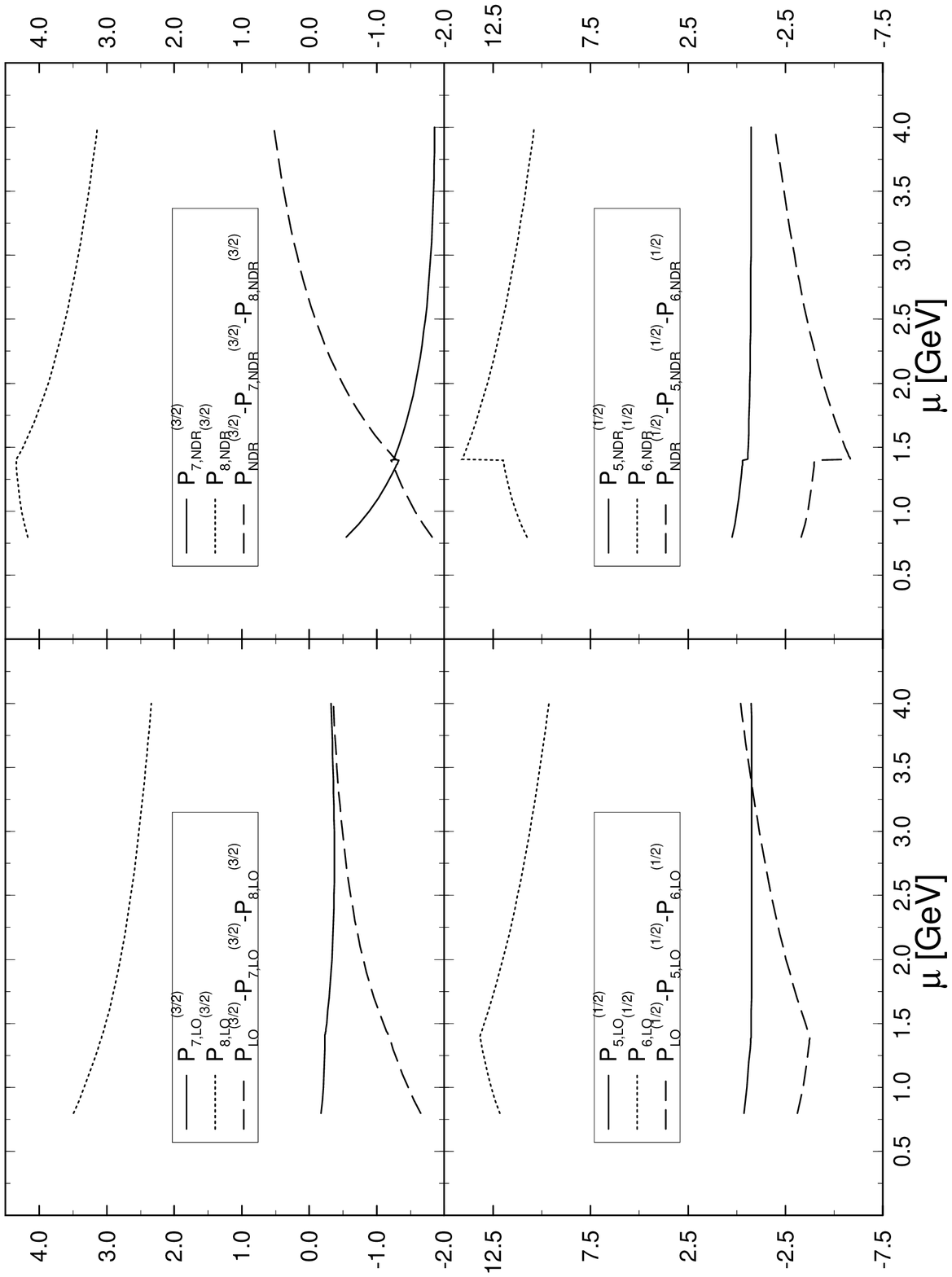}
}
\vspace{0.15in}
\caption[]{
\label{fig:16}}
\end{figure}

\clearpage

\begin{figure}[h]
\vspace{0.15in}
\centerline{
\epsfysize=7in
\rotate[r]{
\epsffile{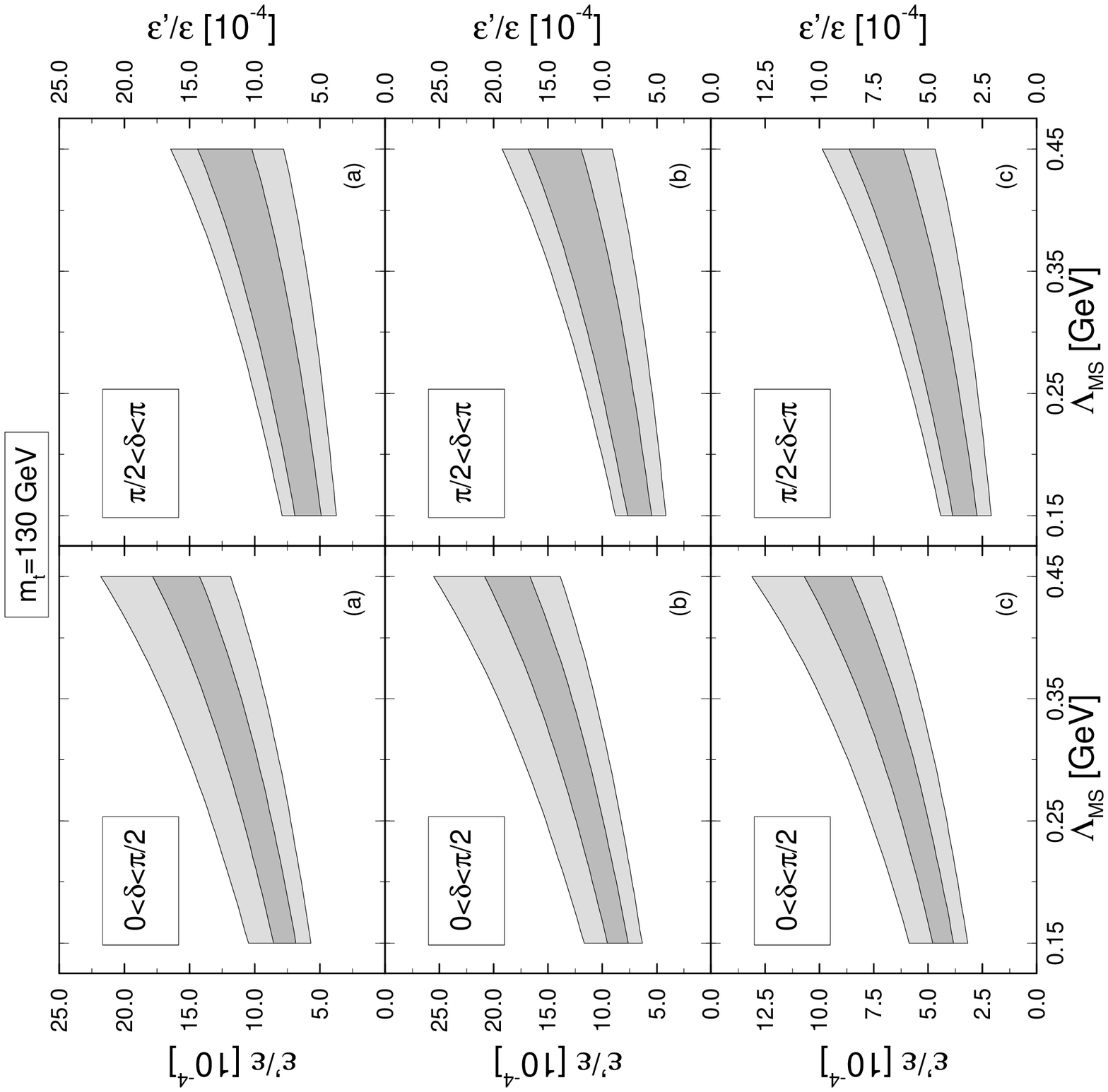}
}}
\vspace{0.15in}
\caption[]{
\label{fig:17}}
\end{figure}

\begin{figure}[h]
\vspace{0.15in}
\centerline{
\epsfysize=7in
\rotate[r]{
\epsffile{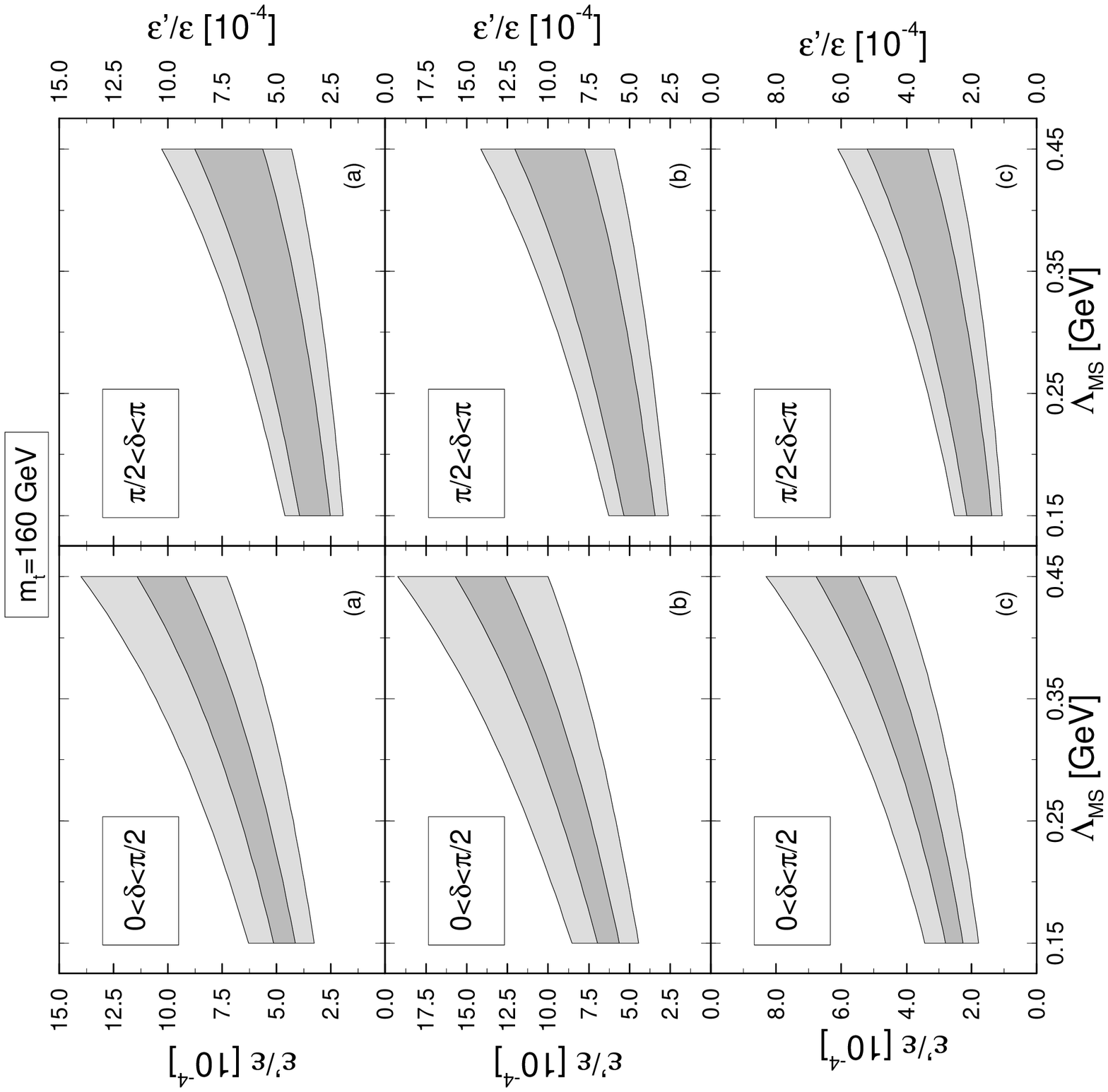}
}}
\vspace{0.15in}
\caption[]{
\label{fig:18}}
\end{figure}

\begin{figure}[h]
\vspace{0.15in}
\centerline{
\epsfysize=7in
\rotate[r]{
\epsffile{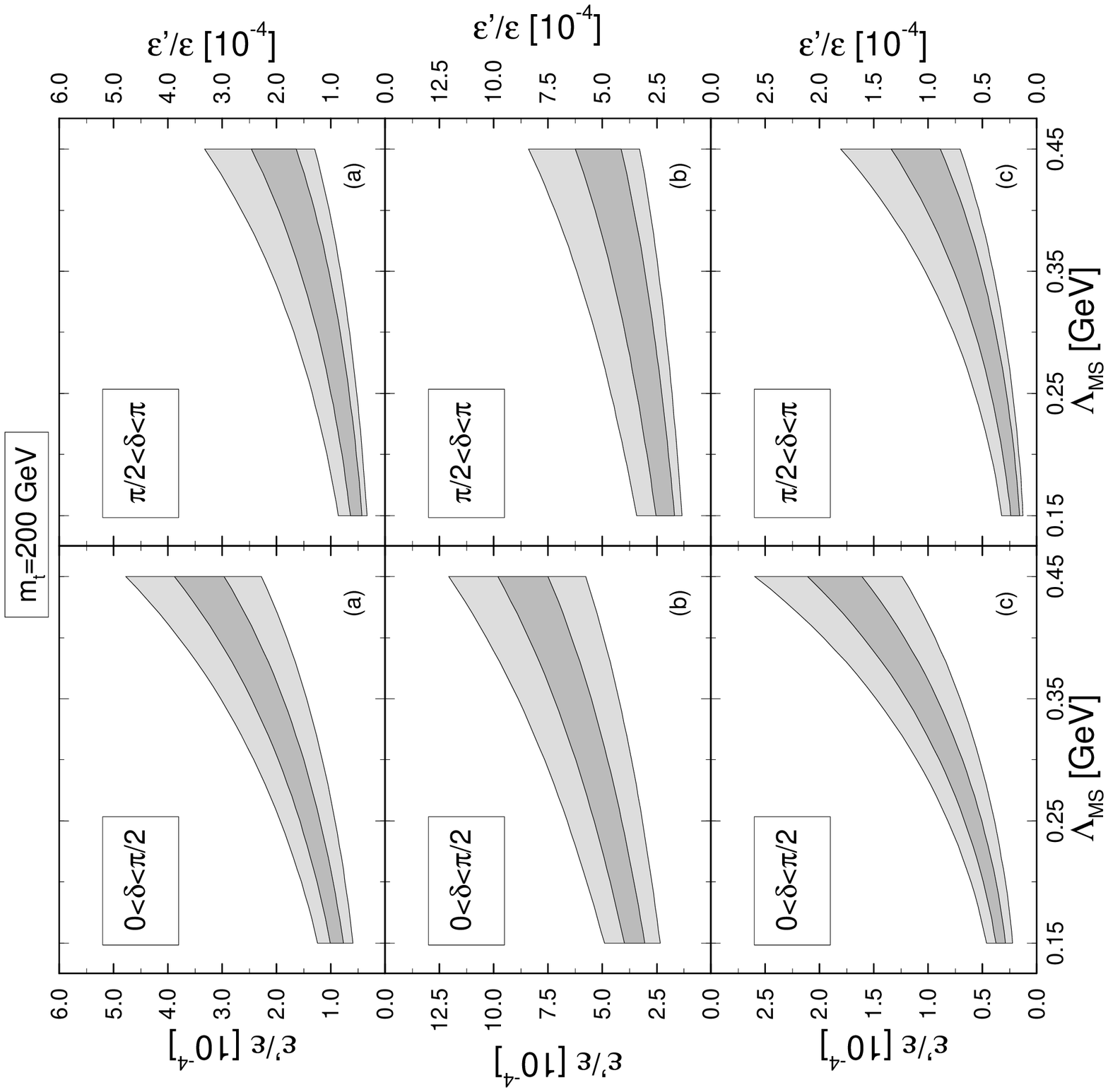}
}}
\vspace{0.15in}
\caption[]{
\label{fig:19}}
\end{figure}

\clearpage
\centerline{\Large\bf Figure Captions}

\bigskip

\begin{description}
\item[Figure~\ref{fig:1a}:]
Current-current and penguin 1--loop diagrams in the full theory. The
wavy lines labeled with ``W'' denote $W^\pm$ bosons, the unlabeled wavy
lines denote either a gluon or a photon.

\item[Figure~\ref{fig:1b}:]
Current-current and penguin 1--loop diagrams in the effective theory. The
unlabeled wavy lines denote either a gluon or a photon.

\item[Figure~\ref{fig:1}:]
$z_\pm(\mu)$ for various schemes and $\Lms=300\mev$.

\item[Figure~\ref{fig:2}:]
$y_6(\mu)$ and $y_8(\mu)/\aem$ for various schemes and $\Lms=300\mev$.

\item[Figure~\ref{fig:3}:]
$y_7(\mc)/\aem$ and $y_8(\mc)/\aem$ as functions of $\mt$ for various
schemes and $\Lms=300\mev$.

\item[Figure~\ref{fig:4}:]
$y_9(\mc)/\aem$ and $y_{10}(\mc)/\aem$ as functions of $\mt$ for various
schemes and $\Lms=300\mev$.

\item[Figure~\ref{fig:12}:]
$B^{(3/2)}(\mu)$ entering $\langle Q_{1,2}(\mu) \rangle_2$ extracted
from $\RE A_2$ for different schemes, $\Lms=300\mev$ and different
cases discussed in the text.

\item[Figure~\ref{fig:13}:]
$B^{(3/2)}(\mu)$ entering $\langle Q_{9,10}(\mu) \rangle_2$ extracted
from $\RE A_2$ for different schemes, $\Lms=300\mev$ and different
cases discussed in the text.

\item[Figure~\ref{fig:6}:]
$B_1^{(1/2)}(\mc)$ as a function of $B_2^{(1/2)}(\mc)$ for $\Lms=300\mev$
and various schemes.

\item[Figure~\ref{fig:14}:]
$B_4^{(1/2)}$ as a function of $B_2^{(1/2)}(\mc)$ for various values of
$B_3^{(1/2)}(\mc)$.

\item[Figure~\ref{fig:9}:]
$B_6^{(1/2)}(\mu)$ for different schemes, $\Lms=300\mev$ and common
normalization $B_6^{(1/2)}(\mc)=1$.

\item[Figure~\ref{fig:11}:]
$B_8^{(1/2)}(\mu)$ for different schemes, $\Lms=300\mev$ and common
normalization $B_8^{(1/2)}(\mc)=1$.

\item[Figure~\ref{fig:8}:]
$B_5^{(1/2)}(\mu)$ for different schemes, $\Lms=300\mev$ and common
normalization $B_5^{(1/2)}(\mc)=1$.

\item[Figure~\ref{fig:10}:]
$B_7^{(1/2)}(\mu)$ for different schemes, $\Lms=300\mev$ and common
normalization $B_7^{(1/2)}(\mc)=1$.


\item[Figure~\ref{fig:7}:]
$B^c_{1,2}(\mu)$ as a function of $B_6^{(1/2)}(\mc)$ in the HV scheme
for $\Lms=300\mev$.

\item[Figure~\ref{fig:5}:]
The ranges of $\IM \lambda_{\rm t}$ as a function of $\mt$ for parameter
ranges I (left) and II (right) given in appendix~C.

\item[Figure~\ref{fig:16}:]
The most important contributions to $P^{(1/2)}$ and $P^{(3/2)}$ as a
function of $\mu$ for LO and the $\ndr$ scheme. $\mt$, $\Lms$ and the
various hadronic parameters $B_i$ were taken at their central value
given in appendix~C.

\item[Figure~\ref{fig:17}:]
The ranges of $\epe$ in the $\ndr$ scheme as a function of $\Lms$ for
$\mt=130\gev$ and parameter ranges I (light grey) and II (dark grey)
given in appendix~C. The three pairs of $\epe$ plots correspond to
hadronic parameters (a) $B_6^{(1/2)}(\mc)=B_8^{(3/2)}(\mc)=1.5$, (b)
$B_6^{(1/2)}(\mc)=1.5$, $B_8^{(3/2)}(\mc)=1$ and (c)
$B_6^{(1/2)}(\mc)=B_8^{(3/2)}(\mc)=1$, respectively.

\item[Figure~\ref{fig:18}:]
Same as fig.~\ref{fig:17} but for $\mt=160\gev$.

\item[Figure~\ref{fig:19}:]
Same as fig.~\ref{fig:17} but for $\mt=200\gev$.
\end{description}

\end{document}